\documentclass[aps,prb,amsmath,amsfonts,floatfix,superscriptaddress,nofootinbib,twocolumn]{revtex4-2} 

\usepackage{mathrsfs} \usepackage{amssymb} \usepackage{graphicx,color} \usepackage{bbm} \usepackage{multirow}
\usepackage{bm} \usepackage{mathrsfs} \usepackage{commath} \usepackage{stmaryrd} \usepackage{color} \usepackage{stmaryrd}
\usepackage{tikz}\usetikzlibrary{shapes,arrows,positioning,calc}\usepackage{comment}\usepackage{dsfont}\usepackage{lpic}
\usepackage{cancel}\usepackage{braket} 
\usepackage{hyperref}
\hypersetup{
    colorlinks,
    linkcolor={blue!70!black},
    citecolor={green!70!black},
    urlcolor={red!70!black}
}

\usepackage{tikz}\usetikzlibrary{shapes,arrows,positioning,calc,shapes.misc}
\usepackage{comment}\usepackage{bbold}
\usepackage{esint}\usepackage{upgreek} 

\tikzset{cross/.style={cross out, draw=black, minimum size=2*(#1-\pgflinewidth), inner sep=0pt, outer sep=0pt},
cross/.default={1pt}}

\let\a=\alpha \let\b=\beta \let\g=\gamma \let\d=\delta
 \let\z=\zeta  \let\k=\kappa
\let\l=\lambda \let\m=\mu \let\n=\nu  \let\p=\pi
\let\s=\sigma   \let\c=\chi
   \let\G=\Gamma
\let\D=\Delta   
   
 \let\r=\rho \let\th=\theta \let\io=\infty
\let\om=\omega
\def\ie{{\textit{i.e.} }}\def\eg{{\textit{e.g.} }}

\def\MM{{\cal M}} 
\def\CC{{\cal C}}\def\FF{{\cal F}} 
\def\TT{{\cal T}} 
  \def \OO{{\cal O}}
\def\DD{{\cal D}}\def\AA{{\cal A}}\def\GG{{\cal G}} \def\SS{{\cal S}}
  
\def\ZZ{{\cal Z}}

\def\Im{{\rm Im}\,}

\def\to{\rightarrow} \def\la{\left\langle} \def\ra{\right\rangle}
\def\RRR{\mathbbm{R}}
\def\ZZZ{\mathbb{Z}}

\def\dd{\mathrm{d}}
\def\id{\mathbbm{1}}
\def\Tr{\mathrm{Tr}}
\def\dd{\mathrm{d}}
\def\sign{\textrm{sign}}

\def\restriction#1#2{\mathchoice
              {\setbox1\hbox{${\displaystyle #1}_{\scriptstyle #2}$}
              \restrictionaux{#1}{#2}}
              {\setbox1\hbox{${\textstyle #1}_{\scriptstyle #2}$}
              \restrictionaux{#1}{#2}}
              {\setbox1\hbox{${\scriptstyle #1}_{\scriptscriptstyle #2}$}
              \restrictionaux{#1}{#2}}
              {\setbox1\hbox{${\scriptscriptstyle #1}_{\scriptscriptstyle #2}$}
              \restrictionaux{#1}{#2}}}
\def\restrictionaux#1#2{{#1\,\smash{\vrule height .8\ht1 depth .85\dp1}}_{\,#2}}

\newcommand{\beq}{\begin{equation}} \newcommand{\eeq}{\end{equation}}
\newcommand{\wh}{\widehat} \newcommand{\wt}{\widetilde}

\newcommand{\argc}[1]{\left[#1\right]}
\newcommand{\arga}[1]{\left\lbrace #1\right\rbrace }
\newcommand{\argp}[1]{\left(#1\right)}

\newcommand{\moy}[1]{\left\langle  #1 \right\rangle }

\def\ux{\underline{x}}
\def\uxu{\underline{u}}

\newcommand{\blue}[1]{\textcolor{blue}{#1}}
\newcommand{\red}[1]{\textcolor{red}{#1}}
\newcommand{\green}[1]{\textcolor{green!70!black}{#1}}
\newcommand{\orange}[1]{\textcolor{orange}{#1}}

\begin{document}

\title{Two-level systems and harmonic excitations in a mean-field anharmonic quantum glass} 

\author{Thibaud Maimbourg}
\affiliation{Universit\'e Paris-Saclay, CNRS, CEA, Institut de physique th\'eorique, 91191, Gif-sur-Yvette, France}

\begin{abstract}
Structural glasses display at low temperature a set of anomalies in thermodynamic observables. A prominent example is the linear-in-temperature scaling of the specific heat, at odds with the Debye cubic scaling found in crystals, due to acoustic phonons. Such an excess of specific heat in amorphous solids is thought of arising from phenomenological soft excitations dubbed tunneling two-level systems (TTLS). Their nature as well as their statistical properties remain elusive from a first-principle viewpoint. In this work we investigate the canonically quantized version of the KHGPS model, a mean-field glass model of coupled anharmonic oscillators, across its phase diagram, with an emphasis on the specific heat. The thermodynamics is solved in a semiclassical expansion. We show that in the replica-symmetric region of the model, up to the marginal glass transition line where replica symmetry gets continuously broken, a disordered version of the Debye approximation holds: the specific heat is dominated by harmonic vibrational excitations inducing a power-law scaling at the transition, ruled by random matrix theory. This mechanism generalizes a previous semiclassical argument in the literature. We then study the marginal glass phase where the semiclassical expansion becomes non-perturbative due to the emergence of instantons that overcome disordered Debye behavior. Inside the glass phase, a variational solution to the instanton approach provides the prevailing excitations as TTLS, which generate a linear specific heat. This phase thus hosts a mix of TTLS and harmonic excitations generated by interactions. 
We finally suggest to go beyond the variational approximation through an analogy with the spin-boson model.
\end{abstract}

\maketitle
\tableofcontents

\section{Introduction}\label{sec:intro}

Uncovering the low-energy excitations of a system is essential to understand its low-temperature physics, as they represent the dominant degrees of freedom. A celebrated example is the success of Debye's theory of crystals~\cite{De12,kittel,AM}, which, by quantizing harmonic vibrations of the lattice, correctly identified  acoustic phonons as the relevant excitations. 
 Such excitations govern the low-temperature behavior of many thermodynamic quantities, such as the cubic temperature dependence of the specific heat $C_V$ and the thermal conductivity $k$. The phonon wavelength being much larger than the lattice spacing, this mechanism is independent of the specific structure of the lattice, hence it is universal. 

One could expect this universality to extend to  amorphous solids (\ie with disordered structure), as acoustic phonons are still present as Goldstone modes stemming from statistical translation invariance on large lengthscales. Experimental measurements by Zeller and Pohl~\cite{ZP71} more than fifty years ago thus came as a surprise, revealing  that the specific heat and the thermal conductivity behave respectively as $C_V\sim T$ and $k\sim T^2$ below 1 K. Many other dielectric, acoustic or thermal properties of glassy materials exhibit remarkable universality, independent of details like chemical composition~\cite{Ph87,Esquinazi,Ramos}. For instance the internal friction (related to sound attenuation) becomes frequency independent and unexpectedly large. This universal `anomalous' (as compared to crystals) character led to the idea that it arises from the amorphous nature itself, which would contain a distinct type of low-energy excitations, besides phonons. Right after Zeller and Pohl's measurements, the standard tunneling model (STM) was developed~\cite{AHV72,Ph72} by considering tunneling two-level systems (TTLS), phenomenological degrees of freedom thought of being single or groups of particles -- electrons, atoms, molecules... -- that live in a metastable state and have another quasi-degenerate configuration to which they can tunnel. Schematically, these fictitious degrees of freedom should effectively lie in a double-well potential (DWP). Postulating the distributions of the relevant parameters controlling the double-well shape allowed to recover many of the aforementioned universal properties~\cite{Ph87,Esquinazi,Ramos}. 

Over the decades many refinements of the STM were proposed~\cite{Esquinazi,Ramos}, mainly to constrain the statistical properties of  the TTLS effective degrees of freedom. The low-temperature scale of 1 K  provides a stringent upper bound for the asymmetry between the two wells of the effective potential, and given the disordered structure, one naturally expects a large number of effective
 degrees of freedom described  by such a potential, dispersed in the random matrix of the amorphous solid. This naturally yields a very diverse distribution of effective potentials, including both DWP and single-well potentials (SWP). SWP turned out to be the vast majority of effective potentials, generating additional soft harmonic excitations \cite{kumar2021density}. 
An extension of the STM based on this idea, more capable to quantitatively adapt to experimental data notably at higher temperatures, was devised through the Soft Potential Model (SPM)~\cite{KKI82,IKP87}. It describes low-energy modes in glasses by a collection of \textit{independent} soft anharmonic oscillators ${v(x)\propto h x+\eta x^2+x^4}$ with an ad-hoc distribution $P(h,\eta)$~\cite{Pa94,RB98}.

Despite the great achievements of these models, they fail to explain a number of outstanding issues~\cite[Chap.4]{Ramos}. 
First, the microscopic nature of the TTLS remains a mystery, except in a few
cases such as $\textrm{(KBr)}_{(1-x)}\textrm{(KCN)}_x$ or $\textrm{KBr}_{(1-x)}\textrm{(CN)}_x$ in which they are thought to be cyanide (CN) molecules~\cite{DYKMP86,MDA84}. 
Computer simulations dealing with microscopic modeling have been used to reveal them but the low density of TTLS  requires a large number of atoms, a daunting numerical challenge which began forty years ago~\cite{He98}. The most recent simulations~\cite{KSBRZ20,MBCKRSZ22,CKBMRSZ23} have found TTLS candidates in double-well potentials formed by pairs of minima of the energy landscape, with reasonable energy scale. 
 These very localized transitions would involve a few atoms, although more rarely, in samples quenched from high temperature, up to hundreds of atoms would be involved in such a tunneling. 
Note that direct experimental evidence of actual tunneling motion is still lacking~\cite{YL88,LV13}. Finally, TTLS couple to the strain field in the glassy structure and can interact with each other by exchanging phonons. This interaction was argued to be significant~\cite{YL88} and a great number of subsequent experimental works~\cite{SEH98,NRO98,CBEH00,LCPTA03} confirmed it in acoustic and dielectric properties, especially below the 100 mK range, along with  theoretical developments~\cite{EH97,BNOK98,Wu98}. 
Crucially, the strain field depends on the amorphous structure, which itself gives rise to the density and properties of TTLS. Therefore interactions are not only important for the dynamics of TTLS but essential to understand their nature. 
Qualitative changes such as collective effects are expected as a consequence. 

Given the complexity of glassy phenomena, minimal models have been designed over the decades, with the advantage of being amenable to an analytical solution in the mean-field limit, where interactions can be accounted for~\cite{CM22}. The main observable under scrutiny as a hallmark quantity of low-temperature anomalies has been the specific heat, since unlike most of the other observables it is an internal property of the system, independent on further modeling of external degrees of freedom. The first mean-field models in which the specific heat in the quantum regime has been investigated came from the spin-glass literature, hoping to get insights into the elusive TTLS physics from a first-principle approach. Oddly, they do not exhibit a linear specific heat. The Sherrington-Kirkpatrick  and multicomponent quantum rotor models in a transverse field were analyzed close to the quantum critical point~\cite{YSR93,MH93,RSY95} and yielded, approaching the transition from the quantum paramagnetic side,  scalings consistent with a specific heat $C\sim T^3$.
However these negative results were reconsidered in~\cite{GPS01}, arguing that replica-symmetry breaking (RSB), that characterizes the glassy phase in these models, requires to consider additional dangerously irrelevant terms in the Landau free energy close to the critical point. These induce a linear specific heat, in agreement with a similar behavior obtained in the bosonic SU$(M)$ Heisenberg spin glass~\cite{GPS00,GPS01} and the quantum spherical $p$-spin~\cite{CGSS01}. But soon after, Schehr, Giamarchi and Le Doussal (SGLD) analyzed a set of spin-glass models~\cite{SGLD04,SGLD05,S05}. They showed that the marginal stability originating from RSB imposes in fact that the prefactor of the linear term in the low-temperature expansion of the specific heat vanishes, therefore retrieving the cubic scaling in these models. This is compatible with a numerical solution of the SU$(2)$ Heisenberg spin glass~\cite{KMGP23} and a spin spectral density linear in frequency obtained from a $1/M$ expansion of the fermionic SU$(M)$ case~\cite{CHS22}. Note however that an unusual $C\sim T^2$ was found earlier numerically in the bosonic SU$(M\to\io)$ model~\cite{CR03}.
Nevertheless, the models considered in~\cite{SGLD04,SGLD05,S05} are rather far from the original context of the TTLS picture: they are models for quantum degrees of freedom moving in a random high-dimensional landscape.

 In this respect, the study~\cite{AM12} of the glass phase of the Sherrington-Kirkpatrick model in a transverse field (TFSK) could be more appealing (see~\cite{TSS23,KZL24} for a broader parameter range in the phase diagram). Within a Debye approximation of the Thouless-Anderson-Palmer free energy~\cite{TAP77,BC01} it however suggests the same cubic scaling~\cite{CM22}, at variance with the TTLS mechanism. This is linked to the linear-in-frequency spin susceptibility found in both~\cite{AM12,KZL24}. 
 
 Concerning structural glass models, so far only a surrogate mean-field model, the spherical perceptron, has been investigated in the quantum regime, resorting to the SGLD expansion~\cite{FMPS19}. It describes a tracer moving in a quenched liquid, with an emphasis on jamming physics~\cite{FPUZ15,FP16,FPSUZ17,ABPS21}. As we shall see this model has a structure of the impurity problem which does not allow TTLS excitations. Yet a linear scaling of the specific heat is present but argued to be cut off at very low temperatures. Its mechanism is instead rooted in the proximity of a jamming transition and the ensuing modes stemming from isostaticity~\cite{FMPS19}. 
 
To get closer to the TTLS physics, in a series of works K\"uhn and collaborators built a mean-field model inspired from the SPM adding quenched spin-glass interactions $J_{ij}x_ix_j$, where the degrees of freedom $\arga{x_i}$ may be regarded as fluctuations of positions around a local equilibrium, expanded up to quartic order~\cite{KH97,Ku00,Ku03}. The classical model allows to derive an effective single-site problem which is \textit{assumed} to give access to the statistical distribution
of TTLS. Quantum fluctuations are added a posteriori by quantizing this single effective degree of freedom. While microscopic details and distributions of relevant parameters differ, the ensuing phenomenology is akin to the SPM. In~\cite{BK10} instead, the full quantum thermodynamics was analyzed through the so-called static    approximation~\cite{BM80b} and subsequent field-theoretical approximations. In this way it is not clear whether the linear specific heat is recovered. 

Nonetheless, from the classical viewpoint, low-energy excitations in the K\"uhn-Horstmann model~\cite{KH97} were shown to be delocalized vibrational modes with a low-frequency density of states $D(\om)\sim\om^2$~\cite{BLRUZ21}, typical of the above-mentioned mean-field spin-glass models~\cite{SYM16} and of a certain class of mean-field structural glasses~\cite{DGLFDLW14,FPUZ15,CCPPZ16,SMBI20,IS22,MZ22}. This spectrum misses, apart from trivial acoustic phonons which are also delocalized with $D(\om)\sim\om^{d-1}$ ($d$ being the space dimension) and purposely absent from quenched mean-field models, localized vibrational modes ruled by the scaling $D(\om)\sim\om^4$ found in numerical simulations~\cite{LS91,LDB16,MSI17,KBL18,WNGBSF19,RGLKPVBL20,BGMPZ20,DP21,LB21,GBPZ22} (although some debate persists~\cite{WSF21,WFN22,SPMKSZR23}). The situation is similar in soft spin glasses, see \eg~\cite{BJMMPPG15,SYM16}. An appealing minimal argument for such a scaling was given in~\cite{GC03}. 
 
 A mechanism for the emergence of these low-energy modes was put forward by Gurevich, Parshin and Schober (GPS)~\cite{GPS03,PSG07,DHLP20}. They proposed a glass model in terms of anharmonic oscillators 
interacting with each other through the surrounding elastic medium (the glass). The interaction decays with distance $r$ as $r^{-d}$. The combined effects of the interaction and the anharmonicity induce an instability characterized by new frequencies and minima of the oscillators. The instability affects the density of states, which is shown both numerically and through phenomenological arguments to reproduce the scaling $D(\om) \sim \om^4$. 

For many years it was thus believed that mean-field models cannot develop such localized excitations~\cite{DGLFDLW14,FPUZ15,CCPPZ16,SMBI20,IS22,MZ22,Ik19}, as one may expect fully-connected models to generate only delocalized modes. The situation changed recently: building on the ideas of GPS and K\"uhn and Horstmann, the KHGPS model~\cite{RUZLB21,BLRUZ21}, was put forward and solved in the zero-temperature classical regime. The model features two qualitatively different glassy phases described by RSB. The first corresponds to a usual mean-field glassy  phase created by a de Almeida-Thouless instability~\cite{AT78,MPV87} where the spin-glass susceptibility diverges. It hosts delocalized modes with density of states $D(\om) \sim \om^2$ and displays  single-site effective potentials consisting in SWP only. The second is induced by a different type of RSB with finite spin-glass susceptibility at the transition, which rules out the SGLD mechanism for the specific heat. The low-frequency vibrations are \textit{localized} with $D(\om) \sim \om^4$. Importantly, although the bare potentials of the effective degrees of freedom are all SWP, the single-site effective potentials are either SWP or DWP. In other words, double wells appear as an effect of the interactions, that destabilize soft enough SWP. 
SWP bring soft vibrational modes while DWP produce pseudogapped non-linear excitations~\cite{FU22}. Interestingly the KHGPS model possesses a spin-glass transition in a field at $T=0$, a much desired feature to set up a renormalization group analysis of the fate of such transitions in finite dimension~\cite{Ur22}. 
This model could also be useful to describe a Gardner transition in molecular glasses (if it exists), as~\cite{ABLTU21} proposed it would be described by a spin-glass one arising from the interaction between local excitations. 

The purpose of this work is to investigate the effect of quantum fluctuations in the KHGPS model across the different phases, with a particular emphasis on the RSB phase driven by DWP. In~\secref{sec:qtd} we derive the generic quantum-thermodynamical equations of the model. We specialize them in the convex replica-symmetric phase in~\secref{sec:semi} to solve them in a semiclassical expansion $\hbar\to0$ keeping $\b\hbar$ fixed. Then in~\secref{sec:DWP} we focus analytically and numerically on the regime where DWP populate the impurity problem, through the same semiclassical expansion, solved by a variational method. In~\secref{sec:pseudogap} we unveil the effect of replica-symmetry breaking on the latter variational results. Finally in~\secref{sec:spinboson} we outline a way beyond these variational results through an analogy with the spin-boson model. We draw our conclusions in~\secref{sec:conclu}. 

\begin{figure}[h!]
\centering
 \begin{lpic}[]{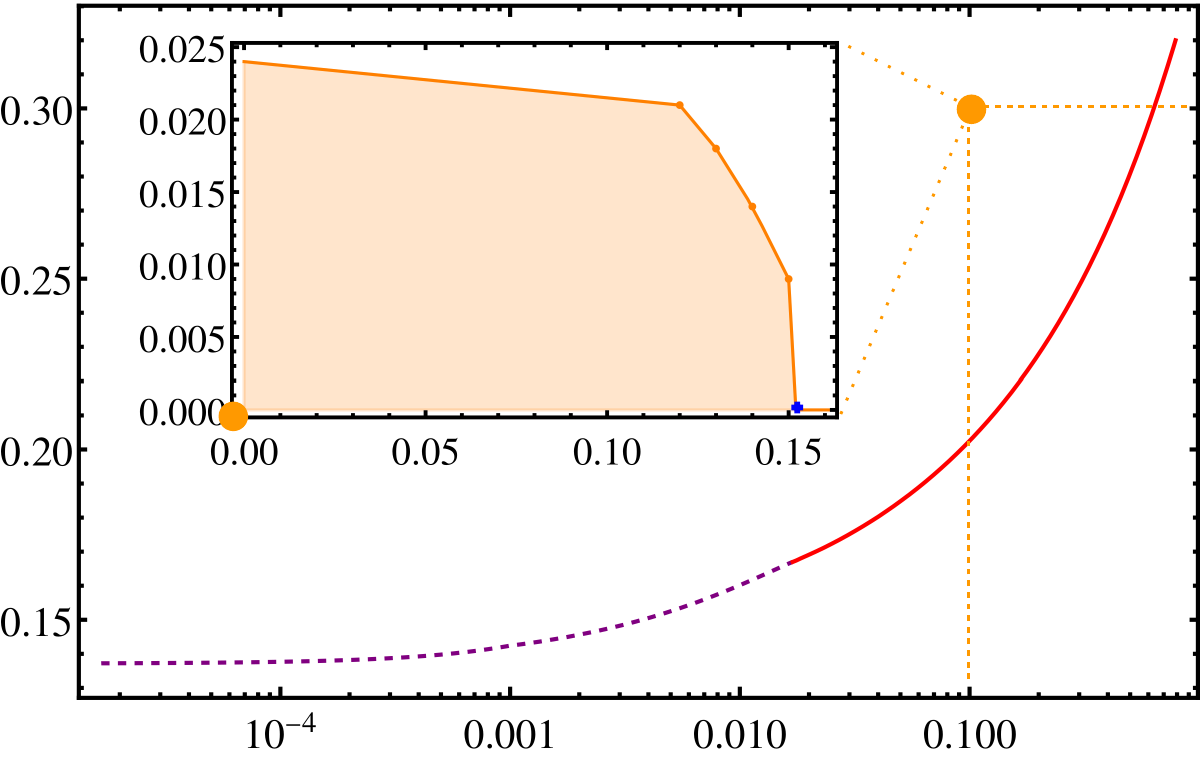(0.4)}
\lbl[]{-8,65;$J$}
\lbl[]{107,-5;$h$}
\lbl[]{90,50;$\hbar$}
\lbl[]{160,30,15;{\bf RS (convex) phase}}
\lbl[]{90,40,3;{\bf FullRSB (glassy) phase}}
\lbl[]{70,23,5;{\color{violet} RSB-$\om^2$ line}}
\lbl[]{180,90,65;{\color{red} RSB-$\om^4$ line}}
\lbl[]{90,80;{\bf \textcolor{orange}{RSB}}}
\lbl[]{132,110;{\bf RS}}
\lbl[]{18,90;$T$}
\end{lpic}
 \caption{Phase diagram of the KHGPS model. (main) Classical zero-temperature phase diagram~\cite{BLRUZ21}. At large enough interaction $J$ there is a transition from a convex phase described by a replica-symmetric ansatz for the order parameter $Q_{ab}$ to a fullRSB glassy phase. At low enough external field $h$ the transition is characterized by a diverging spin-glass susceptibility. The low-energy vibrational excitations are obtained from the spectrum of the Hessian of the Hamiltonian in the energy minimum,  $\partial^2H/\partial x_i\partial x_j$. At the transition the spectrum becomes gapless and displays a low-frequency density of variational modes scaling as $D(\om)\sim\om^2$. Hereafter we thus refer to the purple dashed line as the RSB-$\om^2$ transition line. Instead at large enough $h$ the transition changes nature: the spin-glass susceptibility is finite and the Hessian spectrum scales as $D(\om)\sim\om^4$. It is represented by the red solid line hereafter called RSB-$\om^4$ line. In the RS phase and above the RSB-$\om^2$ line, the effective potential $v_{\k,z}(x)$ has a single minimum, while above the RSB-$\om^4$ line it develops a double-well shape. (inset) Quantum phase diagram at $(J,h)=(0.3,0.1)$. The RSB transition ends in a quantum critical point at $T=0$ (blue cross). The line is a guide to the eye connecting the classical finite-temperature RSB transition~\cite{FU22} to the dots provided by the solution of the model within the low-temperature approximation of~\secref{sub:lowtemp}-\secref{sec:num}.  In both phase diagrams (main and inset), the orange dot signals the point $(J,h,T,\hbar)=(0.3,0.1,0,0)$.}
 \label{fig:qphd}
\end{figure}
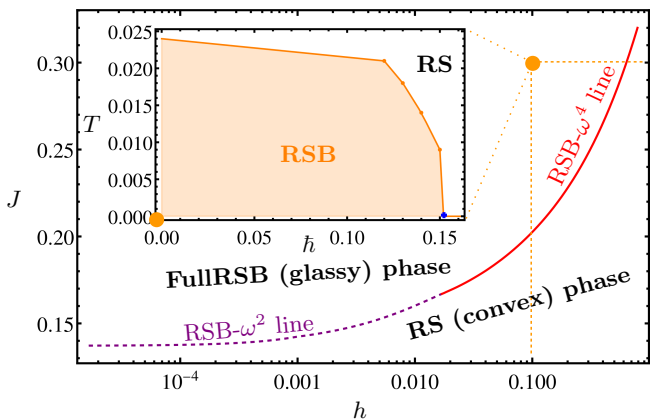

\section{Quantum thermodynamics of the KHGPS model}\label{sec:qtd}

The version of the classical KHGPS model that we consider in this work was introduced in~\cite{BLRUZ21}. It consists in a set of $N$ soft spins, whose coordinates are denoted by $x_i\in \RRR$, being $i=1,\ldots, N$ an index identifying the degrees of freedom. These spins are subjected to a local anharmonic quartic potential and interact with each other with all-to-all random quadratic interactions of the spin glass form. 

We quantize the model by adding conjugate momenta $\arga{p_1,\dots,p_N}$ with the canonical commutation  relation ${\argc{\hat x_i,\hat p_j}=i\hbar\d_{ij}}$. Thus the Hamiltonian of the quantum KHGPS model reads
\begin{equation}\label{eq:H}
\begin{split}
 \hat H = &\sum_{i=1}^N\frac{\hat p_i^2}{2M}+\sum_{i<j}^{1,N}J_{ij}\hat x_i\hat x_j+\sum_{i=1}^Nv_{\k_i}(\hat x_i)\\
v_\k(x)=&\frac{\k}{2}x^2+\frac{x^4}{4!}-h x
\end{split}
\end{equation}
with independent and identically distributed interaction couplings $\overline{J_{ij}}=0$, $\overline{J_{ij}^2}=J^2/N$. The local bare potentials $v_{\k_i}(x_i)$ depend on a set of $N$ quenched random variables, $\k_i$, that are independent and drawn from a uniform distribution $p(\k)$ in $\argc{\k_m,\k_M}$ with $\k_m>0$. They are anharmonic and include a crucial magnetic field $h$ which breaks explicitly the $\ZZZ_2$ symmetry of the model. 
To compute thermodynamic quantities, we need to study the partition function of the model 
\beq
Z=\mathrm{Tr}\, e^{-\beta H}
\eeq
with $\b=1/T$ the inverse temperature (we take $k_B=1$ throughout) as well as the corresponding free energy
\beq
F = -T\lim_{N\to \infty} \overline{\log Z}\:.
\eeq
The overline denotes the average over all sources of disorder, namely the random couplings $J_{ij}$ and the elastic constants $\k_i$. 
In order to analyze the free energy we shall follow closely the methods of~\cite{FMPS19}. The thermodynamics of the system is derived from the Feynman representation of the partition function, which is a path integral over periodic trajectories of the particles $x_i(t)$, being $t$ the imaginary time with Matsubara period $\b\hbar$~\cite{kleinert}.  To easily keep track of $\b$ and $\hbar$ dependences, we use $\b\hbar$ as unit of time and therefore the Matsubara imaginary times are scaled as $t/(\b\hbar)\to t$ in the following expressions, unless otherwise mentioned. In this section we derive the disorder-averaged free energy of the system in general RSB phases and then specialize to the replica-symmetric (RS) case.

\subsection{Replica analysis of the free energy}\label{sub:fRSB}

The bottleneck to compute the free energy of the model lies in the average over the disorder, which we overcome through the replica method. We introduce $n$ replicas of the system and compute the average of the replicated partition function. Similarly to~\cite{BM80b,BLRUZ21,FMPS19}, in the large $N$ limit, one can show that this replicated path integral reads $\overline{Z^n}=\oint \mathrm{D} Q\, e^{N\SS(Q)}$ with
\begin{equation}
 \begin{split}\label{rep_free_energy}
  \SS(Q)=&-\argp{\frac{\b J}{2}}^2\int_0^1\dd t\dd s\sum_{a,b}Q_{ab}(t,s)^2+\ln\ZZ(Q)\\
  \ZZ(Q)=&\int\dd p(\k)\oint\prod_a\mathrm{D} x^a e^{\frac{(\b J)^2}{2}\int_0^1 \dd t\dd s\sum_{a,b}Q_{ab}(t,s)x^a(t)x^b(s)}\\
  &\qquad\times e^{-\b\int_0^1 \dd t\sum_a\argc{\frac{M}{2(\b\hbar)^2}\dot x^a(t)^2+v(x^a(t))}}\:.
 \end{split}
\end{equation}
The order parameter $Q_{ab}(t,s) = \sum_{i}\langle x_i^{a}(t)x_i^{b}(s)\rangle$ is the overlap between two replicas $a$ and $b$ of the system and the brackets denote the average with respect to the replicated path integral.
In the large $N$ limit the overlap concentrates and it is given as the solution of the saddle-point equation $Q_{ab}(t,s)=\moy{x^a(t)x^b(s)}_\ZZ$. The average is now defined by the single-particle generating functional $\ZZ$ defined in Eq.~\eqref{rep_free_energy}. Time-translational invariance and disorder average imply that only the diagonal part of $Q_{ab}(t,s)$ is actually time dependent and a function of $t-s$~\cite{CGSS01,FMPS19}. The integral symbol $\oint$ in Eq.~\eqref{rep_free_energy} reminds that the path integral is done over a closed periodic contour and that the dynamical variable $x(t)$ is periodic with Matsubara period equal to 1 in our time unit. 

Since we eventually take the limit $n\to 0$, we need to consider a sensible ansatz for the form of the solution of the saddle point equations. We therefore consider the following ansatz
\begin{equation}
 Q_{ab}(t,s)=q_d(t-s)\d_{ab}+Q^*_{ab}
\end{equation}
where $Q_{ab}^*$ is a hierarchical static matrix with zero elements on the diagonal, parametrized by a function $q(y)$ with $y\in\argc{0,1}$ \cite{MPV87}. This ansatz is the most general one capable of describing all phases of the model. In the RS phase, $q(y)$ is a constant while in the RSB phase one expects that $q(y)=q_m$ for $y\in\argc{0,y_m}$, $q(y)=q_M$ for $y\in\argc{y_M,1}$ and a non-trivial monotonously increasing curve for $y\in[y_m,y_M]$ \footnote{The shape of $q(y)$ depends on the form of RSB the model has. In \cite{FU22} it has been argued that at the classical level, $q(y)$ is also a continuous function for $y$ sufficiently close to $y_M$.}. 
The derivation and formalism is akin\footnote{For the quantum case a similar derivation is in~\cite[Secs.2-3]{FMPS19}; see also~\cite{AM12} for the related TFSK model.} to the one in~\cite{MPV87,PUZ20}. We get 
\begin{widetext}
\begin{equation}\label{eq:FAio}
\begin{split}
 -\frac{\b F}{N}=&\overbrace{\argp{\frac{\b J}{2}}^2\argc{\int_0^1\dd y\, q(y)^2-\int_0^1\dd t\, q_d(t)^2}}-\int\dd p(\k)\dd H\,P_\k(1,H)\argc{ f_\k(1,H)-\ln\oint\mathrm{D}x\,e^{\AA[x]}}\\
&+\int\dd p(\k)\dd y\dd H\,P_\k(y,H)\underbrace{\argc{\dot f_\k(y,H)+\frac{J^2}{2}\dot q(y)\argp{f''_\k(y,H)+x\argp{f'_\k(y,H)}^2}}}+\int\dd p(\k)\,e^{\frac{J^2}{2}q(0)\partial_h^2}f_\k(0,h)
\end{split}
\end{equation}
\end{widetext}
with the dynamical action
\begin{equation}\label{eq:actionAfull}
\begin{split}
 \AA[x]=&\frac{(\b J)^2}{2}\int_0^1\dd t\dd s\,x(t)G(t-s)x(s)\\
 &-\b\int_0^1\dd t\argc{\frac{M}{2(\b\hbar)^2}\left(\frac{\dd x}{\dd t}\right)^2+v_\k(x)}\:.
\end{split}
\end{equation}
In Eq.~\eqref{eq:FAio} we have indicated with dots the derivative with respect to $y$ and with primes the derivative with respect to the effective field $H$. We introduced the Lagrange multiplier $P_\k(x,H)$ to enforce the Parisi differential equation ruling $f_\k(y,H)$, given by the underbraced term (set to zero). This equation is thus obtained by optimizing over the Lagrange variable $P_\k(y,H)$. In turn, the variational equation over $f_\k(y,H)$ gives the equation for the Lagrange multiplier
\begin{equation}\label{eq:varP}
 \begin{split}
  \dot P_\k(y,H)=&\frac{J^2}{2}\dot q(y)\argc{P''_\k(y,H)-2y\argp{P_\k(y,H)f'_\k(y,H)}'}\\
  P_\k(0,H)=&\g_{J^2q(0)}(h-H)\equiv\frac{e^{-\frac{(h-H)^2}{2J^2q(0)}}}{\sqrt{2\p J^2q(0)}}
 \end{split}
\end{equation}where the last boundary condition is a Gaussian with variance $J^2q(0)$.  
Finally, in Eq.~\eqref{eq:actionAfull} we employed the definition
\begin{equation}\label{eq:G}
 G(t)=q_d(t-s)-q_M\:.
\end{equation}
Note that $G(t)=G(-t)$.
It is useful to define Fourier components:
\begin{equation}
  G(t)=\sum_{n\in\mathbbm Z}\wt{G}(\om_n) e^{i\om_n t} \ ,\ 
  \wt{G}(\om_n)=\int_0^{1}\dd t\,G(t)e^{-i\om_n t}\:.
\end{equation}
With our time unit the Matsubara frequencies are (unless stated otherwise) $\om_n=2\p n$.

The saddle-point equations on $q(x)$ and $q_d(t)$ read
\begin{equation}\label{eq:varq}
\begin{split}
  q(x) = &\int\dd p(\k)\dd H\, P_\k(x,H)\argc{\frac{f'_\k(x,H)}{\b}}^2\\
  q_d(t-s)=& \int\dd p(\k)\dd H\, P_\k(1,H)\moy{x(t)x(s)}\:.
\end{split}
\end{equation}
The thermal averages are defined through the effective action:
\begin{equation}\label{eq:aver}
 \moy{\bullet}\equiv\frac{\oint\mathrm{D}x\,\bullet e^{\AA[x]}}{\oint\mathrm{D}x\,e^{\AA[x]}}\:.
\end{equation}
 Importantly, $f_\k(1,H)$ may be regarded as the logarithm of an effective partition function for the single degree of freedom $x(t)$, that generates the effective average:
 \begin{equation}\label{eq:f1def}
 f_\k(1,h)=\ln\oint\mathrm{D}x\,e^{\AA[x]}
\end{equation}
also known as an impurity problem~\cite{GKKR96}. 

An important difference with respect to the standard spin-glass case~\cite{MPV87} is that the partial differential equations for $P_\k$ and $f_\k$ are stochastic, in the sense that they depend on the random variable $\k$ that is extracted with uniform measure in the support of $p(\k)$.

From these equations, one can derive the marginal stability condition (see \cite{MPV87, PUZ20,FMPS19}) that signals a RSB transition:
\begin{equation}\label{eq:replicon0}
 1= \argp{\frac J \b}^2 \int\dd p(\k)\dd H\,P_\k(1,H)f_\k''(1,H)^2\:.
\end{equation}
This condition is known as the vanishing of the \textit{replicon} eigenvalue $\l_R$ of the stability (Hessian) matrix in replica space. Stability only requires an inequality $\geqslant$ in~\eqref{eq:replicon0}.

When replica symmetry is preserved $q(y)$ is constant, which implies from~\eqref{eq:varP} that $P_\k(1,H)=P_\k(0,H)$ is Gaussian. Furthermore, if the bound in \eqref{eq:replicon0} is violated, the RS solution is no longer valid and one needs to integrate the coupled PDEs for $f_\k$ and $P_k$ to get the form of $q(y)$.
 
We now write the thermodynamic energy through~\eqref{eq:FAio}, using saddle-point relations (\ie the derivatives of variational quantities cancel, only explicit $\b$ derivatives are needed~\cite{CGSS01,FMPS19} -- noted `ex' below):
\begin{equation}\label{eq:UfullRSB}
\begin{split}
 \frac U N =&\frac{\b J^2}{2}\argc{\int_0^1\dd t\, q_d(t)^2-\int_0^1\dd x\, q(x)^2}\\
 &-\int\dd p(\k)\dd H\,P_\k(1,H)\restriction{\frac{\partial}{\partial \b}}{\rm ex}f_\k(1,H)\:.
\end{split}
\end{equation}
The latter expression is the starting point to analyze the specific heat at low temperature. 

\subsection{Replica-symmetric phase}

We start the analysis of the saddle-point solution of the RSB equations from the simplest case where the solution is replica symmetric. This amounts to specialize the above equations to the RS ansatz $q(x)=q_M=q$ \ie $Q_{ab}(t,s)=G(t-s)\d_{ab}+q$, which is at the basis of the developments of the next section. Using a Hubbard-Stratonovitch transformation, we introduce a Gaussian variable $z$ with average 0 and variance 1, whose measure is noted $\DD z$. The RS free energy is then, after $n\to0$,
\begin{equation}\label{eq:RSF}
\begin{split}
 -\frac{\b \FF_{\rm RS}}{N}=&\restriction{\frac{\partial_n\overline{Z^n}}{N}}{n\to0}=\argp{\frac{\b J}{2}}^2\argc{q^2-\int_0^1 q_d^2}\\
 &+\int\dd p(\k)\DD z\,\ln\oint\mathrm{D}x\, e^{\AA_{\k,z}[x]}\\
 \AA_{\k,z}[x]=&\frac{(\b J)^2}{2}\int_0^1\dd t\dd s\,x(t)G(t-s)x(s)\\
 &-\b\int_0^1 \argc{\frac{M}{2(\b\hbar)^2}\dot x^2+v_{\k,z}(x)}\\
 v_{\k,z}(x)=&\frac{\k}{2}x^2+\frac{x^4}{4!}-\argp{h+zJ\sqrt q}x\:.
\end{split}
\end{equation}
Thermal averages are defined as in~\eqref{eq:aver}. The saddle-point equations read:
\begin{subequations}\label{eq:RSsp}
 \begin{align}
  q_d(t-s)=&\int\dd p(\k)\DD z\,\moy{x(t)x(s)}\\
  q=&\int\dd p(\k)\DD z\,\moy{x(t)}^2\label{eq:RSspq}\\
  G(t-s)=&\int\dd p(\k)\DD z\,\moy{x(t)x(s)}_c\:.\label{eq:RSspG}
 \end{align}
\end{subequations}
The index $c$ stands for connected correlation function, see~\eqref{eq:G}. 
Taking the limit $\hbar\to0$ for all equations above \eqref{eq:RSF}-\eqref{eq:RSsp} one recovers the corresponding RS classical expressions of~\cite{BLRUZ21}.

Finally, the RS expression of the thermodynamic energy is derived from~\eqref{eq:UfullRSB}:
\begin{equation}\label{eq:URS}
\begin{split}
 \frac{U_{\rm RS}}{N}=&-\frac{\b J^2}{2}\int_0^1\dd t\, G(t)^2\\
 &+\int\dd p(\k)\DD z\,\moy{-\frac{M}{2(\b\hbar)^2}\dot x^2+v_{\k,z}(x)}\:.
 \end{split}
\end{equation}

\section{Semiclassical analysis and Debye physics}\label{sec:semi}

\subsection{Semiclassical expansion in the replica-symmetric phase}\label{sec:SGLD}

In this section we analyze the RS phase with the idea that it supplies the starting point to approach the RSB phases (see Fig.~\ref{fig:qphd}).  
Eqs.~\eqref{eq:RSsp} are sensibly more manageable than the RSB ones but remain difficult to solve analytically. Therefore we analyze them within the perturbative scheme studied in~\cite{SGLD04,SGLD05,S05,FMPS19}, which provides an expansion in the small $\hbar$ limit at fixed $\b\hbar$. We are thus interested in expressing average values of any observable (noted $\OO$) as a perturbative series of the form 
\beq
\OO=\sum_{n=0}^{\io}\hbar^n \OO_{n}(\b\hbar)\:.
\eeq
The lowest order and reference point of the expansion is thus the classical $T=0$ model. The model parameters $(J,h)$ are thus chosen here within the RS phase of the classical $T=0$ phase diagram of Fig.~\ref{fig:qphd}. 
As we shall see, this semiclassical approach is a well-suited tool to investigate the low-temperature phase once it is dressed by quantum fluctuations. 
The technical details on how to perform such an expansion are very close to what has been developed in Ref.~\cite[Appendix]{FMPS19}. 
In particular, effective averages~\eqref{eq:aver} are computed order by order through an asymptotic expansion around saddle-point trajectories: the dynamical action appearing in the exponent has indeed a diverging factor $\b\sim1/\hbar$ as in a standard semiclassical limit.

As the lowest order of the expansion is the classical $T=0$ model, we get: $q\to q_0$ (classical $T=0$ overlap), $q_d(t)\to q_0$, thus $G_0=0$ and $\b\wt G(\om_n)\to \b\hbar\wt G_1(\om_n)$ (notice the corresponding $\b^2$ dependence in the first term of $\AA_{\k,z}$ in~\eqref{eq:RSF}).  Note also that in the different limit of \textit{first} taking $\hbar\to0$ \textit{then} $T\to0$, we get $\b\wt G(\om_n)\sim\chi_{\rm cl}\d_{n0}$, where $\c_{\rm cl}$ is the  classical $T=0$ linear magnetic susceptibility of~\cite{BLRUZ21}. We thus expect $\c_{\rm cl}$ to be the zero mode $\b\wt G(0)\sim\b\hbar\wt G_{1}(0)$ at lowest order. 
For this reason we here define instead
\begin{equation}
 \chi=\b\wt G(0)\:.
\end{equation}
Its lowest order $\chi_0$ will be later identified as $\c_{\rm cl}$, the classical $T=0$ linear magnetic susceptibility of~\cite{BLRUZ21}.

\subsubsection{Saddle-point solutions: fixed trajectories or instantons}

As in~\cite[Sec.5.B]{FMPS19}, in the limit we consider we must expand the effective averages around their saddle-point trajectories. 
These verify
\begin{equation}\label{eq:spsingle}
 \frac{M}{(\b\hbar)^2}\ddot{x}(t)=-J^2\int_0^1\dd s\,\b G(t-s) x(s)+v_{\k,z}'(x(t))
\end{equation}
which is a Newtonian motion in a potential $-v_{\k,z}$ with a non-local stiffness (or memory term). 

Constant (classical) solutions $x^*$ verify
\begin{equation}\label{eq:mincl}
 v_{m,z}'(x^*)=0 \qquad \textrm{with}\qquad m=\k-J^2\chi
\end{equation}
which is akin to the classical RS $T=0$ saddle-point~\cite{BLRUZ21}. Note that in the present expansion at fixed Matsubara period we have to expand $q=q_0+\hbar q_1+\dots$ and $\c=\c_0+\hbar \c_1+\dots$, consequently $x^*=x^*_0+\hbar x^*_1+\dots$. 
At lowest order, the constant solution is the classical $T=0$ one $x^*_0$, which is unique in the RS phase~\cite{BLRUZ21}. In this phase the quadratic coefficient $m_0=\k-J^2\c_0\geqslant0$ so that the potential is a single well, with $\sign\argp{x_0^*}=\sign\argp{h+zJ\sqrt{q_0}}$. Neglecting for a moment the memory term, if there is a unique constant solution, no instanton (time-dependent solution with finite action) exists~\cite{coleman_erice,rajaraman}, the particle can only go forever downhill if it departs from $x^*_0$, yielding an infinite action. 
%
Including now the memory term, we can at small deviations around a constant solution $x(t)=x^*+\d x(t)$ and linearize~\eqref{eq:spsingle} ($\forall n\in\ZZZ$):
\begin{equation}\label{eq:hessiancond}
 \wt{\d x}(\om_n)\argc{M\argp{\frac{\om_n}{\b\hbar}}^2+v_{\k,z}''(x^*)-J^2\b \wt G(\om_n)}=0\:.
\end{equation}
Strictly speaking in our expansion the above quantities should be the lowest order ones (\ie $q\to q_0$, $x^*\to x^*_0$, $\b G\to\b\hbar G_1$). 
No periodic mode can develop unless the bracketted term is zero for some $n$. We shall see self-consistently in the following that this is precisely the same condition as requiring that the Hessian around the saddle-point solution develops a zero-mode, signaling a breakdown of the approach. Indeed, as in the classical case, the mass term $v_{\k,z}''(x^*_0)-J^2\c_0=v_{m,z}''(x^*_0)$, \ie \eqref{eq:hessiancond} at $\om_n=0$, then vanishes for $(m,z)=(0,h/(J\sqrt{q_0}))$ (purely quartic potential). This happens along the classical RSB-$\om^4$ line, where several constant solutions $x^*_0$ appear: one thus needs to consider instanton solutions and possibly RSB equations. 
In the following we start from the RS part of the classical phase diagram in Fig.~\ref{fig:qphd} and from the above discussion no instantons exist, the saddle-point solution is unique and constant.

\subsubsection{Asymptotic expansion around the saddle-point solution}

In order to compute observables such as the energy~\eqref{eq:URS}, we need to perform an asymptotic expansion of averages around the constant saddle point. We set the fluctuation around the saddle point ${ \r(t)=x(t)-x^*}$
and expand the exponent $\AA_{\k,z}[x]=-\b E[x]$~\cite[Sec.5.B]{FMPS19}. In this exponent, orders in $\r$ greater than the quadratic one are subdominant; due to the quartic action, there are only cubic and quartic vertices. Diagrammatically we thus write 
$\r(t)=\begin{tikzpicture}[baseline=-1mm]  \draw (0:0) node[cross=4pt,line width=1pt,rotate=0] {}; \draw(0,-0.3) node {$t$}; \draw[line width=1pt] (0:0) -- (0:0.7); \end{tikzpicture}$ and
\begin{equation}\label{eq:perturbth}
\begin{split}
\la \r(t)\r(t')\ra_\r&=\frac{1}{\b}\left(\restriction{\frac{\d^2 E}{\d x(t)\d x(t')}}{x^*}\right)^{-1}=
\begin{tikzpicture}[baseline=-1mm]  \draw (0:0) node[cross=4pt,line width=1pt,rotate=0] {}; \draw(0,-0.3) node {$t$}; 
\draw[line width=1pt] (0:0) -- (0:1);
 \draw (0:1) node[cross=4pt,line width=1pt,rotate=0] {}; \draw(1,-0.3) node {$t'$};
\end{tikzpicture}=O(\hbar)\\
-\b E[x]&=-\b E_{\rm G}+
\begin{tikzpicture}[baseline=-1mm]  \draw[fill=black] (0:0) circle (0.1) {}; 
\draw[line width=1pt] (0:0) -- (60:0.5);
\draw[line width=1pt] (0:0) -- (120:0.5);
\draw[line width=1pt] (0:0) -- (0,-0.5);
\end{tikzpicture}
+
\begin{tikzpicture}[baseline=-1mm]  \draw (0:0) circle (0.1) {}; 
\draw[line width=1pt] (60:0.1) -- (60:0.5);
\draw[line width=1pt] (120:0.1) -- (120:0.5);
\draw[line width=1pt] (-60:0.1) -- (-60:0.5);
\draw[line width=1pt] (-120:0.1) -- (-120:0.5);
\end{tikzpicture}
\\
E_{\rm G}[\r]&=E[x^*]+\frac12\iint_0^1\dd t\dd t'\,\r(t)\restriction{\frac{\d^2 E}{\d x(t)\d x(t')}}{x^*}\r(t')
\end{split}
\end{equation}
For convenience we define 
\begin{equation}
 v_{\k,z}^*=v_{\k,z}(x^*)\ , \quad v_{\k,z}^0=v_{\k,z}(x^*_0)
\end{equation}
and similarly for derivatives. 
Cubic vertices carry a factor $\begin{tikzpicture}[baseline=-1mm]  \draw[fill=black] (0:0) circle (0.1) {}; \end{tikzpicture}=-\b{v_{\k,z}^{(3)*}}/{3!}=-\b{x^*}/{3!}$  while quartic vertices $\begin{tikzpicture}[baseline=-1mm]  \draw (0:0) circle (0.1) {}; \end{tikzpicture}=-\b{v_{\k,z}^{(4)*}}/{4!}=-{\b}/{4!}$. As usual vertices contain a dummy variable that is integrated over, in contrast with fixed variables in outer lines represented by $\begin{tikzpicture}[baseline=-1mm]  \draw (0:0) node[cross=4pt,line width=1pt,rotate=0] {}; \end{tikzpicture}$. We shall write explicitly symmetry factors. 

The propagator in Fourier modes reads
\begin{equation}\label{eq:diagramexpr}
 \begin{split}
   \begin{tikzpicture}[baseline=-1mm]  \draw (0:0) node[cross=4pt,line width=1pt,rotate=0] {}; \draw(0.5,0.3) node {$n$}; 
\draw[line width=1pt] (0:0) -- (0:1);
 \draw (0:1) node[cross=4pt,line width=1pt,rotate=0] {}; 
\end{tikzpicture}
=&\int_0^1\dd (t-s)\,e^{i\om_n(t-s)}\frac{1}{\b}\left(\restriction{\frac{\d^2 E}{\d x(t)\d x(s)}}{x^*}\right)^{-1}\\
=&\frac1\b\frac{1}{M\argp{\frac{\om_n}{\b\hbar}}^2-J^2\b\wt G(\om_n)+v_{\k,z}''^*}\:.
\end{split}
\end{equation}

From~\eqref{eq:RSspG} we can readily get at lowest order in the expansion from the connected correlation function (Dyson equation) 
\begin{equation}\label{eq:modenG}
 \b\hbar\wt G_1(\om_n)=\int\frac{\dd p(\k)\DD z}{M\argp{\frac{\om_n}{\b\hbar}}^2-J^2\b\hbar\wt G_1(\om_n)+v_{\k,z}''^0}
\end{equation}
which is also valid for $n=0$, giving:
\begin{equation}\label{eq:c_0}
 \c_0=\int\frac{\dd p(m)\DD z}{m+\frac{(x^*_0)^2}{2}}\quad,\quad m=\k-J^2\c_0
\end{equation}
which is as expected the classical $T=0$ expression for $\b(q_d-q)$~\cite{BLRUZ21}, \ie the classical zero-temperature linear magnetic susceptibility. $p(m)$ is just the uniform distribution on $\argc{\k_m-J^2\c_0,\k_M-J^2\c_0}=\argc{m_m,m_M}$.
We note that~\eqref{eq:modenG} would be exact for the full $\wt G (\om_n)$ if the potential $v_\k(x)$ was at most quadratic, and represents a similar equation to the one for the classical density of the Hessian eigenmodes~\cite{BLRUZ21}.

As in~\cite[Sec.4.C]{FMPS19}, we need to regularize divergences arising from the kinetic energy term in~\eqref{eq:URS}, which can be seen at lowest order from
\begin{equation}
-\frac{M}{2\argp{\b\hbar}^2}\int_0^1\moy{\dot x^2}=-\frac{M}{2\argp{\b\hbar}^2}\sum_{n\in\ZZZ}\om_n^2\,\begin{tikzpicture}[baseline=-1mm]  \draw (0:0) node[cross=4pt,line width=1pt,rotate=0] {}; \draw(0.5,0.3) node {$n$}; 
\draw[line width=1pt] (0:0) -- (0:1);
 \draw (0:1) node[cross=4pt,line width=1pt,rotate=0] {}; 
\end{tikzpicture}\:.
\end{equation}
This is done by subtracting the free-particles expression
\begin{equation}\label{eq:URS2}
\begin{split}
  \frac{U_{\rm RS}}{N}=&\frac{U_{\rm RS}}{N}-\frac{U_{\rm RS}(v_{\k,z}=0)}{N}+\frac{1}{2\b}\\
  =&-\frac{J^2\b\hbar^2}{2}\sum_{n\in\ZZZ}\wt G_1(\om_n)^2+\int\dd p(\k)\DD z\left[\moy{v_{\k,z}(x)}\right.\\
  &+\left.\frac{1}{2\b}\sum_{n\in\ZZZ}\frac{-J^2\b\hbar\wt G_1(\om_n)+v_{\k,z}''^*}{M\argp{\frac{\om_n}{\b\hbar}}^2-J^2\b\hbar\wt G_1(\om_n)+v_{\k,z}''^*}\right]\:.
\end{split}
\end{equation}
The average term is expanded as: 
\begin{equation}\label{eq:expandv}
 \moy{v_{\k,z}(x)}=v_{\k,z}^* +\frac{v_{\k,z}''^*}{2}
\begin{tikzpicture}[baseline=-1mm]
 \draw[line width=1pt] (0:0) circle (0.3) node[cross=4pt,line width=1pt, xshift=-0.3cm]{} ;
\end{tikzpicture}
 +3 v_{\k,z}'^*
 \begin{tikzpicture}[baseline=-1mm]
\draw (-0.7,0) node[cross=4pt,line width=1pt,rotate=0] {};
\draw[line width=1pt] (-0.7,0) -- (0:0);
 \draw[line width=1pt] (0.3,0) circle (0.3) node[circle,fill=black,inner sep=0pt,minimum size=6pt,xshift=-0.3cm]{} ;
\end{tikzpicture}
+O(\hbar^2)\:.
\end{equation}
However we need to take into account the fact that $v_{\k,z}(x^*)$ must also be itself expanded and gives extra terms at order $\hbar$:
\begin{equation}\label{eq:vexpanded}
 v_{\k,z}^*=v_{\k,z}^0+\hbar\argc{v_{\k,z}'^0x_1^*-\frac{zJx_0^*q_1}{2\sqrt{q_0}}}+O(\hbar^2)\:.
\end{equation}
From~\eqref{eq:mincl} one has
\begin{equation}\label{eq:x1}
 x_1^*=\frac{\c_1J^2x_0^*+\frac{zJq_1}{2\sqrt{q_0}}}{{v_{\k,z}''^0}}
\end{equation}
while the saddle-point equation for $q$~\eqref{eq:RSspq} gives
\begin{equation}\label{eq:qexpanded}
 \begin{split}
  q_0=&\int\dd p(\k)\DD z\,(x_0^*)^2\\
  \hbar q_1=&\int\dd p(\k)\DD z\,2x_0^*\argc{\hbar x_1^*+3\, \begin{tikzpicture}[baseline=-1mm]
\draw (-0.7,0) node[cross=4pt,line width=1pt,rotate=0] {};
\draw[line width=1pt] (-0.7,0) -- (0:0);
 \draw[line width=1pt] (0.3,0) circle (0.3) node[circle,fill=black,inner sep=0pt,minimum size=6pt,xshift=-0.3cm]{} ;
\end{tikzpicture}}\:.
 \end{split}
\end{equation}
The first equation coincides with the classical $T=0$ equation for the replica overlap~\cite{BLRUZ21}, as expected. We need one last equation for $\c_1$ in order to compute~\eqref{eq:vexpanded}, given by $\c=\b \wt G(0)$ together with the Dyson equation for $G$~\eqref{eq:RSspG}. We need to go one higher order  in the expansion of the connected correlation function than in \eqref{eq:c_0}:
\begin{equation}\label{eq:ccfhigh}
\begin{split}
 \moy{x(t)x(s)}_c =& \begin{tikzpicture}[baseline=-1mm]  \draw (0:0) node[cross=4pt,line width=1pt,rotate=0] {}; \draw(0,-0.3) node {$t$}; 
\draw[line width=1pt] (0:0) -- (0:0.5);
 \draw (0:0.5) node[cross=4pt,line width=1pt,rotate=0] {}; \draw(0.5,-0.3) node {$s$};
\end{tikzpicture} + 4\cdot3 \begin{tikzpicture}[baseline=-1mm]
\draw (-0.5,0) node[cross=4pt,line width=1pt,rotate=0] {};
 \draw(-0.5,-0.3) node {$t$}; 
\draw[line width=1pt] (-0.5,0) -- (-0.1,0);
\draw (0:0) circle (0.1) {}; 
 \draw (0.5,0) node[cross=4pt,line width=1pt,rotate=0] {};
  \draw(0.5,-0.3) node {$s$}; 
\draw[line width=1pt] (0.1,0) -- (0.5,0);
\draw[line width=1pt,yshift=-0.7] (0.07,0.1) arc
    [
        start angle=-50,
        end angle=230,
        x radius=0.1,
        y radius =0.3
    ] ;
\end{tikzpicture}
+6\cdot3 \begin{tikzpicture}[baseline=-1mm]
\draw (-0.5,0) node[cross=4pt,line width=1pt,rotate=0] {};
 \draw(-0.5,-0.3) node {$t$}; 
\draw[line width=1pt] (-0.5,0) -- (0:0);
 \draw[line width=1pt] (0.3,0) circle (0.3) node[circle,fill=black,inner sep=0pt,minimum size=6pt,xshift=-0.3cm]{} ;
 \draw (0.6,0) node[circle,fill=black,inner sep=0pt,minimum size=6pt] {};
 \draw (1.1,0) node[cross=4pt,line width=1pt,rotate=0] {};
  \draw(1.1,-0.3) node {$s$}; 
\draw[line width=1pt] (0.6,0) -- (1.1,0);
\end{tikzpicture}\\
&+6\cdot3 \begin{tikzpicture}[baseline=-1mm]
\draw (-0.5,0) node[cross=4pt,line width=1pt,rotate=0] {};
 \draw(-0.5,-0.3) node {$t$}; 
\draw[line width=1pt] (-0.5,0) -- (-0.1,0);
\draw (0:0) node[circle,fill=black,inner sep=0pt,minimum size=6pt] {};
 \draw (0.5,0) node[cross=4pt,line width=1pt,rotate=0] {};
  \draw(0.5,-0.3) node {$s$}; 
\draw[line width=1pt] (0.1,0) -- (0.5,0);
\draw (0,0.3) node[circle,fill=black,inner sep=0pt,minimum size=6pt] {};
\draw[line width=1pt] (0,0) -- (0,0.3);
\draw[line width=1pt] (0,0.5) circle (0.2) ;
\end{tikzpicture}
+O(\hbar^3)
 \end{split}
\end{equation}
which provides the equation for $\c_1$, that we will write later in~\eqref{eq:c1} for notational convenience. 

\subsubsection{Low-frequency self-energy and Matsubara sums}\label{sub:lowMat}

The various diagrams can be written with Matsubara sums, whose low-temperature limit depends on the $\om\to0$ behavior of the propagator~\cite{Mahan,Bruus,AS06,coleman2015introduction}. We define the self-energy
\begin{equation}\label{eq:selfdef}
 \wt I(\om_n)=-J^2\b\argc{\wt G(\om_n)-\wt G(0)}\:.
\end{equation}
From the Dyson equation~\eqref{eq:RSspG} we get the self-consistent equation for the lowest order $I_0$:
\begin{equation}\label{eq:eqI0}
 \c_0-\frac{\wt I_0(\om_n)}{J^2}=\int_{\tilde a_m}^{\tilde a_M}\dd p(\tilde a)\frac{1}{M\argp{\frac{\om_n}{\b\hbar}}^2+\tilde a+\wt I_0(\om_n)}
\end{equation}
with 
\begin{equation}\label{eq:tildeamass}
 \tilde a =v_{\k,z}''^0= m + \frac{(x^*_0)^2}{2}\quad ;\quad  \dd p(\tilde a)=\dd p(\k)\DD z\:.
\end{equation}
The analysis of~\eqref{eq:eqI0} proceeds similarly to the classical case~\cite{BLRUZ21}, from which we know that the distribution of the mass $\tilde a$ (whose support is noted $\argc{\tilde a_m,\tilde a_M}$) is gapped everywhere in the RS phase except at the RSB-$\om^4$ transition line. Let us note 
\begin{equation}
F(z)=\int_{\tilde a_m}^{\tilde a_M}\frac{\dd p(\tilde a)}{\tilde a+z}\quad 
\textrm{and}\quad\overline{\,\bullet\,}=\int\dd p(\tilde a)\,\bullet
\end{equation}
such that $F(0)=\overline{\tilde a^{-1}}=\c_0$, $F'(0)=-\overline{\tilde a^{-2}}=\argp{\l_R^0-1}/{J^2}$ with $\l_R^0$ the classical $T=0$ replicon, $F''(0)=2\overline{\tilde a^{-3}}$. For reference we write the classical $T=0$ replicon equation~\cite{BLRUZ21} which is obtained through the lowest-order expansion of~\eqref{eq:replicon0}:
\begin{equation}\label{eq:repliconcl}
 \l_R^0=1-J^2\overline{\tilde a^{-2}}\:.
\end{equation}
 We come back to the conventional time units for the rest of this subsection~\ref{sec:SGLD}, $\om_n=2\p n/(\b\hbar)$. We set below $\om=\om_n$ and look at the $\om\to 0$ limit where by definition $\wt I_0\to0$. Then \eqref{eq:eqI0} gives
 \begin{equation}\label{eq:Fequ}
 \begin{split}
   0=& \frac{F''(0)}{2}\argp{M\om^2}^2+\frac{\l_R^0-1}{J^2}M\om^2\\
   &+\wt I_0(\om)\argp{\frac{\l_R^0}{J^2}+F''(0)M\om^2}+\wt I_0(\om)^2 \frac{F''(0)}{2}+\dots
     \end{split}
 \end{equation}
\textit{(i) In the bulk of the RS phase:} $\l_R^0\neq0$ so we get the dominant $\om\to0$ behavior which is analytic
\begin{equation}\label{eq:I0bulk}
\wt I_0(\om)\underset{\om\to0}{\sim} \frac{1-\l_R^0}{\l_R^0}M\om^2
\end{equation}
\textit{(ii) On the RSB-$\om^2$ transition line:} $\l_R^0=0$ but $F''(0)$ is finite ($\tilde a$ is gapped \ie $\tilde a_m>0$)~\cite{BLRUZ21}. 
Therefore the self-energy becomes non analytic and conforms to the Schehr-Giamarchi-Le Doussal arguments~\cite{SGLD04,SGLD05,S05,FMPS19}
\begin{equation}\label{eq:I02}
 \wt I_0(\om)\underset{\om\to0}{\sim}B\abs{\om}\ ,\qquad B=\sqrt{\frac{M}{J^2\overline{\tilde a^{-3}}}}
\end{equation}
\textit{(iii) On the RSB-$\om^4$ transition line:} $\tilde a_m=0$, \ie $\tilde a$ becomes gapless and we know the asymptotics near the edge $\tilde a=0$: $p(\tilde a)\propto \tilde a^{3/2}$~\cite{BLRUZ21}. This means $F(0)$ and $F'(0)$ are finite but $F''(0)$ diverges. \\We study further~\eqref{eq:eqI0} 
by defining 
\begin{equation}\label{eq:Zdef}
 J^2Z(\om)=M\om^2+\wt I_0(\om)\to 0\:.
\end{equation}
We have $F\argp{J^2Z(\om)}\underset{\om\to0}{\sim}\c_0+F'(0)J^2Z(\om)+R(\om)$ with
\begin{equation}
 \begin{split}
  R(\om)&=\int_0^{\tilde a_M}\dd p(\tilde a)\argc{\frac{1}{\tilde a +J^2Z(\om)}-\frac{1}{\tilde a}+\frac{J^2Z(\om)}{\tilde a^2}}\\
  &=\argp{J^2Z(\om)}^2\int_0^{\tilde a_M}\dd \tilde a \,\frac{\tilde a^{3/2}P(\tilde a)}{\tilde a^2(\tilde a+J^2Z(\om))}\\
  &\underset{\om\to0}{\sim}A\argp{J^2Z(\om)}^{3/2}\underbrace{\int_0^\io\frac{\dd x}{(1+x)\sqrt x}}_{=\p}
 \end{split}
\end{equation}
where $P(\tilde a)=p(\tilde a)/\tilde a^{3/2}\underset{\tilde a\to0}{=}A$. Thus~\eqref{eq:eqI0} becomes 
\begin{equation}
 M\om^2\underset{\om\to0}{\sim}\l_R^0J^2Z(\om)+J^2A\p\argp{J^2Z(\om)}^{3/2}+\dots
\end{equation}
By inverting this relationship, we get the non-analytic expression for the self-energy
\begin{equation}\label{eq:I04}
\wt I_0(\om)\underset{\om\to0}{=} \frac{1-\l_R^0}{\l_R^0}M\om^2-\underbrace{\frac{J^2A\p}{(\l_R^0)^{5/2}}\argp{M\om^2}^{3/2}}_{\rm non-analytic}+\dots
\end{equation}
All in all, due to the diagrams, one has to compute three Matsubara sums. Diagrams in~\eqref{eq:expandv},\eqref{eq:qexpanded},\eqref{eq:ccfhigh} contain the first sum, another comes from the kinetic term in~\eqref{eq:URS2} and the third will appear in~\eqref{eq:2x2system}:
\begin{equation}\label{eq:S2}
\begin{split}
 \SS_1(\tilde a)=&\frac{1}{\b\hbar}\sum_{n\in\ZZZ}\frac{1}{M\om_n^2+\tilde a+\wt I_0(\om_n)}\\
 \SS_2(\tilde a)=&\frac{1}{\b\hbar}\sum_{n\in\ZZZ}\frac{\tilde a+\wt I_0(\om_n)}{M\om_n^2+\tilde a+\wt I_0(\om_n)}\\
 \SS_3(\tilde a)=&\frac{1}{\b\hbar}\sum_{n\in\ZZZ}\frac{1}{\argc{M\om_n^2+\tilde a+I_0(\om_n)}^2}\:.
\end{split}
\end{equation}
Finally we may write more explicitly the equation determining $\c_1$ by expanding $\c=\b\wt G(0)$. Gathering the expansion of the connected correlation function~\eqref{eq:ccfhigh} and the equation for $\c_0$~\eqref{eq:c_0}, we get from the Dyson equation~\eqref{eq:RSspG}:
\begin{equation}\label{eq:c1}
\begin{split}
 \c_1=&-\frac12\overline{\frac{\SS_1(\tilde a)}{\tilde a^2}}+\frac12\overline{\argp{\frac{x_0^*}{\tilde a}}^2\SS_3(\tilde a)}+\frac12\overline{\frac{(x_0^*)^2}{\tilde a^3}\SS_1(\tilde a)}\\
 &+J^2\c_1\overline{\tilde a^{-2}}-\overline{\frac{x_0^*x_1^*}{\tilde a^2}}\:.
 \end{split}
\end{equation}
Eqs.~\eqref{eq:x1},\eqref{eq:qexpanded},\eqref{eq:c1} combined allow to determine $(x_1^*,\c_1,q_1)$ which enter in the energy through~\eqref{eq:URS2}-\eqref{eq:vexpanded}. $x_1^*$ is directly computed from $(\c_1,q_1)$ through~\eqref{eq:c1}; with the other two equations we get
\begin{equation}\label{eq:2x2system}
\begin{split}
  \begin{pmatrix}  \c_1\\q_1 \end{pmatrix}&=\MM^{-1}\begin{pmatrix}  \frac12 \overline{\argp{\frac{(x_0^*)^2}{\tilde a^3}-\frac{1}{\tilde a^2}}\SS_1(\tilde a)}+\frac12 \overline{\argp{\frac{x_0^*}{\tilde a}}^2\SS_3(\tilde a)}\\
   \overline{\argp{\frac{(x_0^*)^2}{\tilde a}}\SS_1(\tilde a)} \end{pmatrix} \\
  \MM&=\begin{pmatrix} \l_R^0+J^2\overline{\frac{(x_0^*)^2}{\tilde a^3}} & \frac{J}{2\sqrt{q_0}}\overline{\frac{zx_0^*}{\tilde a^3}} \\
       2J^2\overline{\frac{(x_0^*)^2}{\tilde a}}  & \frac{J}{\sqrt{q_0}}\overline{\frac{zx_0^*}{\tilde a}} -1
       \end{pmatrix}\:.
\end{split}
\end{equation}


\subsubsection{Gaussian approximation and the classical limit}\label{sub:Gaussian}

Under the RS assumption we expect the Debye approximation to be valid at lowest order in the present semiclassical expansion~\cite{FMPS19}, that is, the low-temperature physics is dominated by quantizing the harmonic modes around the energy minimum, provided by the Hessian of the classical energy landscape. 
As a hint we obtained the lowest-order Dyson equation~\eqref{eq:modenG} that is purely determined by the Gaussian part of the action and formally identical to the resolvent equation for the classical Hessian spectrum~\cite{BLRUZ21}. 

Let us then first analyze~\eqref{eq:URS2} keeping only the contribution from Gaussian terms of the action (\ie the non-Gaussian vertices are set to zero). The equations for $(\c_1,q_1)$ simplify as 
\begin{equation}
 \left\{\begin{split}
  \l_R^0\c_1=&-\overline{\frac{x_0^*x_1^*}{\tilde a^2}}\\
  q_1=&2\overline{x_0^*x_1^*}
 \end{split}\right.
\end{equation}
and $x_1^*$ still set by~\eqref{eq:x1}; the only solution in this case is $(x_1^*,\c_1,q_1)=(0,0,0)$. 
The expansion~\eqref{eq:expandv} without the non-Gaussian vertices then reads at lowest order
\begin{equation}\label{eq:expandvGaussian}
 \moy{v_{\k,z}(x)}=
v_{\k,z}^0 +\frac{1}{2\b}\sum_{n\in\ZZZ}\frac{v_{\k,z}''^0}{M\om_n^2-J^2\b\hbar\wt G_1(\om_n)+v_{\k,z}''^0}
\end{equation}
so that all $O(\hbar)$ terms in the energy~\eqref{eq:URS2} can be combined into a single one with the above denominator: indeed using~\eqref{eq:modenG} 
\begin{equation}
\begin{split}
  (J\b\hbar)^2\sum_{n\in\ZZZ} \wt G_1(\om_n)^2=
  \sum_{n\in\ZZZ} \int\frac{\dd p(\k)\DD z\,J^2\b\hbar \wt G_1(\om_n)}{M\om_n^2-J^2\b\hbar\wt  G_1(\om_n)+v_{\k,z}''^0}
\end{split}
\end{equation}
In the end, the above two terms give the same contribution as the kinetic term. In conclusion
\begin{equation}\label{eq:UG}
  \restriction{\frac{U_{\rm RS}}{N}}{\rm Gaussian}=\overline{v_{\k,z}^0}+\hbar\overline{\SS_2(\tilde a)}+O(\hbar^2)
\end{equation}
$\overline{v_{\k,z}^0}$ is the classical ground-state energy. 

Note that in the purely classical limit $\hbar\to0$ (\textit{at fixed} $T$), which in the present expansion translates into keeping only $n=0$ modes in the Matsubara sums~\cite{FMPS19} (as $\abs{\om_n}\to\io$ for $n\neq0$ in this limit), we directly get from~\eqref{eq:UG}
\begin{equation}
 \frac{U_{\rm RS}}{N}\underset{\hbar\to0}{=}\overline{v_{\k,z}^0}+T+O(T^2)
\end{equation}
\ie Dulong \& Petit's law, as expected from the equipartition theorem~\cite{kittel,AM}. Classically the excitations around the energy minimum seem thus harmonic. 
The same result holds performing a similar direct calculation including the non-Gaussian vertices in the classical limit. This hints at a cancellation  of the non-Gaussian contributions, which will be proven in Sec.~\ref{sub:NG}.

\begin{figure}[h!]
\centering
 \includegraphics[width=0.6\linewidth,height=0.3\linewidth]{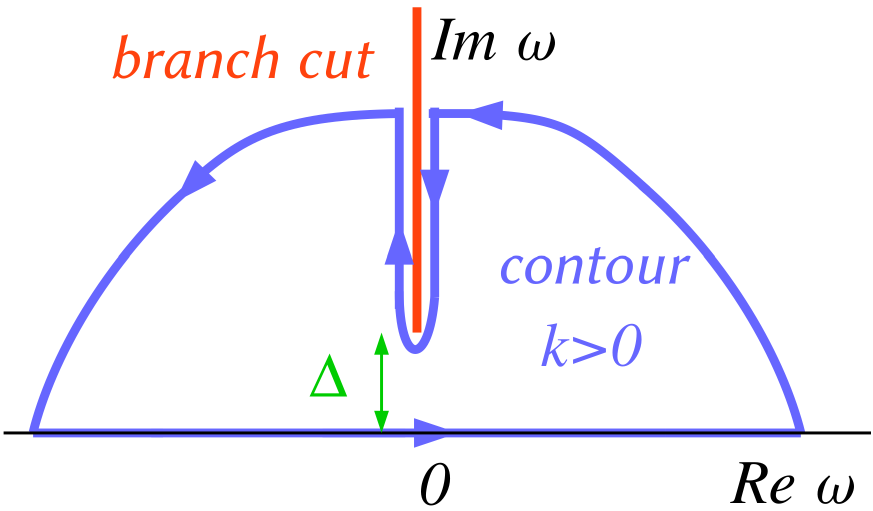}\\
 \includegraphics[width=0.6\linewidth,height=0.3\linewidth]{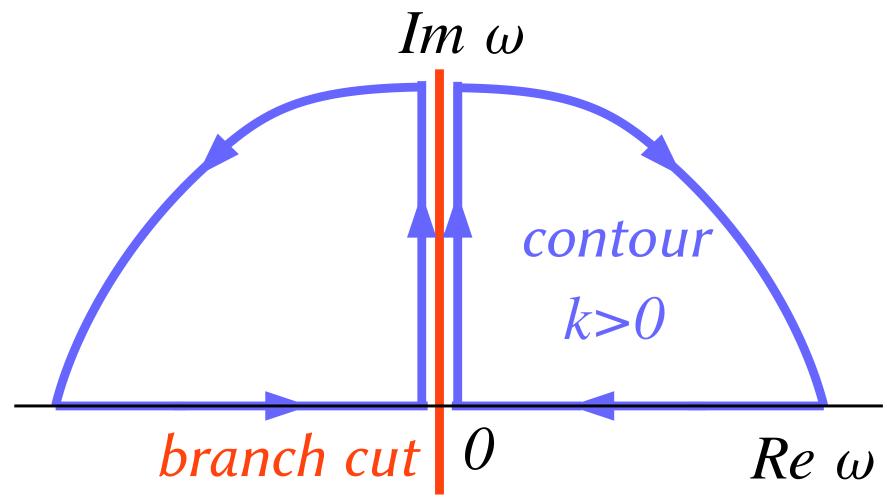}\\
 \caption{(top) Contour for the bulk RS phase. From~\eqref{eq:I0bulk} we know there are no singularity around $\om=0$. (bottom) Contour used on both RSB-$\om^2$ and RSB-$\om^4$ transition lines. A branch cut extends to $\om=0$, from the singular behavior of~\eqref{eq:I02} or~\eqref{eq:I04}. }
 \label{fig:contours}
\end{figure}

Going back to the quantum case~\eqref{eq:UG}, the Matsubara sums at low temperature are calculated through standard contour integral methods~\cite{Mahan,Bruus,AS06,coleman2015introduction}. The main idea\footnote{See~\cite[App.B]{FMPS19} for more details.} is to transform the sums in integrals through the Poisson formula~\cite[Chap.11]{Appel}
$  \frac{2\p}{\b\hbar}\sum_{n\in\ZZZ}\d(\om-\om_n)=\sum_{k\in\ZZZ}e^{i\b\hbar k\om}\ 
  \Rightarrow \ \SS_i=\sum_{k\in\ZZZ}\SS_{ik} $ where $\SS_{ik}$ is an integral over ${\om\leftarrow2\p n/(\b\hbar)}$, and use contours drawn in Fig.~\ref{fig:contours}.  

In the bulk of the RS phase, the behavior of the Matsubara integrals follows closely the discussion of the edge of the classical Hessian spectrum~\cite{BLRUZ21}. To make closer contact with the notations of~\cite{BLRUZ21} we write the Dyson equation~\eqref{eq:eqI0} as 
\begin{equation}\label{eq:forbidden}
\begin{split}
  -M\om^2=&\FF\argp{g(\om)}\\
  \FF(g)\equiv&-J^2\int_{a_m}^{a_M}\dd p(a)\argc{g+\frac{1}{a+J^2g}}
\end{split}
\end{equation}
with 
\begin{equation}
 g(\om)\equiv Z(\om)-\c_0
\end{equation}
and $a\equiv\tilde a +J^2\c_0$ corresponds to the distribution of $v_{\k,z}''^0$. 
There is a band of forbidden real values $g\in\argc{-a_M/J^2,-a_m/J^2}$ and there are two possibilities for $g>-a_m/J^2$ : \textit{(i)} $\FF$ has a maximum in $g_m>-a_m/J^2$ (spectrum dominated by the GOE part of the Hessian $J_{ij}+\d_{ij}\argp{\k_i+x_i^2/2}$) \textit{(ii)} $\FF$ has a maximum in $g_m=-a_m/J^2$ (spectrum dominated by the diagonal part of the Hessian). The lower edge is then in either case $\l_m=\FF(g_m)>0$; a similar discussion holds for $g<-a_M/J^2$ which defines the upper edge $\l_M=\FF(g_M)$ with $g_M$ the minimum~\cite{BLRUZ21}. The forbidden band on real $g$ corresponds through~\eqref{eq:forbidden} to two forbidden bands on the purely imaginary frequency $\om=i\Im\om$ with $\abs{\Im\om}\in\argc{\sqrt{\l_m/M},\sqrt{\l_M/M}}$. One of these branch cuts is drawn in Fig.~\ref{fig:contours}(top), with a finite gap to $\om=0$ noted $\D$.  The latter gap reads from~\eqref{eq:forbidden} $\D=\sqrt{\l_m/M}$, implying analycity around $\om=0$. This gap is therefore independent of any value of $\tilde a$ that is a variable in most of the Matsubara sums. 
All sums at low temperature are thus $\SS_i\propto e^{-\b\hbar\D}$ in this regime, and so are both the energy and specific heat, \ie a \textit{gapped} scaling with the Debye spectral energy gap $\hbar\D$ at this lowest semiclassical order. 

On the RSB transition lines instead, the gap closes and $\wt I_0(\om)$ is non-analytic around $\om=0$ (\secref{sub:lowMat}). The contour is then of the form of Fig.~\ref{fig:contours}(bottom). We thus derive the following expression: 
\begin{equation}\label{eq:S2G}
 \hbar\overline{\SS_2(\tilde a)}=
 \int_0^\io\frac{\dd\om}{\p}\hbar \om \argp{f_{\rm B}(\om)+\frac12} \frac{2M\om}{J^2} \,\Im\wt I_0(-i\om+0^+)
 \end{equation}
with $f_{\rm B}(\om)=\argp{e^{\b\hbar  \om}-1}^{-1}$ the Bose-Einstein factor. The  $1/2$ term gives the vacuum energy\footnote{It corresponds to the $k=0$ term of the Poisson summation defined below~\eqref{eq:2x2system}, while all $k\neq0$ bring the Bose-Einstein factor. }. 

We next compare the direct Gaussian result~\eqref{eq:UG} to the Debye approximation.


\subsubsection{Debye approximation}\label{sub:Debye}

The Debye approximation for the energy is
\begin{equation}\label{eq:Debye}
 \frac{U_{\rm D}}{N}=e_{\rm GS}+\int_0^\io D(\om)\dd\om\,  \hbar \om \argp{f_{\rm B}(\om)+\frac12}
\end{equation}
where $e_{\rm GS}$ is the classical ground-state energy, $D(\om)\dd\om=\r(\l)\dd\l$ is the density of classical vibrational modes, related to the Hessian spectral density 
\begin{equation}\label{eq:spHess}
 \r(\l)=(1/\p)\Im g(\l-i0^+)
\end{equation} obtained in~\cite{BLRUZ21}, with the correspondence $\l=M\om^2$.

\textit{(i)~In the bulk of the RS phase:} the support of $D(\om)$ is bounded from below by the spectral gap $\hbar\D$, providing the above-mentioned gapped scaling for $T\to0$. 

When the gap closes we can read off~\eqref{eq:S2G} the direct relationship between the two approaches
\begin{equation}\label{eq:relDebyeMatsubara}
 D(\om)\underset{\om\to0}{=}\frac{2M\om}{\p J^2} \,\Im\restriction{\wt I_0(\om_n)}{\om_n\to-i\om+0^+}
\end{equation}
which is easily related to the Hessian spectral density above: the factor $1/\p$ comes from its definition~\eqref{eq:spHess}, $2M\om$  from the change of variable $\l\leftrightarrow\om$ and $\Im\wt I_0/J^2$ is directly related to $\Im g$ defined below~\eqref{eq:forbidden}. 

\textit{(ii)~On the RSB-$\om^2$ transition line:} ${\r(\l)\underset{\l\to0}{\sim}\left.\sqrt{{\l}}\Big/\argp{\p J^3\sqrt{\overline{\tilde a^{-3}}}}\right.}$~\cite{BLRUZ21} which yields the same result as~\eqref{eq:relDebyeMatsubara} with~\eqref{eq:I02}. 
For completeness we give the final low-temperature behavior:
\begin{equation}\label{eq:cubicCV}
  \frac{C_V^{\rm RS}}{N}\sim \frac{8\p^3}{15}\frac{M^{3/2}}{J^3\sqrt{\overline{\tilde a^{-3}}}}\argp{\frac T\hbar}^{3}
\end{equation}
a similar scaling as in the marginal (fullRSB) phase in~\cite{FMPS19,AM12,S05,CM22}. The linearity in frequency of the spectral density $\r(\om^2)$, or equivalently of the self-energy $\Im \wt I_0(\om)$, is akin to the same frequency behavior found in the spin susceptibility of quantum spin-glass models such as the TFSK~\cite{AM12,CM22,KZL24} or in structural glass models~\cite{FMPS19,ABPS21}. 

\textit{(iii)~On the RSB-$\om^4$ transition line:} ${\r(\l)\underset{\l\to0}{\sim}{A}\l^{3/2}/{(\l_R^0)^{5/2}}}$~\cite{BLRUZ21} which gives the same result as~\eqref{eq:relDebyeMatsubara} with~\eqref{eq:I04}. 
The final low-temperature behavior is
\begin{equation}
  \frac{C_V^{\rm RS}}{N}\sim \frac{32\p^6}{21}A\argp{\frac{M}{\l_R^0}}^{5/2} \argp{\frac T\hbar}^{5}
\end{equation}
To show the latter scalings are exact, we prove independently in the next section that non-Gaussian terms vanish at $O(\hbar)$.

\subsubsection{Cancellation of the non-Gaussian terms}\label{sub:NG}

Let us demonstrate that non-Gaussian terms vanish at $O(\hbar)$ for any value of $\b\hbar$. We examine the terms  we left out from~\eqref{eq:URS2} in the Gaussian case~\eqref{eq:UG}:
\begin{equation}\label{eq:decomposeG}
\begin{split}
  &\frac{U_{\rm RS}}{N}-\restriction{\frac{U_{\rm RS}}{N}}{\rm Gaussian}=\,\begin{tikzpicture}[baseline=-1mm] \draw[fill=white] (0:0) circle (0.2) node {1};\end{tikzpicture}+\begin{tikzpicture}[baseline=-1mm] \draw[fill=white] (0:0) circle (0.2) node {2};\end{tikzpicture}\\
  &\begin{tikzpicture}[baseline=-1mm] \draw[fill=white] (0:0) circle (0.2) node {1};\end{tikzpicture}=-\frac\hbar2 \overline{\frac{v_{\k,z}'^0x_0^*}{\tilde a}\SS_1(\tilde a)}\\
&  \begin{tikzpicture}[baseline=-1mm] \draw[fill=white] (0:0) circle (0.2) node {2};\end{tikzpicture}=\hbar J^2\c_1\overline{\frac{v_{\k,z}'^0x_0^*}{\tilde a}} +\hbar q_1\frac{J}{2\sqrt{q_0}}\overline{\frac{zv_{\k,z}'^0}{\tilde a}-zx_0^*}
\end{split}
\end{equation}
The only missing piece is to know how to compute the different averages over $\tilde a$. This is done as in~\cite[App.D]{BLRUZ21} by rewriting the averages and enforcing the value of $x^*_0$~\eqref{eq:mincl} through a Dirac delta function
\begin{equation}
\begin{split}
 \overline{\frac{f(m,z,x_0^*)}{\tilde a^n}}=&\int \dd x\dd p(m)\DD z\frac{f(m,z,x)}{\argp{m+\frac{x^2}{2}}^{n-1}}\\
 &\qquad\times\d\argp{mx+\frac{x^3}{3!}-h-zJ\sqrt{q_0}}\\
 =&\int \dd x\dd p(m)\frac{\dd \hat x}{2\p}\DD z\frac{f(m,z,x)}{\argp{m+\frac{x^2}{2}}^{n-1}}\\
 &\qquad\times e^{i\hat x\argp{mx+\frac{x^3}{3!}-h-zJ\sqrt{q_0}}}
\end{split}
\end{equation}
Depending on the actual form of $f(m,z,x_0^*)$, one can integrate over the Gaussian $z$ then $\hat x$ variables. 
Through integrations by parts we get the following relations covering all types of such averages encountered in the energy calculation:
\begin{equation}\label{eq:relations}
 \begin{split}
 \arga{\bullet}\equiv& \int\frac{\dd x\dd p(m)}{\sqrt{2\p J^2q_0}}\bullet e^{-\frac{\argp{mx+\frac{x^3}{3!}-h}^2}{2J^2q_0}}\\
  \overline{\frac{f(x_0^*)}{\tilde a^n}}=&\arga{\frac{f(x)}{\argp{m+\frac{x^2}{2}}^{n-1}}}\\
  \overline{z\frac{f(x_0^*)}{\tilde a^n}}=&\frac{1}{\sqrt{J^2q_0}}\arga{\frac{f(x)\argp{mx+\frac{x^3}{3!}-h}}{\argp{m+\frac{x^2}{2}}^{n-1}}}\\
  \overline{z^2\frac{f(x_0^*)}{\tilde a^n}}=&\frac{1}{{J^2q_0}}\arga{\frac{f(x)\argp{mx+\frac{x^3}{3!}-h}^2}{\argp{m+\frac{x^2}{2}}^{n-1}}}
 \end{split}
\end{equation}
Note that $\c_0=\arga{1}$ in this notation. These relations are always convergent for the RSB-$\om^2$ line. For the RSB-$\om^4$ they are \textit{formally} valid\footnote{Due to divergences one may find more appropriate
to reduce the powers of $\tilde a$ in the denominators, as there $\overline{\tilde a^{-3}}=\arga{\argp{m+\frac{x^2}{2}}^{-2}}$ becomes the first divergent moment (all higher powers diverge). For this purpose one can once again use integrations by parts to prove the formal relationship
\begin{equation}\label{eq:formalrel}
 \begin{split}
 \arga{\frac{f(x)}{\argp{m+\frac{x^2}{2}}^{n}}}=&\frac{1}{n-1} \arga{\frac{\frac{\dd}{\dd x}\argp{\frac{f(x)}{x}}}{\argp{m+\frac{x^2}{2}}^{n-1}}}\\
 &-\frac{1}{J^2q_0}\arga{\frac{f(x)}{x}\frac{mx+\frac{x^3}{3!}-h}{\argp{m+\frac{x^2}{2}}^{n-2}}}
  \end{split}
\end{equation}
useful only if $f(x)\underset{x\to0}{\to}0$ at least linearly with $x$ to get less diverging expressions.}. 
Solving the linear system~\eqref{eq:2x2system} we get
\begin{equation}
 \begin{tikzpicture}[baseline=-1mm] \draw[fill=white] (0:0) circle (0.2) node {2};\end{tikzpicture}
 =\hbar\frac{J^2\c_0}{2}\arga{x^2\SS_1(\tilde a)}
\end{equation}
which is easily seen to cancel the other term $\begin{tikzpicture}[baseline=-1mm] \draw[fill=white] (0:0) circle (0.2) node {1};\end{tikzpicture}$ using the above formulas~\eqref{eq:relations}-\eqref{eq:formalrel}.
We conclude that at the lowest order in the present semiclassical expansion, in the RS phase including the transition lines, the Debye approximation and the direct calculation coincide, \ie non-Gaussian vertices do not contribute.

\subsection{The RSB-SWP phase}

Now let's briefly consider the starting point of the expansion within the marginal phase, close enough to the RSB-$\om^2$ line. There the saddle-point equation~\eqref{eq:spsingle} has a single constant solution $x^*$. In the following we thus call RSB-SWP phase such a state of the system. 
As $x^*$ is unique and the mass $\tilde a=v_m''(x_0^*)$ is non-zero, no other instanton solution can develop and the situation is analog to the RSB phase of the quantum spherical perceptron~\cite{FMPS19}:  most of the expansion in the RS phase worked out in~\secref{sec:SGLD} is readily translated to the RSB phase, provided we replace the Gaussian weight $\DD z$  by $\dd H\,P_\k^{(0)}(1,H)$ (with $h+zJ\sqrt{q_0}\to H$). In particular one retrieves the Dyson equation
\begin{equation}\label{eq:Dysonfull}
 \b\hbar\wt G_1(\om_n)=\int\frac{\dd p(m)\dd H\,P_\k^{(0)}(1,H)}{M\argp{\frac{\om_n}{\b\hbar}}^2-J^2\b\hbar\wt G_1(\om_n)+v_m''(x_0^*)}
\end{equation}
Combined with the marginal stability condition ${\l_R=0}$~\eqref{eq:replicon0} at lowest order, \eqref{eq:Dysonfull} brings the usual non-analytic self-energy
\begin{equation}\label{eq:selffull}
 \wt I_0(\om)\underset{\om\to0}{\sim}B\abs{\om}
\end{equation}
with $B$ given as in~\eqref{eq:I02}, replacing ${\overline{\tilde a^{-3}}\to \int\dd p(m)\dd H\,P_\k^{(0)}(1,H)/[m+(x_0^*)^2/2]^3}$. 
Regarding now the energy~\eqref{eq:UfullRSB}, doing the semiclassical expansion one has to consider the perturbation $P_\k(1,H)=P_\k^{(0)}(1,H)+\hbar P_\k^{(1)}(1,H)+O(\hbar^2)$. Similarly to~\eqref{eq:UG} and~\eqref{eq:decomposeG}, at $O(\hbar)$ one gets three contributions: \textit{(i)} the one from the Gaussian part of the imaginary-time action coming from the potential average $\partial_\b f_\k(1,H)\sim\moy{v_m(x)}$, weighted by $P_\k^{(0)}(1,H)$ \textit{(ii)}~the non-Gaussian contribution from the vertices of the same average and weight as in \textit{(i)} \textit{(iii)} the contribution from the pertubation of $P_\k(1,H)$, which simply reads $\hbar\int \dd p(m)\dd H\,P_\k^{(1)}(1,H)v_m(x_0^*)$. The Gaussian part \textit{(i)} yields the Debye approximation corresponding to the self-energy~\eqref{eq:selffull} (\ie $D(\om)\sim\om^2$), as in~\secref{sub:Debye}, inducing the specific heat $C_V\sim T^3$ -- \ie \eqref{eq:cubicCV} with the above replacement for the prefactor. The other contributions \textit{(ii)}-\textit{(iii)} were shown to cancel in the RS phase (\secref{sub:NG}), but we could not check it without explicit knowledge of $P_\k^{(1)}(1,H)$, which should be obtained by expanding the partial differential equation~\eqref{eq:varP}. Anticipating the cancellation mechanism suggested by Schehr-Giamarchi-Le Doussal, one can expand the marginal stability condition $\l_R=0$ up to $O(\hbar)$, which provides a sum rule satisfied by $P_\k^{(1)}(1,H)$, in terms of non-Gaussian contributions (coming from the ones in~\eqref{eq:ccfhigh}). Unfortunately this sum rule is unhelpful to prove the cancellation of \textit{(ii)}-\textit{(iii)}. This is the same situation as in the RSB phase of the quantum spherical perceptron~\cite{FMPS19}. 

\newpage

\subsection{Generic mechanism with replica symmetry: loop expansion and Debye approximation}

The structure of the expansion around the saddle-point for $\hbar\to0$ at fixed Matsubara period~\eqref{eq:perturbth} shows that the fluctuation $x-x^*$ around the saddle-point value typically scales as $\sqrt\hbar$; besides, the fact that non-Gaussian terms do not contribute at $O(\hbar)$ lead to the following generic argument. 

We come back to standard notations in this section. Noting the canonical position variables by $\ux=\argp{x_1,\dots,x_N}$, 
the partition function is 
\begin{equation}
 Z=\oint \mathrm{D}\ux\,e^{-\frac1\hbar\int_0^{\b\hbar}\dd t\argc{\frac{\dot{\ux}(t)^2}{2M}+V\argp{\ux(t)}}}
\end{equation}
Assuming there is a unique global minimum $\ux^*$ of the $N$-dimensional energy landscape $V(\ux)$, we expand at all orders around the minimum $\ux=\ux^*+\uxu\sqrt\hbar$. The factor $\sqrt\hbar$ accounts for the above-mentioned fluctuation scaling. 
\begin{widetext}
 \begin{equation}\label{eq:expgen}
 \ln Z= -\b V(\ux^*)+\overbrace{\ln\oint \mathrm{D}\uxu\,e^{-\frac12\int_0^{\b\hbar}\dd t\dd s\,u_i(t)\GG^{-1}_{ij}(t-s)u_j(s)}}^{\begin{tikzpicture}[baseline=-1mm] \draw[fill=white] (0:0) circle (0.2) node {D};\end{tikzpicture}}
 +\ln\moy{e^{-\int_0^{\b\hbar}\dd t\argc{\frac{\sqrt{\hbar}}{3!}u_iu_ju_k\partial_{ijk}^3V\argp{\ux^*}+  \frac{\hbar}{4!}u_iu_ju_ku_l\partial_{ijkl}^4V\argp{\ux^*}+\dots}}}
\end{equation}
\end{widetext}
Repeated indices are summed over, and we noted
 \begin{equation}
\begin{split}
 \GG^{-1}_{ij}(t-s)=&-M\d_{ij}\d(t-s)\frac{\partial^2}{\partial s^2} +\partial_{ij}^2V\argp{\ux^*}\\
 \moy{\d x_i(t)\d x_j(s)}=&\GG_{ij}(t-s)
\end{split}
\end{equation}
In~\eqref{eq:expgen} the first term is the classical ground-state energy and $\begin{tikzpicture}[baseline=-1mm] \draw[fill=white] (0:0) circle (0.2) node {D};\end{tikzpicture}$ is the partition function of harmonic oscillators with propagator $\GG$, \ie the Debye contribution. The last term is an average with respect to the Gaussian saddle-point action (defined by the latter propagator $\GG$) which groups all non-Gaussian perturbations. These can be evaluated through a loop expansion~\cite[Chap.7]{Zinn-Justin}.
Fixing $\b\hbar$ and taking $\hbar\to0$ makes clear that at lowest order (only) in this expansion, one must recover the Debye approximation:
\begin{equation}\label{eq:Debyeloop}
 \begin{split}
  \overline{\begin{tikzpicture}[baseline=-1mm] \draw[fill=white] (0:0) circle (0.2) node {D};\end{tikzpicture}}=&-\frac12\Tr\,\overline{\ln\argp{-M\partial_t^2\id+\partial^2V(\ux^*)}}\\
  =&N\int\dd\r(\l) \argc{-\frac{1}{2}\sum_{n}\ln\argp{M\om_n^2+\l}}\\
  =&N\int\dd\r(\l)\,\ln\Tr\,e^{-\b\argp{\frac{\hat p^2}{2M}+\frac{\l}{2}\hat x^2}}\\
  =&-N\int\dd\r(\l)\,\ln\argc{2\sinh\argp{\frac{\b\hbar}{2}\sqrt{\frac{\l}{M}}}}
 \end{split}
\end{equation}
The overline stands for a disorder average if there is disorder; $\r(\l)$ is the (disorder-averaged) density of eigenvalues of the Hessian evaluated at the minimum $\partial^2V(\ux^*)$. The contribution of this term to the energy is 
\begin{equation}\label{eq:UD}
  U_{\rm D}=-\frac{\partial  \overline{\begin{tikzpicture}[baseline=-1mm] \draw[fill=white] (0:0) circle (0.2) node {D};\end{tikzpicture}}}{\partial\b}=
N\int\dd\r(\l)\,\hbar\Omega(\l)\argc{\frac12+f_{\rm B}(\Omega(\l))}
\end{equation}
where one recognizes the usual Debye expression with the zero-point energy  and the Bose-Einstein factor at frequency $\Omega(\l)=\sqrt{\l/M}$.

This loop expansion is in fact the one we performed on the disorder-averaged energy, where the action is advantageously reduced to a single degree of freedom. 
However, starting  from the energy rather than the free energy, we had to push to one higher order in the loop expansion than needed to check that non-Gaussian terms do not contribute to the energy at the lowest non-trivial order $O(\hbar)$. The loop expansion from the free energy thus gives a more direct result. Yet when disorder averaging is done first, the replicated action becomes less convenient to work with: one has to introduce extra replicas\footnote{In an analogous way to the Franz-Parisi scheme~\cite{FP95}. } to handle the unknown disorder-dependent minimum $\ux^*$. It is so far unclear to us whether such a formalism would provide an easier way to compute semiclassical corrections to the Debye formula.

\subsection{Conclusion}

In this section, we have studied a semiclassical expansion ($\hbar\to0$ with $\b\hbar$ fixed) to solve the quantum thermodynamics of the model and get analytically important quantities such as the self-energy and the specific heat, starting from a known solution of the classical model in the zero-temperature limit. By first simplifying the quantum thermodynamics through the large $N$ limit, yielding an effective single-particle impurity problem, one can perturbatively compute thermal observables order by order in $\hbar$ via an asymptotic expansion around the semiclassical saddle point. 

At first order in this expansion, in line with the earlier works by Schehr-Giamarchi-Le Doussal (SGLD) and on the spherical quantum perceptron~\cite{SGLD04,SGLD05,S05,FMPS19}, a gapped scaling of thermodynamic quantities arises in the bulk of the RS phase, while at the RSB transition lines where the system becomes marginal, the self-energy develops a singularity and the specific heat is a power law in temperature. Due to the marginal condition~\eqref{eq:replicon0}, a cancellation occurs in the first power in temperature, which impedes a linear scaling of the specific heat. This was the mechanism put forward by SGLD to explain the cubic scaling in the models they had analyzed. The interest of the KHGPS model is that there is a RSB transition line without a vanishing replicon (RSB-$\om^4$ line), thus precluding the SGLD mechanism, in addition to the usual de Almeida-Thouless transition line (RSB-$\om^2$ line) where $\l_R=0$. The latter displays a cubic scaling of the specific heat while the former is quintic. We showed that the origin of the power-law exponent is actually not the vanishing of the replicon eigenvalue. 
The mechanism is criticality, which brings gaplessness, combined with a Debye analysis provided by random matrix theory applied to the Hessian of the classical energy landscape. This is fine for glass models with continuous degrees of freedom, as in structural glasses or continuous spin glasses. For discrete degrees of freedom, it is less clear how to define a Debye approximation. In the Sherrington-Kirkpatrick model in a transverse field~\cite{AM12} it was done by approximating the TAP free energy, an argument made generic in~\cite{CM22}. We comment further on this point in~\secref{sec:conclu}.

We have thus shown that the first order in the semiclassical expansion coincides with such a ``disordered Debye'' approximation, \ie setting the specific results of the KHGPS model and of~\cite{FMPS19} on a general ground. This order is given by the Gaussian action around the semiclassical saddle-point. Technically,  the assumption of a single minimum in the energy landscape  in the RS phase is akin to the single constant saddle-point solution $x^*$~\eqref{eq:mincl} found in the impurity problem. Both reflect the replica-symmetric structure. The perturbative expansion can be seen as a semiclassical scheme that generalizes Debye's approximation, as in principle one can compute higher-order corrections that are by construction absent from Debye's approximation (where one cannot have energies other than linear in $\hbar$; in principle higher orders may be expected). These higher orders include contributions from the non-Gaussianity of the action around the semiclassical saddle point. 

Based on the analysis of the multiple Matsubara sums that appear for the higher orders in perturbation, SGLD argued that, at criticality\footnote{Similarly one can argue that in a gapped phase, the energy gap gets perturbative corrections in $\hbar$, as the position of the lower boundary of the branch cut in the self-energy does.}, only the prefactor of the specific heat gets perturbatively renormalized, but the power-law exponent remains unchanged. In the KHGPS model and in~\cite{FMPS19}, these higher orders are more involved, as, due to the self-consistent structure, one has to investigate a higher number of loops. This point deserves more examination in these models. Through a scaling argument and a conflict with Heisenberg inequality, it was argued in~\cite{FMPS19} that these higher orders may modify the power-law exponent, \ie leading to a breakdown of Debye's approximation, in the case of the jamming transition. Another point which would deserve a careful assessment is how the transition lines get shifted by the quantum fluctuations, and if this is accessible through the same perturbative strategy \cite{markland2011quantum, urbani2023quantum, winer2023glass}. 

We next looked at the RSB-SWP phase. Right at the RSB-$\om^2$ line the Debye approximation holds at first order. It remains to be understood whether it is as well true within the bulk of this usual spin-glass marginal phase where the replicon vanishes. Here the above generic argument fails because one cannot assume a single global minimum of the energy landscape. Instead the direct calculation hints at the validity of the Debye approximation, brought by a SGLD cancellation in the self-energy. However the cancellation of the related non-Gaussian terms at the energy level --thus whether or not the specific heat is actually determined by this disordered Debye scaling $C_V\sim T^3$-- could not be checked, exactly as in the analysis of the spin-glass phase in the quantum spherical perceptron~\cite{FMPS19}. 

What happens when several semiclassical saddle points are present, \ie in the case of the appearance of DWP in the impurity problem, is the subject of the next section. 

\section{Away from Debye behavior: tunneling physics}\label{sec:DWP}

A very interesting feature of the  KHGPS model is that, while it couples together a collection of SWP $(\k_m>0)$, the interaction may create effective potentials that are either SWP or DWP, thus destabilizing SWP.  This is akin to the GPS vibrational instability~\cite{GPS03,PSG07}, or to the mass renormalization felt by a particle coupled to harmonic oscillators, usually compensated by the introduction of counter-terms~\cite{CL83a,LCDFGZ87,weiss} (see~\secref{sec:spinboson}). In the following we consider the semiclassical analysis of~\secref{sec:SGLD}, but the starting point $(J,h)$ of the analysis lies inside the classical $T=0$ RSB phase where effective double-well potentials appear (see Fig.~\ref{fig:qphd}), contrary to the previous section. When $\hbar$ is increased, we expect replica symmetry to be restored. Starting close enough to the classical RSB-$\om^4$ line, the saddle-point equation~\eqref{eq:spsingle} has now  two constant solutions corresponding to minima of an effective DWP.  
We dub this a RS(B)-DWP phase. We now explore the influence of these DWP. 

A semiclassical computation of the partition function therefore requires to consider instanton solutions~\cite{coleman_erice,rajaraman,Zinn-Justin}. These are imaginary-time-dependent saddle-point solutions in the limit $\hbar\to0$ with a finite action; all quantities can be computed by asymptotic expansion around these solutions. At variance with quantum field theory, in statistical mechanics $\hbar\to0$ also appears in the upper boundary of the imaginary time integrals and as a result one must consider either $T\to 0$ first or $\b\hbar$ fixed in order to extremize a well-defined action. In other words this is the semiclassical expansion we have explored so far. Most importantly, it is able to capture non-perturbative effects  such as tunneling, while the analysis of the previous section was purely perturbative. Indeed, when restricting to single-particle quantum mechanics it becomes equivalent to the WKB method~\cite{coleman_erice,GP77,rajaraman,Zinn-Justin}, and the $\hbar$ dependence cannot be perturbative anymore. The instanton trajectories are nevertheless still ruled by the same equation that extremizes the action~\eqref{eq:spsingle}. 
Here there are two specific analytic obstacles: \textit{(i)} one should solve this equation in a self-consistent manner with respect to the variational equation for $G(t)=q_d(t)-q(1)$~\eqref{eq:varq}; without further approximations this cannot be done but numerically. \textit{(ii)} The presence of a non-local (memory) term complexifies the task compared to instantonic solutions of local-in-time models, obtained analytically~\cite{rajaraman}. Once equipped with a kink solution going from one well to the other, one usually has to resort to approximation schemes to resum the vastly many possible kink-antikink trajectories on the whole imaginary time interval $\argc{0,\b\hbar}$. 

This is a very difficult task and in this section we perform a simpler self-consistent variational computation. We expect that, starting from a point $(J,h)$ within the classical $T=0$ RSB-DWP phase, upon increasing temperature or $\hbar$ one should hit a boundary where replica symmetry is restored (see Fig.~\ref{fig:qphd}). So, \textit{we consider $\hbar$ finite so that we are in such a phase and take the limit $T\to0$}. We still solve the problem using the instanton method to uncover TTLS physics and perform a variational approximation 
that avoids the non-local analysis, turning the problem into an effective single-particle quantum-mechanical one. The approximation can be viewed as a self-consistent quantum version of K\"uhn and collaborators' analysis in~\cite{Ku97,HK99,Ku00,Ku03}. 
The main objectives here are to probe the phase diagram of the system in this regime, uncover the dominant low-temperature excitations and derive the corresponding behavior of the specific heat, both numerically and analytically. 

\subsection{A variational expression for the energy}
\label{sub:varen}

We define as before $\c=\b\wt G(0)$. The energy~\eqref{eq:UfullRSB} depends in particular on the effective partition function $f_\k(1,H)$~\eqref{eq:f1def}. 
The latter is difficult to handle due to the memory term in the action $\AA$~\eqref{eq:actionAfull}. In the following we will resort to a variational approximation that simplifies this memory term. As our approach is devised for low temperature, we will right away consider only dominant term in the free energy for $T\to0$. In this limit $\b G$ is naturally of order one from~\eqref{eq:actionAfull}, meaning $q_d\to q(1)$. Separating $q(x)^2-q_d(t)^2=q(x)^2-q(1)^2+q(1)^2-q_d(t)^2$, we rewrite the overbraced free energy term  in~\eqref{eq:FAio} as
\begin{equation}\label{eq:lowTvar}
\begin{split} 
 &\argp{\frac{\b J}{2}}^2\int_0^1\dd x \argc{q(x)^2-q(1)^2}\\&\qquad-\frac{\b J^2}{2}q(1)\c-\argp{\frac{J}{2}}^2\int_0^1 \dd t\,\argc{\b G(t)}^2
\end{split}
 \end{equation}
For $\b\to\io$ the last term $(\b G)^2$ in~\eqref{eq:lowTvar} is order one and thus subdominant: we discard it from the start\footnote{This subdominant term would pointlessly complicate the variational approximation we study here (formally divergent). }. 
Next, we consider a ``Markovian'' simplification by using a variational parametrization 
\begin{equation}\label{eq:varapprox}
 \b G(t)=\c\d(t)
\end{equation}
which gives the low-temperature variational approximation of the free energy~\eqref{eq:FAio} parametrized by $\arga{q(x),\c}$:
\begin{widetext}
\begin{equation}\label{eq:FAioa}
\begin{split}
 -\frac{\b F^a}{N}=&\argp{\frac{\b J}{2}}^2\int_0^1\dd x\, \argc{q(x)^2-q(1)^2}
  -\frac{\b }{2}J^2 q(1)\c-\int\dd p(\k)\int\dd H\,P_\k(1,H)\argc{ f_\k(1,H)-\ln\Tr \, e^{-\b \hat H_{\rm eff}}}\\
&+\int\dd p(\k)\,e^{\frac{J^2}{2}q(0)\frac{\partial^2}{\partial h^2}}f_\k(0,h)+\int\dd p(\k)\int_0^1\dd x\int\dd H\,P_\k(x,H)\argc{\dot f_\k(x,H)+\frac{J^2}{2}\dot q(x)\argp{f''_\k(x,H)+x\argp{f'_\k(x,H)}^2}}
\end{split}
\end{equation}
\end{widetext} 
Indeed, this approximation neglects the non-locality through ${\int_0^1\dd s\,\b G(t-s)x(s)=\c x(t)}$; thus $\AA$ becomes the standard quantum-mechanical action of a particle in a potential $v_m(x)$~\eqref{eq:H} with $m=\k-J^2\c$, meaning $f_\k(1,H)$ is approximated as 
\begin{equation}\label{eq:fm1}
\begin{split}
 f_\k(1,H) \to f_m(H)=&\ln\Tr\,e^{-\b \hat H_{\rm eff}}\,,\ 
 \hat H_{\rm eff}=\frac{\hat p^2}{2M}+ v_m(\hat x)
\end{split}
\end{equation}
Averages over the impurity problem become standard single-particle averages:
\begin{equation}
 \moy{\bullet}\to\ \frac{\Tr\,\bullet\,e^{-\b \hat H_{\rm eff}}}{\Tr\,e^{-\b \hat H_{\rm eff}}}
\end{equation}
The saddle-point equations are obtained through extremization of the approximate free energy~\eqref{eq:FAioa}.  
The fullRSB equations for $P_\k$ and $f_\k$ remain unchanged except for the boundary condition $f_\k(1,H)$~\eqref{eq:fm1}. The exact remaining equations from Sec.~\ref{sub:fRSB} are
\begin{subequations}\label{eq:approxfRSB}
 \begin{align}
  q(x) = &\int\dd p(\k)\dd H\, P_\k(x,H)\argc{\frac{f'_\k(x,H)}{\b}}^2\label{eq:frsbq}\\
  \c=&\b\wt G(0)=\int\dd p(\k)\dd H\, P_\k(1,H)\frac{f''_\k(1,H)}{\b}\label{eq:frsbc}\\
  1=&J^2 \int\dd p(\k)\dd H\,P_\k(1,H)\argc{\frac{f''_\k(1,H)}{\b}}^2\label{eq:frsbrep}
 \end{align}
\end{subequations}
the last one being the marginal stability condition associated to the replicon. \eqref{eq:frsbq} and \eqref{eq:frsbc} are indeed obtained extremizing with respect to $\c$ and $q(x)$, except that the equation for $q(x)$ is slightly different\footnote{The extremization over $\c$ indeed yields
\begin{equation}
  q(x) = \int\dd p(\k)\dd H\, P_\k(x,H)\moy{x^2}
\end{equation}
with $\moy{x^2}=\argc{f'_\k(1,H)^2+f''_\k(1,H)}/\b^2$. We expect the difference with~\eqref{eq:frsbq} to be subdominant, owing to ${f''_\k(1,H)/\b^2\ll \argc{f'_\k(1,H)/\b}^2}$ as the logarithm of the effective partition function $f_\k(1,H)\underset{T\to0}{\sim}-\b E_g(\k,H)$, $E_g$ being the ground-state energy of the impurity problem. This is self-consistently confirmed by the low-temperature expressions~\eqref{eq:fHO} and~\eqref{eq:fTLS}.}. However at dominant order in temperature it is identical to~\eqref{eq:frsbq} and one may consider either one, as expected from the present low-temperature simplification.
 

From the variational principle, this approximate free energy bounds from above the correct one~\cite{FHS10}.  The energy~\eqref{eq:UfullRSB} becomes
\begin{equation}\label{eq:URSBvar}
\begin{split}
 \frac{U^a} N =&\frac{\b J^2}{2}\int_0^1\dd x\, \argc{q(1)^2- q(x)^2}+\frac{J^2q(1)\c}{2}\\
 &+\int\dd p(\k)\dd H\,P_\k(1,H)\moy{\hat H_{\rm eff}}
\end{split}
 \end{equation} 
The fullRSB equations will be useful in~\secref{sec:pseudogap} but in this section, as previously mentioned,  we shall specialize to a RS phase taking $\hbar$ sufficiently large and then $T\to0$. Here the only parameters are the overlap $q(x)=q$ and $G(t)=q_d(t)-q$. The RS equations for $(\c,q)$ and the marginal stability equations are given by~\eqref{eq:approxfRSB} with $x=1$ and $P_\k(x,H)\to\g_{J^2q}(H-h)$~\eqref{eq:varP}. This simply means that we integrate with the Gaussian measure ${\DD_{J^2q} H=\g_{J^2q}(H-h)\dd H}$. Taking the classical limit $\hbar\to0$ one recovers\footnote{For $\hbar\to0$, the free energy~\eqref{eq:FAioa} becomes the exact classical one apart from a term $-(J\c_{\rm cl}/2)^2$ that we dropped in~\eqref{eq:lowTvar}, which for $T\to0$ is only an irrelevant constant in the free energy.} the classical equations of~\cite{BLRUZ21} for $T\to0$, apart from the kinetic term $p^2/(2M)$ -- the present approximation thus does not spoil this aspect. The RS low-$T$ variational approximation for the energy reads
\begin{equation}\label{eq:Uvar}
  \frac{U_{\rm RS}^a}{N}=\frac{J^2q\c}{2}+\int\dd p(m)\DD_{J^2q} H\,\moy{\hat H_{\rm eff}}
\end{equation}

\subsection{Low-energy excitations}\label{sub:lowtemp}

The crucial input is the partition function $f_m(H)$. The potential $v_m(x)$, depending on $(m,H)$ values, displays either a single well (SWP) or double well (DWP). The model is now very reminiscent of the soft-potential model~\cite[Chap.9]{Esquinazi}: we have a collection of independent particles in either a SWP or a DWP, although here the original interparticle interaction is  manifesting through self-consistent determination of the potential parameters and their distribution.  At low temperature for SWP the partition function can be replaced by the one of a harmonic oscillator in the bottom of the well, while for some DWP tunneling must be taken into account. We do so in the simplest manner restricting ourselves to the first two levels and their tunnel amplitude. Both regimes break down close to the origin in the $(m,H)$ plane, corresponding to the purely quartic potential. At low temperature this region is truncated to the first two eigenlevels, computed independently and exactly for this special case. In this subsection we first study the associated low-energy single-particle excitations of each region, which as an aside allows to fix the regions' boundaries. Then we will proceed in~\secref{sub:numerapprox} to numerical checks of the approximations. 

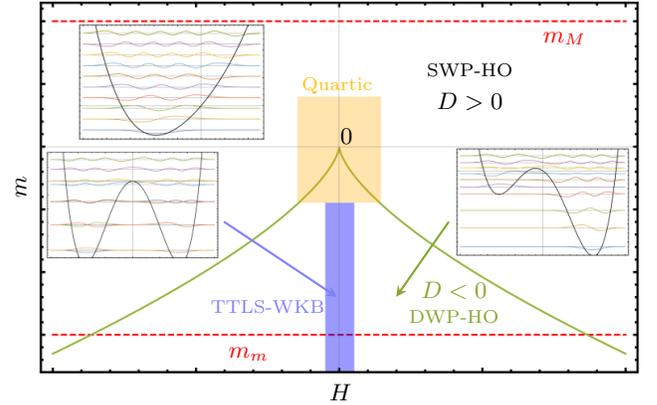
\begin{figure}[h!]
\centering
\begin{lpic}[]{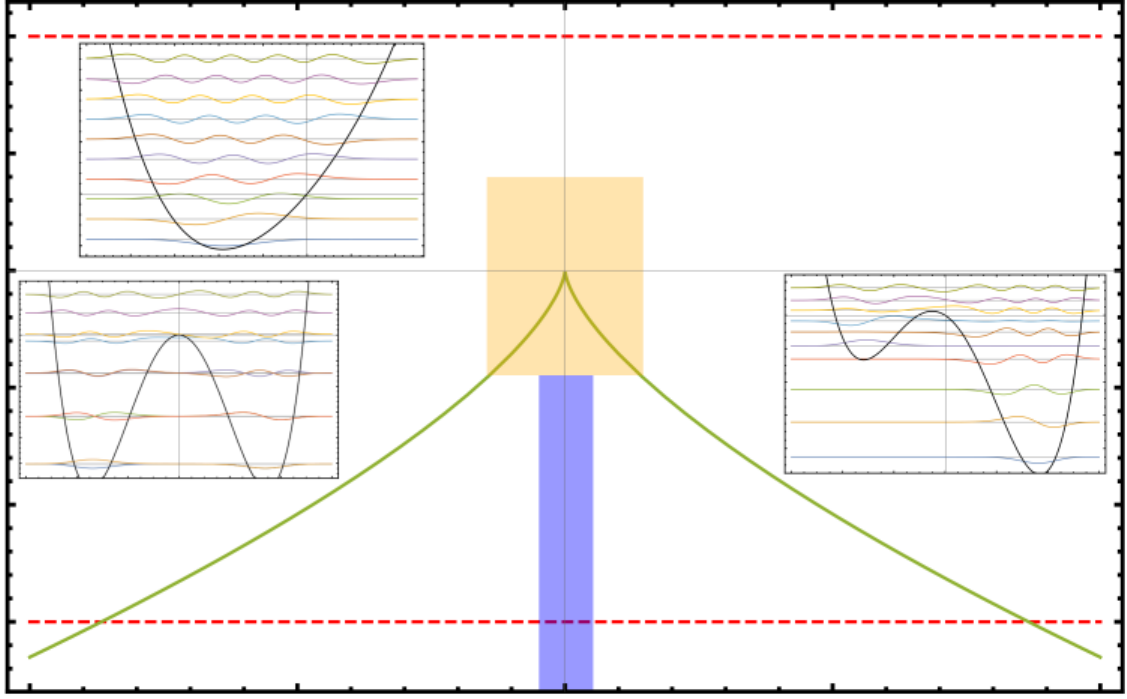(0.42)}
\lbl[]{-5,60,90;$m$}
\lbl[]{96,-5;$H$}
\lbl[]{132,19;\textcolor{green!30!brown}{{\scriptsize DWP-HO}}}
\lbl[]{73,22;\textcolor{blue!50!white}{{\scriptsize TTLS-WKB}}}
\lbl[]{95,92;\textcolor{yellow!50!orange}{{\scriptsize Quartic}}}
\lbl[]{137,97;{\scriptsize SWP-HO}}
\lbl[]{67,7;$\red{m_m}$}
\lbl[]{167,107;$\red{m_M}$}
\lbl[]{98,76;$0$}
\lbl[]{132,27;$\textcolor{green!30!brown}{D<0}$}
\lbl[]{137,87;$D>0$}
\lbl[]{77,37;\begin{tikzpicture}\draw[-stealth, thick,color=blue!50!white] (-0.5,0) -- (1,-1);\end{tikzpicture}}
\lbl[]{122,37;\begin{tikzpicture}\draw[-stealth, thick,color=green!30!brown] (0.2,0) -- (-0.5,-1);\end{tikzpicture}}
     \end{lpic}
 \caption{Nature of the potential ${v_m(x)=\frac{m}{2}x^2+\frac{x^4}{4!}-H x}   $ as a function of $(m,H)$ with boundaries $m\in\argc{m_m,m_M}$ (dashed red) for $m_m<0$ and $m_M>0$. A second minimum appears (DWP) when the discriminant $D= m^3+\frac98H^2<0$  while for $D>0$ there is a single minimum of the potential (SWP). Both regimes are separated by the green line. The yellow center zone is the quartic zone $\abs{m}\lesssim\hbar^{2/3}$, $\abs H\lesssim\hbar$, see~\eqref{eq:quartic}. The thin blue zone below is where TTLS are found, see~\eqref{eq:WKBvalid} and~\secref{sec:pseudogap}. Outside these zones, for low temperature the harmonic approximation around the (lowest) minimum is valid. In inset are shown potentials $v_m(x)$  with their first ten energy levels (horizontal lines) and corresponding wavefunctions for the Hamiltonian $\hat H_{\rm eff}$~\eqref{eq:fm1}. Tunneling is only observed in the nearly symmetric DWP ($m\ll-\hbar^{2/3}$, $H\simeq0$), the other DWP has same value of $m$ but larger $H>0$ (still within $D<0$). The inset SWP has $m>0$ and $H<0$. For a nearly-symmetric DWP, the energy levels within the wells come in pair, unlike the well-separated energy levels of the harmonic oscillator.}
 \label{fig:DWP}
\end{figure}

The extrema of the potential can be obtained analytically as $v_m'(x)=0$ is a cubic depressed equation~\cite{cubic}. We note $x_a$ the absolute minimum and $x_s$ the secondary one when it exists (only when the discriminant $D<0$):
\begin{equation}\label{eq:cubicdepressed}
 \begin{split}
  x_a(m,H)=&2\sign(H)\sqrt{2\abs{m}}\FF_{\sign(m)}(y)\ ,\\
  \FF_+(y)=&\sinh\argc{\frac13\textrm{arcsinh}(y)}\,, \ y\equiv\frac{3\abs{H}}{\argp{2\abs{m}}^{3/2}}\\
   \FF_-(y)=&\left\{ 
   \begin{split}
   &\cos\argc{\frac13\arccos(y)}\,,\ y<1 \\
&\cosh\argc{\frac13\textrm{arccosh}(y)}\,,\ y>1
   \end{split}\right.\\
  x_s(m,H)=&2\sign(H)\sqrt{2\abs{m}}\GG(y)\,,\\
  \GG(y)=&\cos\argc{\frac13\arccos(y)-\frac{4\p}{3}}\\
  y=&1\quad\Leftrightarrow \quad D\equiv   m^3+\frac98H^2=0
 \end{split}
\end{equation}

\subsubsection{Harmonic oscillators}

When the potential is a SWP ($D>0$) or if it is DWP ($D<0$) with one well much lower than the other (precised later on), one can approximate the low-energy spectrum by the one of a harmonic oscillator (HO) around the global minimum, whose partition function\footnote{At low temperatures only the gap is important and we could have truncated the harmonic oscillator to its two lowest levels.}  is
\begin{equation}\label{eq:fHO}
\begin{split}
  f_m^{\rm HO}(H)=&
  -\b v_a-\ln\argc{2\sinh\argp{\frac{\b\hbar\om_a}{2}}}\\
   \textrm{with}\ 
  v_*=&v_m(x_*)\ ,\ \om_*=\sqrt{\frac{v''_m(x_*)}{M}}\ ,\ *=a,s
\end{split}
\end{equation}
The harmonic approximation fails for $(m,H)$ close to the origin $(0,0)$ (purely quartic potential). In this case we consider a two-level truncation (valid for low temperatures) similar to tunneling DWP, discussed below. 

\subsubsection{Tunneling two-level systems}

For DWP with significant tunneling, we instead approximate using the two-level system (TLS)  model~\cite{AHV72,Ph72,Ph87,Esquinazi}. At each bottom of a well, the Hamiltonian $\hat H_{\rm eff}$ is approximated by the corresponding harmonic oscillator $\hat H_*=\frac{\hat p^2}{2M}+v_*+\frac{M\om_*^2}{2}(\hat x-x_*)^2$, $*=a,s$. The ground state  $\ket{\phi_*}$ of each oscillator, satisfying 
$H_*\ket{\phi_*}=E_*\ket{\phi_*}$, and the transition amplitude between them, provide an approximate low-energy truncation of the Hamiltonian. The transition amplitude gives rise to tunneling and is assessed through a WKB method~\cite[Eq.(39)]{So15}-\cite{Ga00,So08} matching the wavefunction within the barrier with the one of the harmonic wells beyond. As mentioned earlier the WKB result is equivalent to the instanton one for single-particle quantum mechanics~\cite{coleman_erice,GP77,rajaraman,Zinn-Justin}, consistently with our strategy\footnote{An instanton trajectory from $a$ to $b$ has the constant of motion $E=\frac{\dot x^2}{2}-v_m(x)$, from which the action reads $\int_0^{\b\hbar} \dd t \argc{\frac{\dot x^2}{2}+v_m(x)}=-\b\hbar E+\int_0^{\b\hbar} \dd t\, \dot x\sqrt{2\argp{v_m(x)-E}}=-\b\hbar E+\int_{a}^{b}\dd x\,\sqrt{2\argp{v_m(x)-E}}$. This is where the exponent of the tunneling amplitude $\D_0$~\eqref{eq:TLS} comes from; the exponential (in $\hbar$) form stems from summing over many kink-antikink solutions in the dilute approximation (well-separated kinks)~\cite{coleman_erice}. Instead, HO potentials do not involve any instanton, \ie Debye approximation is valid for them, resulting in a very different (linear) scaling in $\hbar$ of the gap. \label{footinst} }. We note this tunneling amplitude $\D_0/2$. 
The Hamiltonian is therefore restricted to a $2\times2$ low-energy sector $\arga{\ket{\phi_a},\ket{\phi_s}}$
\begin{equation}\label{eq:TLS}
 \begin{split}
 \hat H_{\rm eff}\approx&\begin{pmatrix}
                    E_a & \bra{\phi_a}\hat H_{\rm eff}\ket{\phi_s}\\
                    \bra{\phi_s}\hat H_{\rm eff}\ket{\phi_a} & E_s
                   \end{pmatrix}
                   =\begin{pmatrix}
                    E_a & \D_0^*/2\\
                    \D_0/2 & E_s
                   \end{pmatrix}\\
E_*=&v_*+\frac{\hbar\om_*}{2}\ ,\ *=a,s\\
\D_0=&\hbar\sqrt{\frac{\om_a\om_s}{{\p e}}}\exp\argp{-\frac1\hbar\int_{a}^{b}\dd x\,\sqrt{2\argp{v_m(x)-E}}}
 \end{split}                   
\end{equation}
with $a(m,H)$ and $b(m,H)$ respectively the left and right classical turning points at energy $E$ (\ie the solutions of $v_m(x)=E$ within the barrier).
$E$ is in fact $E_s$ due to the matching with harmonic wells~\cite{So15} and is obtained explicitly from the latter quartic equation\footnote{The quartic equation $v_m(x)=E$ is written in depressed quartic form as $x^4+\a x^2+\b x+\g=0$~\cite{quartic} with $\a=12 m$, $\b=-4!H$, $\g=-4!E$, providing 4 solutions
\begin{equation*}
x=\frac{\pm_1W\pm_2\sqrt{-\argp{3\a+2y\pm_1\frac{2\b}{W}}}}{2} 
\end{equation*}
with $W=\sqrt{\a+2y}$, 
\begin{equation*}
 y=-\frac56\a+\left\{
 \begin{split}
 -\sqrt[3]{Q}\ ,\quad U=0\\
 U-\frac{P}{3U}\ ,\quad U\neq0
 \end{split}\right. \ ,
\end{equation*}
with $P=-\frac{\a^2}{12}-\g$, $Q=-\frac{\a^3}{108}+\frac{\a\g}{3}-\frac{\b^2}{8}$, $U=\sqrt[3]R$, $R=-\frac{Q}{2}+\sqrt{\frac{Q^2}{4}+\frac{P^3}{27}}$.  
Fixed sign $\pm_1$ solutions have same sign; $a(m,H)$ is obtained for $(\pm_1,\pm_2)=(-,+)$ whereas  $b(m,H)$ is obtained for $(\pm_1,\pm_2)=(+,-)$.}. 
One requirement of the latter formula is that the wells are deep enough (\eg $\abs m$ large enough). Thus $E_s-v_s=\hbar\om_s/2\ll \abs{v_s}$. This means we rather set $E=v_s$ which ensures positivity inside the square root. Note indeed that in the instanton approach, any small in $\hbar$ deviation of $E$ from $v_s$ is a perturbation, and would only change the prefactor of the exponential -- a particularly weak effect. We shall instead see that when $\hbar$ becomes large, this TTLS-WKB approximation is not anymore the correct regime, replaced by the quartic potential scalings, which are studied below. Finally, these choices do not preserve analyticity in $H$, which we will come back to in \secref{sub:analytic}.
Except these precautions, the particular choice of $E$ has negligible impact on our numerics in~\secref{sec:num}. \\
In the following we will be interested only in the eigenvalues of the Hamiltonian, so that $\D_0$ can be taken real. One has in the pseudospin representation
\begin{equation}\label{eq:HeffTLS}
 \begin{split}
  \hat H_{\rm eff}&= \bar\varepsilon\,\hat\id+\frac{\D_0}{2}\hat\s_x+\frac{\D}{2}\hat\s_z\ , \quad \D=E_s-E_a\ ,\\
  \bar\varepsilon&=\frac{E_a+E_s}{2}\ ,\quad \varepsilon=\sqrt{\D^2+\D_0^2}\quad \textrm{(gap)}
 \end{split}
\end{equation}
where $\hat\s_i$ are the Pauli matrices, generating two levels at energies $\bar\varepsilon\pm\varepsilon/2$. 
To fix ideas, simple expressions are obtained for $E=v_s$ in the symmetric case $H\to0$, which maximizes tunneling at fixed $m$. One has $x_{a,s}=\pm\sqrt{6\abs m}+O(H)$, $\om_{a,s}=\sqrt{2\abs m}+O(H)$, $\D=O(H)$ (see also Figs.~\ref{fig:Debye_gap}(a)-\ref{fig:D}(a) graphically) and the WKB integral can be computed analytically, providing
\begin{equation}\label{eq:D0H0}
 \varepsilon(H=0)=\D_0(H=0)=\frac{2\hbar}{\sqrt{e \p}}\abs m e^{-4\sqrt 2 \,{\abs{m}^{3/2}}/{\hbar}}
\end{equation}
The corresponding logarithm of the partition function is
\begin{equation}\label{eq:fTLS}
 f_m^{\rm TLS}(H)
 =-\b \bar\varepsilon+\ln\argc{2\cosh\argp{\frac{\b\varepsilon}{2}}}
\end{equation}
Notice further that all formulas 
depend on $M$ through $\hbar/\sqrt{M}$. We now set $M=1$ for convenience. Moreover, the Hamiltonian possesses the symmetry $\arga{H,x}\to\arga{-H,-x}$ so that we can always restrict to $H\geqslant0$ integrations.

\subsubsection{Quartic oscillators}\label{sub:quartic}

Close to the quartic region $(m,H)\simeq(0,0)$, we use the same truncation to two levels~\eqref{eq:fTLS} parametrized by $(\bar\varepsilon,\varepsilon)$, except that in practice these parameters are not assessed by a semiclassical WKB method but by solving the static Schr\"odinger equation for the two lowest  levels $\bar\varepsilon\pm\varepsilon/2$. Numerically we perform it in Mathematica using the matrix algorithm described in~\cite{Ko02,codeJN} for polynomial potentials in one dimension, writing position and momentum operators in the basis of harmonic-oscillator eigenfunctions. \\
A sketch of the three different regimes in the $(m,H)$ plane is displayed in Fig.~\ref{fig:DWP}. 

\subsubsection{Boundaries of the three low-energy excitation regimes}\label{sub:boundaries}

We now need to delimit the different regimes (HO, TTLS and quartic) in the $(m,H)$ diagram. The HO potentials are either SWP or DWP with negligible tunneling, typically very asymmetric. TTLS potentials are typically almost symmetric DWP with moderately high barrier to ensure significant tunneling~\cite{cohen}. Both regimes fail close to the quartic point $(m,H)=(0,0)$, see below. To gain insight on these regimes, in the rest of the section we compare them to a numerical solution of the static Schr\"odinger equation for $\hat H_{\rm eff}$~\eqref{eq:fm1}, as described in the previous subsection. 

We seek a HO/TTLS cutoff line within the DWP region $D<0$, yet both regimes must be bounded from above on the $m$ axis: indeed the purely quartic potential $x^4/4!$ has a gap (for $\hbar/\sqrt M=1$, we compute it numerically as 0.598) whereas both HO/TTLS predict a vanishing gap ($\om_a=\D=\D_0=0$). This large gap brings an exponentially small specific heat, making such values of $m$ negligible for this quantity. At small $\abs m$ with $H=0$, the levels given by the static Schr\"odinger equation are 
\begin{equation}
 \argp{-\frac{\hbar^2\nabla^2}{2}+\frac{x^4}{4!}}\psi(x)=E\psi(x)
\end{equation}
We look for a scaling in $\hbar$ of the energy $E$. As $x_a\to0$, scaling $x\sim\hbar^\a$  implies that both left-hand-side terms are of the same order iff $\a=1/3$, thus 
\begin{equation}\label{eq:numtruegap}
 E\sim\hbar^{4/3}
\end{equation}
is a typical scaling of the gap around the origin $(m,H)=(0,0)$. Close to the purely quartic potential, requesting that $\frac m2 x^2\sim x^4/4!$, \ie $x\sim\sqrt{\abs m}$, one gets the harmonic gap 
\begin{equation}\label{eq:harmgap}
 \hbar\om=\hbar\sqrt{m+\frac{x^2}{2}}\sim\hbar\sqrt{\abs m}
\end{equation}
Numerically we see that the above harmonic gap is a good approximation at large $m>0$, but as we lower $\abs m$ the harmonic gap $\om$ decreases down to $0$, departing from the true finite gap as soon as the correct gap scaling~\eqref{eq:numtruegap} becomes comparable to the harmonic one~\eqref{eq:harmgap}, \ie when
\begin{equation}\label{eq:lowerm}
\abs m\sim\hbar^{2/3}
\end{equation}
\noindent This is valid for either $m>0$  (SWP) or $m<0$ (DWP) regimes and comes from the above fact that in the quartic region $m x^2/2 \sim \hbar^{4/3}$ with $x\sim \hbar^{1/3}$ . 

\begin{figure}[h!]
\centering
\begin{lpic}[]{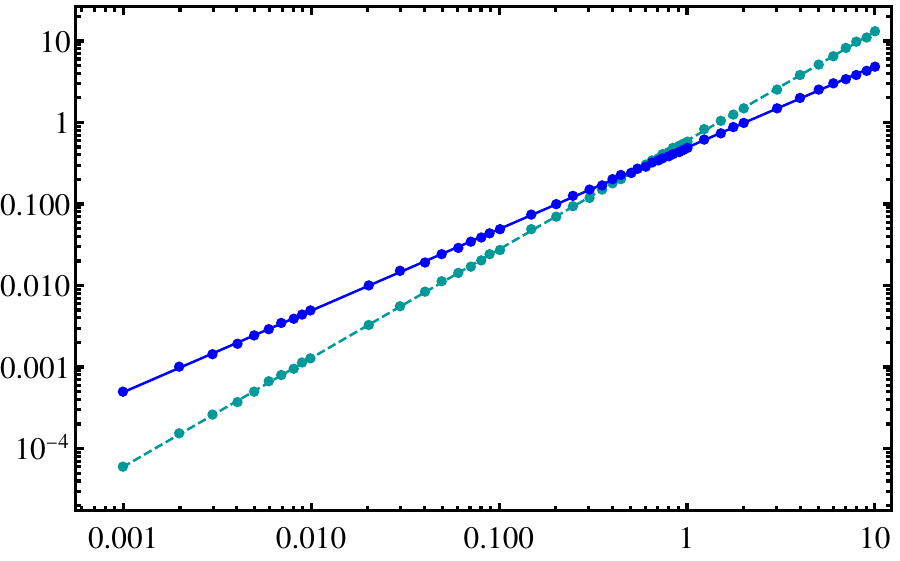(0.42)}
\lbl[]{80,100;$\hbar$}
\lbl[]{-10,50,90;{\scriptsize\textcolor{blue}{$H_c(m=-\hbar^{2/3})$}, \textcolor{green!60!blue}{$\varepsilon(m=0,H=0)$}}}
     \end{lpic}\\
     \begin{lpic}[]{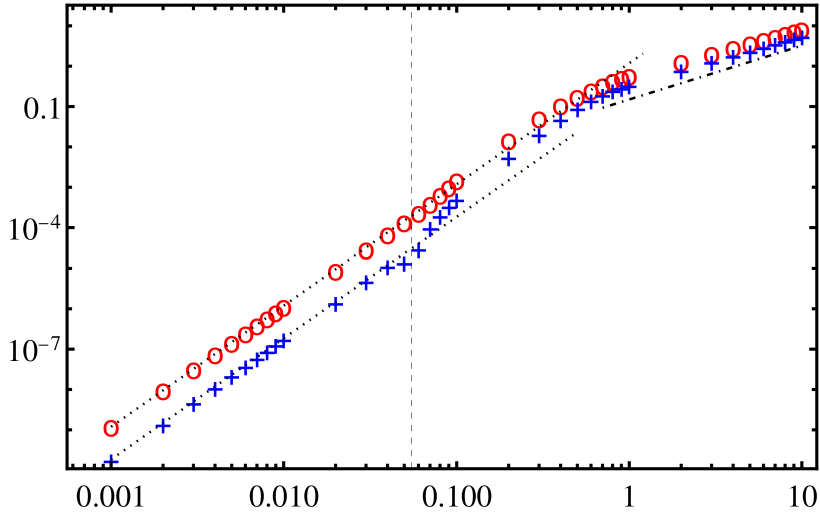(0.46)}
\lbl[]{74,-5;$\hbar$}
\lbl[]{-10,50,90;${\abs{U_Q}}/{N}$}
     \end{lpic}
 \caption{(top) $\hbar$ scaling of (solid line) the boundary point of the quartic region $H=H_c(m=-\hbar^{2/3})$ (computed according to the criterion Eq.~\eqref{eq:HOTLScriterion}), (dashed) the purely quartic potential's gap. Dots represent the numerical data while the lines are power-law fits. The $H_c$ line is $H_c(m=-\hbar^{2/3})=0.491\dots\hbar$ (Eq.~\eqref{eq:Hscal}) and the gap line is $\varepsilon(m=0,H=0)=0.598\dots\hbar^{4/3}$  (Eq.~\eqref{eq:numtruegap}). 
  (bottom) $\hbar$ scaling  of the quartic region (delimited by~\eqref{eq:quartic}) contribution, noted $U_Q$, to the absolute value of the full energy, Eq.~\eqref{eq:Uvar}, for two temperatures. The self-consistent values of $\c(T),q(T)$ have been input (see~\secref{sec:num}). Circles are for $\b\hbar=1$, + signs for $\b\hbar=10^5$. The former have $U_Q>0$, while the latter has  $U_Q<0$ for $\hbar\lesssim0.05$ and $U_Q>0$ above (delimited by the vertical dashed line). Dotted lines represent a power law $\propto \hbar^3$ as $U_Q/N\sim \D m\D H\,\varepsilon\, f(\b\varepsilon) $ combined with the scalings~\eqref{eq:numtruegap},\eqref{eq:Qext} and that for low enough $\hbar$ the function $f$ should not vary much, as $\b\varepsilon\sim(\b\hbar)\hbar^{1/3}$ with $\b\hbar$ fixed here. For large $\hbar$ instead the whole fixed interval $\argc{m_m(\c),m_M(\c)}$ is completely included in the quartic window (see Fig.~\ref{fig:T=0}). Therefore  $U_Q/N\sim\varepsilon\sim\hbar^{4/3}$ (dot-dashed line). }
 \label{fig:hbarscaling}
\end{figure}

Furthermore, this lower bound on $\abs m$ is recovered consistently from the validity of the WKB approximation for the TTLS expressions~\eqref{eq:TLS}. Semiclassicality requires that the action $\int\abs{p}\dd x\gg\hbar$, which is the same condition as having an instanton solution (finite action with respect to $\hbar$). From~\eqref{eq:D0H0} this condition is equivalent to large values of the exponential argument, which breaks down when the condition~\eqref{eq:lowerm} is satisfied\footnote{One consequence is that the WKB formula~\eqref{eq:D0H0} incorrectly predicts that the gap vanishes for $\abs m\to0$, \ie outside its validity domain.}. 

Besides, reinstating the field term $-Hx\psi(x)$ in the Schr\"odinger equation, the same scaling argument provides the scaling 
\begin{equation}\label{eq:Hscal}
 H\sim\hbar
\end{equation}
in this  region around the purely quartic potential $(m,H)=(0,0)$. Therefore this quartic region has extension 
\begin{equation}\label{eq:Qext}
 \D m\D H\sim \hbar^{5/3}
\end{equation}
These $\hbar$ scalings can be successfully tested numerically, see Fig.~\ref{fig:hbarscaling}.

\begin{figure}[!htbp]
\centering
\begin{tabular}{cc}
\includegraphics[width=7cm]{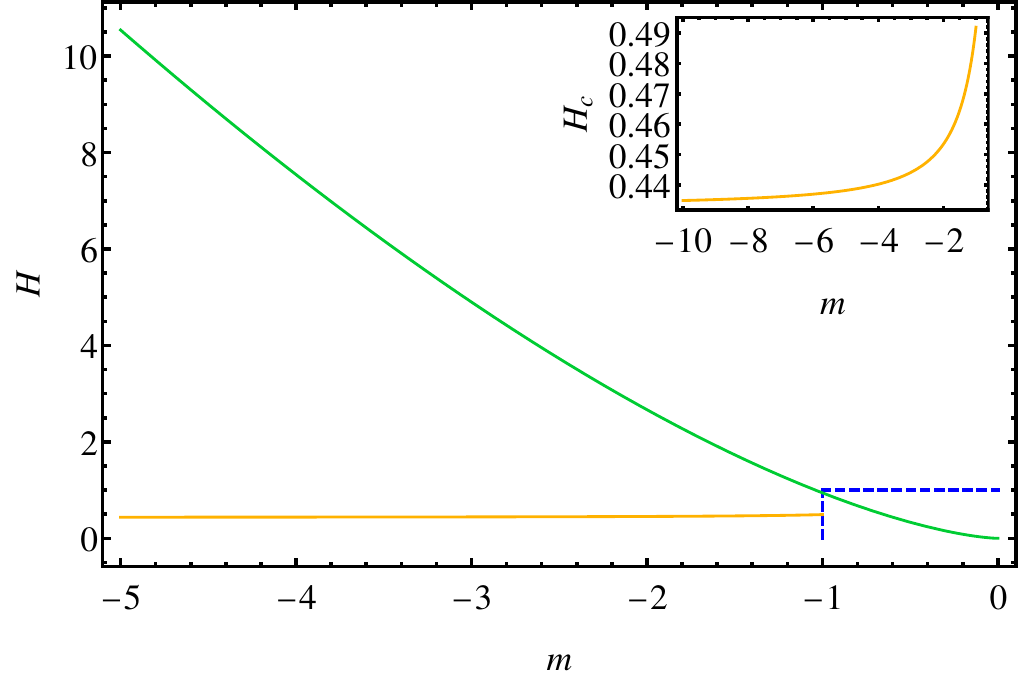}
\end{tabular} 
 \caption{ Green solid line: boundary SWP/DWP. Yellow solid line:  HO/TTLS boundary  from criterion~\eqref{eq:HOTLScriterion}. Dashed lines delimit the quartic region at $m<0$. Here $\hbar=1$ but others values only rescale the axes, see~\eqref{eq:quartic} and Fig.\ref{fig:hbarscaling}(top). Inset: HO/TTLS boundary for a larger range of $m$. It is almost a constant in $m$ and in our numerics is approximated by $H_c=0.491\dots\hbar$ (Fig.~\ref{fig:hbarscaling} top). }  
 \label{fig:mHlowT}
\end{figure}

Let us come back to the HO/TTLS cutoff in the DWP region (fixing $m\ll-\hbar^{2/3}$). It is in principle temperature dependent: the crucial variables scaling with temperature are $\b\hbar\om_a$ for HO and $\b\varepsilon$ for TTLS. As soon as these variables become large we get ${f_m^{\rm TLS}(H)\sim f_m^{\rm HO}(H)\sim-\b E_a}$; either regime is valid. Thus the cutoff is allowed to be quantitatively imprecise at low enough temperature. It is sufficient that only TTLS with vanishing gap $\varepsilon\sim T$ be counted as TTLS, the rest can be assigned as HO. With this in mind, numerics (see~\secref{sub:numerapprox}) validate the following approximate criterion to separate HO from TTLS potentials: HO potentials need localized energy levels in the deepest well, which cannot be true if the two wells' levels start hybridizing, giving rise to tunneling. Thus the first excited energy level within the deepest well must not exceed the classical energy of the secondary minimum:
\begin{equation}\label{eq:HOTLScriterion}
 \textrm{HO}\qquad\Rightarrow \qquad E_a+\hbar\om_a<v_s
\end{equation}
This criterion gives a line noted $\pm H_c(m)$ depending on the sign of $H$ in the DWP sector, see Fig.~\ref{fig:mHlowT}.
In other words, if the classical energy difference $v_s-v_a$ is smaller than the harmonic gap, we assign the potential to be TTLS. 

\subsection{Numerical assessment of the WKB approximation}\label{sub:numerapprox}

\begin{figure}[!htbp]
\centering
 \includegraphics[width=7cm]{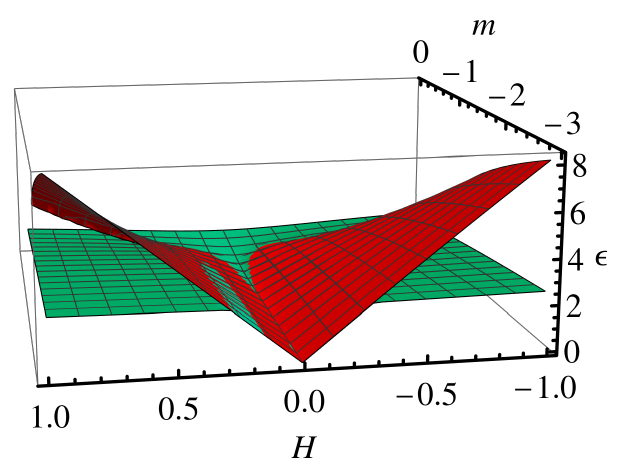}\\
\includegraphics[width=7cm]{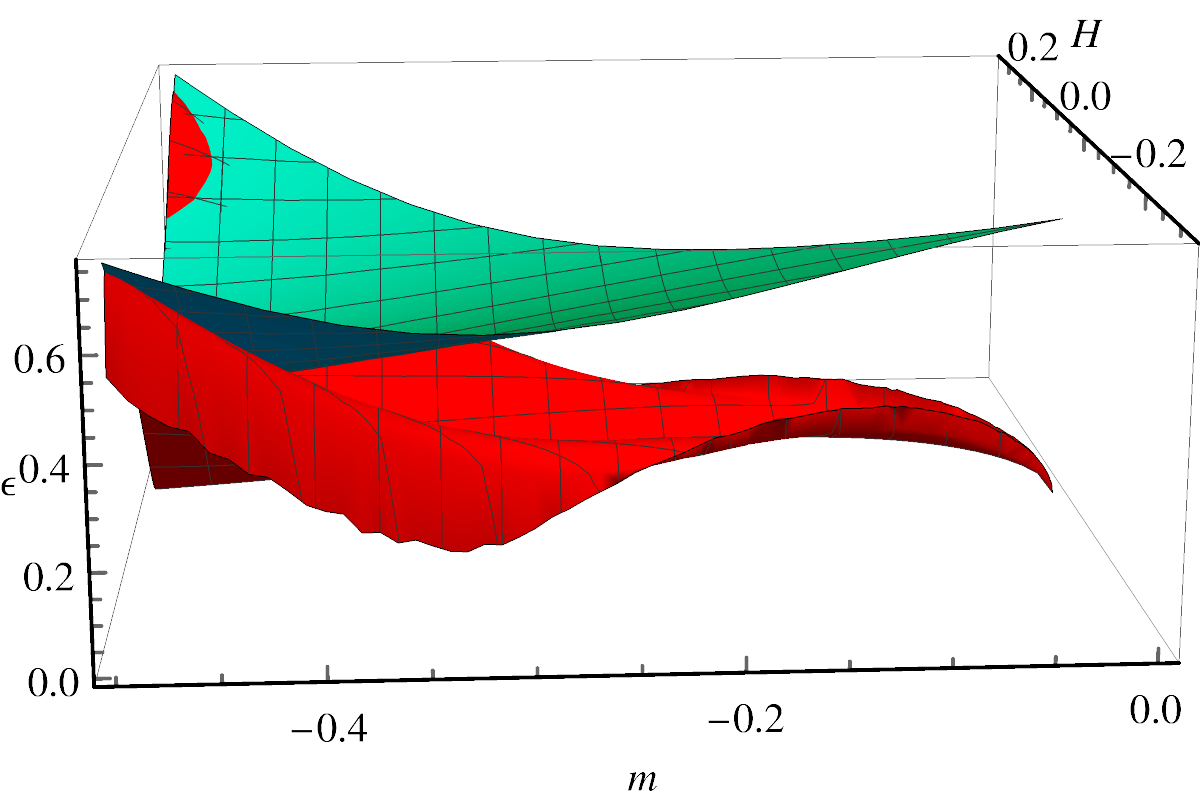}
 \caption{(top) TTLS gap $\varepsilon$~\eqref{eq:HeffTLS} via WKB approximation (red) against ``exact'' numerical gap (cyan) for $\hbar =1$. The gap has a linear $H$ dependence close to $H=0$, captured by both curves which yet differ at larger $H$ (as WKB breaks down). The WKB approximation also fails around the origin $(m,H)=(0,0)$. (bottom) is a zoom into this discrepancy, with data only for DWP parameters ($D<0$). }
 \label{fig:WKB_vs_num}
\end{figure}

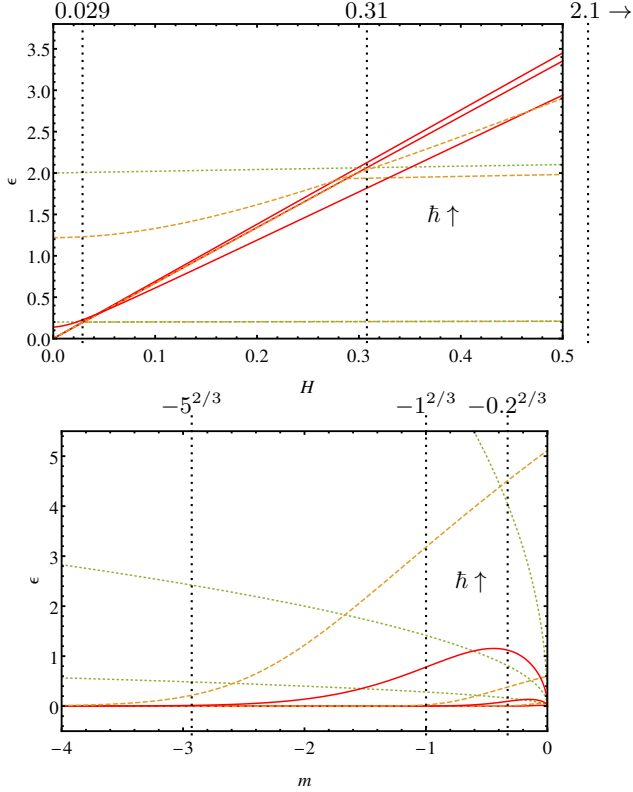
\begin{figure}[!htbp]
\centering
\begin{lpic}[]{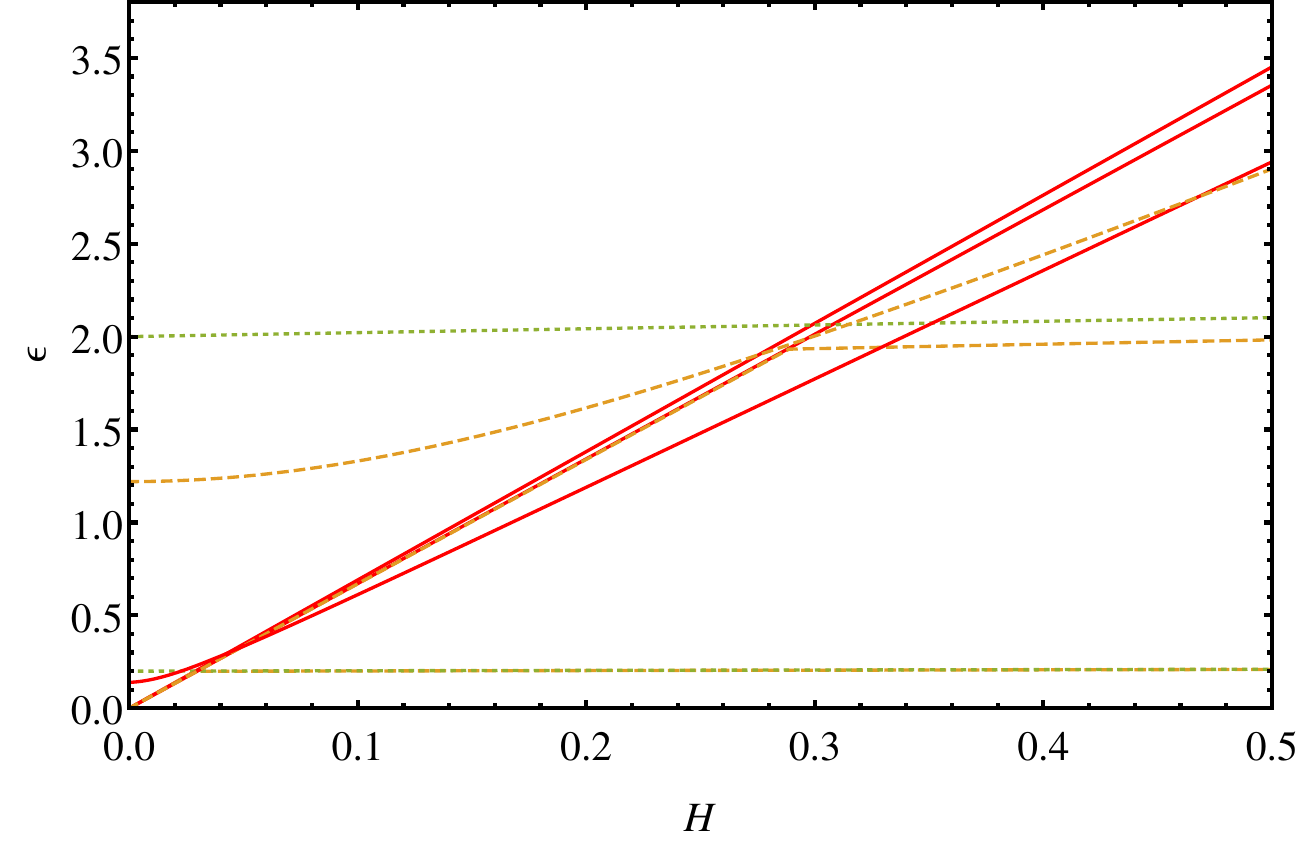(0.35)}
\lbl[]{33,83;\begin{tikzpicture}\draw[-,thick,dotted,color=black] (0,0) -- (0,4.3);\end{tikzpicture}}
\lbl[]{33,148;$0.029$}
\lbl[]{141,83;\begin{tikzpicture}\draw[-,thick,dotted,color=black] (0,0) -- (0,4.3);\end{tikzpicture}}
\lbl[]{141,148;$0.31$}
\lbl[]{225,83;\begin{tikzpicture}\draw[-,thick,dotted,color=black] (0,0) -- (0,4.3);\end{tikzpicture}}
\lbl[]{230,148;$2.1\to$}
\lbl[]{170,70;$\hbar\uparrow$}
\end{lpic}
\\~~\\
\begin{lpic}[]{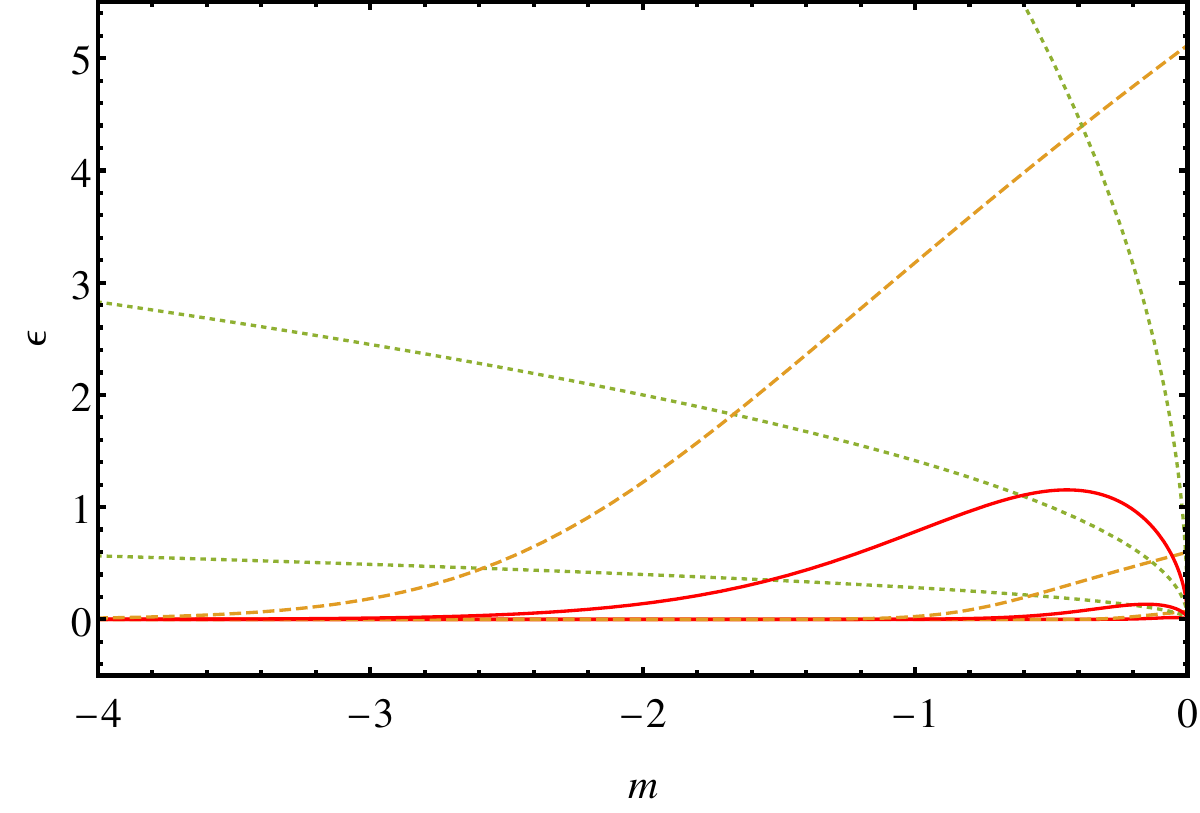(0.35)}
\lbl[]{172,80;$\hbar\uparrow$}
\lbl[]{66,83;\begin{tikzpicture}\draw[-,thick,dotted,color=black] (0,0) -- (0,4.3);\end{tikzpicture}}
\lbl[]{66,148;$-5^{2/3}$}
\lbl[]{155,83;\begin{tikzpicture}\draw[-,thick,dotted,color=black] (0,0) -- (0,4.3);\end{tikzpicture}}
\lbl[]{155,148;$-1^{2/3}$}
\lbl[]{186,83;\begin{tikzpicture}\draw[-,thick,dotted,color=black] (0,0) -- (0,4.3);\end{tikzpicture}}
\lbl[]{186,148;$-0.2^{2/3}$}
 \end{lpic}
 \caption{Gap $\varepsilon$~\eqref{eq:HeffTLS} via WKB approximation (solid red line), ``exact'' numerical gap (dashed yellow), Debye harmonic gap $\hbar\om_a$~\eqref{eq:fHO} (dotted green). In each plot $\hbar$ increases from bottom to top.  (top) $m=-2$, $\hbar=0.1,1,5$.  Vertical dotted lines exhibit the crossover between TTLS and HO regimes as given by the criterion~\eqref{eq:HOTLScriterion},\eqref{eq:WKBvalid} for each value of $\hbar$, increasing from left to right. The gap has a linear regime in $H$ close to $H=0$, but for $\hbar$ large enough ($\hbar=5$ curve) this regime is cut off and seems quadratic very close to $H=0$ (see discussion in~\secref{sub:hbarscal}).  (bottom) $H=0$, $\hbar=0.2,1,5$. The Debye gap is clearly off. Here vertical dotted lines locate the threshold $-\hbar^{2/3}$, which separates well the departure of the WKB approximation from the exact gap.}
 \label{fig:various_hbar}
\end{figure}

In Figs.~\ref{fig:WKB_vs_num}-\ref{fig:various_hbar} we contrast the WKB approximation with a numerical solution of the static Schr\"odinger equation for a particle in a DWP.
WKB computations are done with $E=v_s$.  

 In Figs.~\ref{fig:WKB_vs_num}(top) and~\ref{fig:various_hbar}(top), at fixed $m$  there is a linear dependence of the gap $\varepsilon\propto H$ close to $H=0$ well captured by the WKB expression. This regime extends linearly as $\hbar$ is increased, as the TTLS region is upper bounded by $H_c(m)\propto\hbar$, see~\eqref{eq:Hscal},\eqref{eq:HOTLScriterion} and Fig.~\ref{fig:hbarscaling}(top). Beyond the linear regime, a slightly increasing plateau appears, well approximated by the harmonic gap. 
 Finally for $\hbar=5$ in Fig.~\ref{fig:various_hbar}(top), we uncover another regime before the linear one, in which the gap is quadratic $\varepsilon\propto H^2$. This regime is studied further in~\secref{sub:analytic} and stems from analyticity arguments in $H=0$. It spans however an exponentially tiny range of $\hbar$ as $m$ decreases, hence remains hidden in other plots. 
 
 As anticipated in the previous section, TTLS-WKB approximation is thus useful only close to $H=0$; further the Debye approximation (HO) is valid.
 Besides, one notices that for $\abs m\lesssim\hbar^{2/3}$ the WKB approximation breaks down as discussed above, \eg in Fig.~\ref{fig:WKB_vs_num}(bottom), for $\hbar=5$ in Fig.~\ref{fig:various_hbar}(top), or in Fig.~\ref{fig:various_hbar}(bottom). In the latter the Debye harmonic approximation badly fails for the symmetric DWP and the WKB approximation seems valid only at $\abs m > \hbar^{2/3}$. 
  In numerical computations we hence delimit the quartic region by
 \begin{equation}\label{eq:quartic}
  \abs m\leqslant \hbar^{2/3}\ ,\qquad \abs H\leqslant H_c=0.491\dots\hbar
 \end{equation}
as these scalings and prefactors are adequate from the above numerical tests. 
The TTLS-WKB approximation fits in the region
\begin{equation}\label{eq:WKBvalid}
 m<-\hbar^{2/3}\ ,\qquad \abs H\leqslant H_c=0.491\dots\hbar
\end{equation}
In the other domains, the harmonic approximation (HO) is valid.  

\subsection{Numerical results in the RS-DWP phase}\label{sec:num}

Here we discuss our direct numerical solution of the RS-DWP problem. We fix $\k_m=0.1$, $\k=1$, $M=1$, $J=0.3$, $h=0.1$ (a point in the zero-temperature classical RS-DWP phase~\cite{BLRUZ21}) and vary $T$ and $\hbar$. 
The numerical procedure starts by solving the RS equations for $(\c,q)$~\eqref{eq:frsbc}-\eqref{eq:frsbq} under the low-temperature approximation detailed in~\secref{sub:lowtemp}. We make use of a self-consistent algorithm: the first value of the couple  $(\c,q)$ is the classical one\footnote{These classical values are obtained by a self-consistent algorithm, similar to the one described here, on the $\hbar\to0$ version of~\eqref{eq:frsbq}-\eqref{eq:frsbc}:
\begin{equation}\label{eq:classicalqchi}
  \begin{split}
   q_{\rm cl}=&\int_{m_m^{\rm cl}}^{m_M^{\rm cl}} \dd p(m)\int\DD_{J^2q_{\rm cl}}H\,\argp{\frac{\int\dd x\,x\, e^{-\b v_m(x)}}{\int\dd x\, e^{-\b v_m(x)}}}^2\\
   \c_{\rm cl}=&\int_{m_m^{\rm cl}}^{m_M^{\rm cl}} \dd p(m)\int\DD_{J^2q_{\rm cl}}H\,\frac{\partial_H^2\ln\int\dd x\, e^{-\b v_m(x)}}{\b}
  \end{split}
 \end{equation} Other starting points have been tested, with no impact on the final values. \label{footnote11}} at the corresponding $T$. We then iterate by plugging the $n^{\rm th}$ value $(\c^n,q^n)$ in the rhs integrals~\eqref{eq:frsbc}-\eqref{eq:frsbq}, obtaining a new couple $(\c^{n+1},q^{n+1})$. This is done with a damping \ie the actual output value is shifted  to prevent oscillating behavior: $\c^{n+1}\leftarrow (1-\d)\c^{n+1}+\d\c^{n}$, with $0.01\leqslant\d\leqslant0.1$. This procedure is repeated until convergence \ie both $\abs{\c^{n+1}/\c^n}$ and $\abs{q^{n+1}/q^n}$ are $1$ up to a $\epsilon=10^{-3}$ threshold. 
 Throughout, the integration intervals are fixed using $H\to h+zJ\sqrt q$ so that $\DD_{J^2q}H\to\DD z$ does not depend anymore on $q$. In practice $z\in[-10,10]$ is enough. 
 Integrals are separated into the (up to three) possible domains of Fig.~\ref{fig:DWP} according to the value of $\argc{m_m,m_M}$. In the quartic and TTLS domains, we make a double integration by parts in the expression for $\c$~\eqref{eq:frsbc} in order to least rely on the second derivative of $f_m(H)$, which is less accurately computed numerically. Indeed the TLS parameters $(\varepsilon,\bar\varepsilon)$ and their numerical $H$ derivatives are sampled beforehand in the $(m,H)$ plane to speed the process up, and the higher the $H$ derivative the noisier the data gets. We get these parameters from a numerical solution of the static Schr\"odinger equation for $\hat H_{\rm eff}$~\eqref{eq:fm1} (see~\secref{sub:quartic}) in the quartic domain, and from WKB expressions~\eqref{eq:TLS} in the TTLS domain. The HO domain parameters $(v_a,\om_a)$ are instead analytically known~\eqref{eq:fHO}. In practice the lowest temperature that we regard as close enough to $T=0$ for each $\hbar$ is $T=\hbar/(\b\hbar)$ with $\b\hbar=10^5$. 
 Equipped with the values of $(\c,q)$ we numerically compute the energy~\eqref{eq:Uvar}, specific heat and replicon~\eqref{eq:frsbrep}. In particular, the replicon values give access to the phase diagram shown in Fig.~\ref{fig:qphd}(inset). 
 
 The $\epsilon=10^{-3}$  threshold is good enough for the absolute values but not for a precise assessment of the temperature dependence of $(\c,q)$ at low $T$. It is the main bottleneck: our algorithm is not capable of converging for finer thresholds, which traces back to numerical errors in sampling precisely the $H$ dependence of $H$ derivatives of the TTLS gap $\varepsilon$. Consequently the specific heat dependence on temperature inherited from this $(\c(T),q(T))$ dependence cannot be tracked at the lowest temperatures, see Fig.~\ref{fig:CV}(top).   
 
 \begin{figure}[h!]
\centering
(a)\\
 \includegraphics[width=7cm]{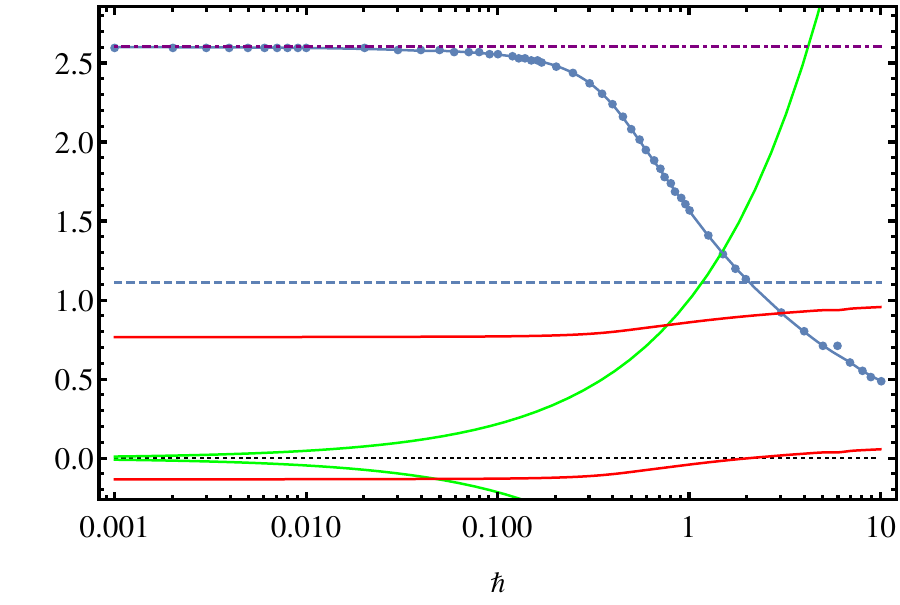}\\
 \includegraphics[width=7cm]{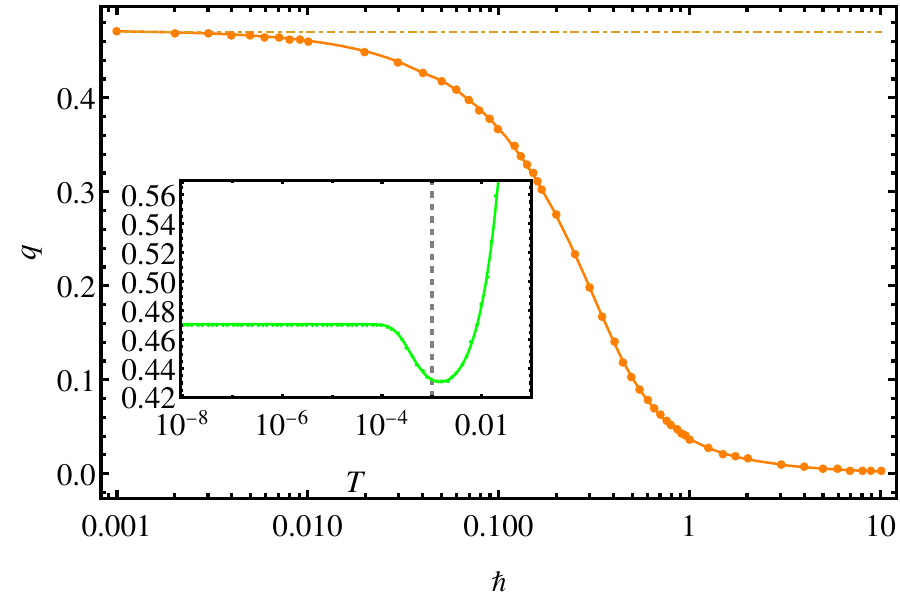}\\
  (b)\\
 \includegraphics[width=7cm]{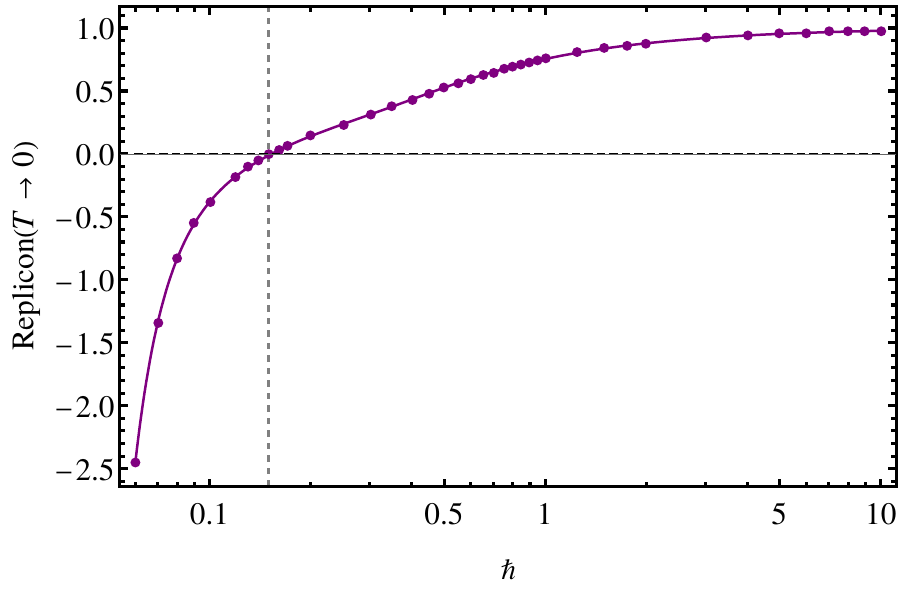}\\
 (c)
 \caption{(a)(b)(c) $T\to0$ values as a function of $\hbar$. 
 (a) (solid blue line) $\c(\hbar)$ (horizontal dot-dashed) classical value of $\c_{\rm cl}(T\to0)$ (horizontal dashed) $\k_m/J^2$,\ie the value of $\c$ for which $m_m=0$ ; if $\c$ is above, there exist DWP, below only SWP. On the same plot is shown (red lines) $\argc{m_m,m_M}$ as a function of $\hbar$ and (green lines) $\pm\hbar^{2/3}$ in order to show the importance of the quartic region. Above $\hbar\approx0.05$ the TTLS-WKB region is overrun by the quartic region. 
 (b) (solid line) $q(\hbar)$  (horizontal dot-dashed) classical value of $q_{\rm cl}(T\to0)$. We note that the lowest value of $\hbar=10^{-3}$ gives the classical $T\to0$ values of these quantities, namely $(\c_{\rm cl},q_{\rm cl})\simeq(2.60,0.47)$, as calculated independently from~\eqref{eq:classicalqchi}. Inset: $q(T)$ for $\hbar=10^{-3}$. The vertical dashed line signals the heuristic value $\b\hbar=1$.   (c) Replicon. The vertical dashed line signals the transition at $\hbar\simeq 0.15$. }
 \label{fig:T=0}
\end{figure}
 
 In Fig.~\ref{fig:T=0}(a)(b) we display the converged values of $(\c,q)$ at $T\to0$ for four decades of $\hbar$. The small $\hbar$ data is in excellent agreement with an independent numerical solution of the classical equations~\eqref{eq:classicalqchi} (see footnote~\footref{footnote11}). As $\hbar$ grows both quantities relax to zero, making the TTLS-WKB region and eventually all DWP disappear. For $\hbar\gtrsim1$ the quartic region dominates entirely the whole interval $\argc{m_m,m_M}$. The behavior is then unusual: only SWP are present but the HO approximation falls short. In the inset of Fig.~\ref{fig:T=0}(b) we pinpoint that for $\b\hbar<1$ (a simple heuristic value, $T\gg E_{\rm gap}$ a typical energy gap of the system would be more accurate)  we usually see signs of the breakdown of the low-$T$ approximation, here manifested by an odd regime of increasing $q$ with temperature, quickly overcoming $1$. This is unphysical and stems from not accounting of higher energy levels that get populated.
 
 \begin{figure}[!htbp]
\centering
 (a)\\
 \includegraphics[width=6.5cm]{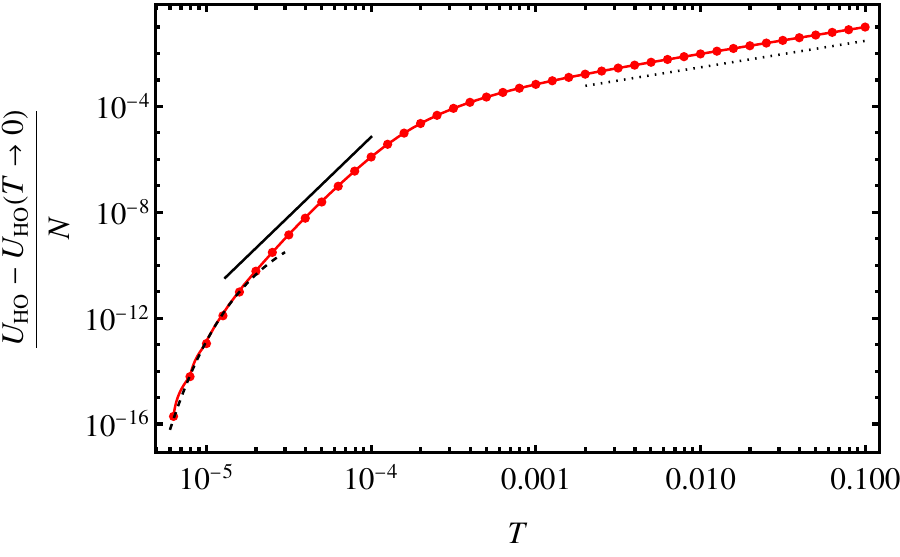}\\
 \includegraphics[width=6.5cm]{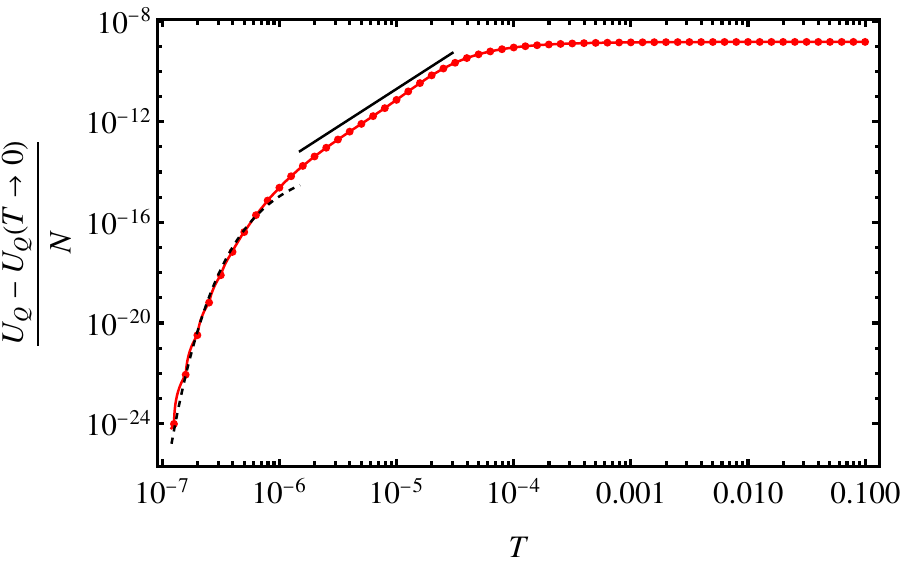}\\
 (b)\\
 \includegraphics[width=6.5cm]{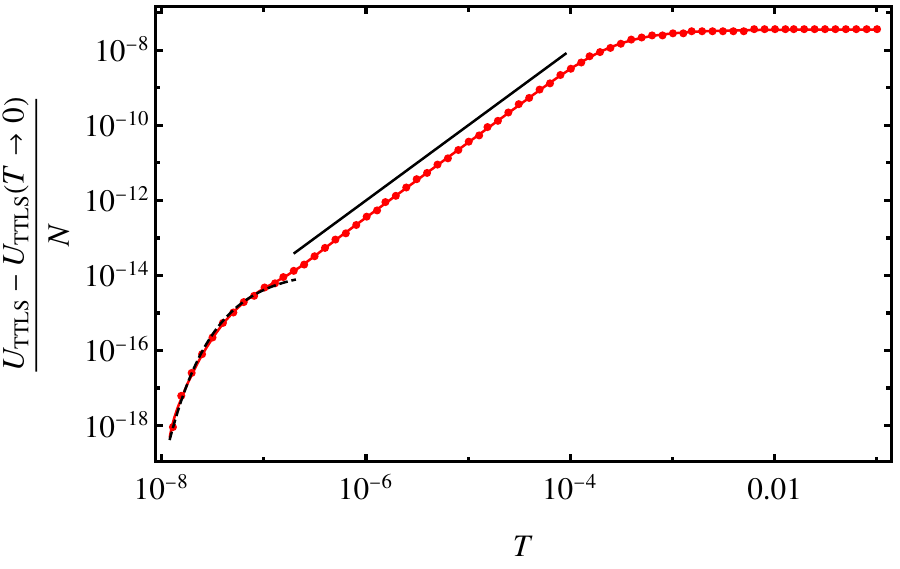}\\
(c)
 \caption{(a)(b)(c) Energies in the different regions as a function of temperature at fixed $(\c,q)=(\c(T\to0),q(T\to0))$ for $\hbar=10^{-3}$. They are shifted with their $T\to0$ limit so that they go to zero in that limit. (a) (red solid line) harmonic domain $U_{\rm HO}/N$, (solid black line) $T^6$, (dotted) $T$ (high-$T$ limit of harmonic oscillators), (dashed line) exponential fit  $C e^{-E_{\rm gap}/T}$, with $C\simeq1.5\cdot10^{-8}$, $E_{\rm gap}\simeq1.2\cdot10^{-4}$. Note that indeed the smallest harmonic gaps are of order $\hbar^{4/3}=10^{-4}$. Due to the large exponent and the rather high gap $\hbar^{4/3}$, it is difficult to see a clean power law on a large interval. Thus for temperatures below this gap the exponential contribution is felt. (b) (red solid line) quartic domain $U_{\rm Q}/N$. (solid black line) $T^3$ as an approximate fit in this quartic region, which is a crossover between TTLS $T^2$ and Debye $T^6$ scalings -- see discussion after~\eqref{eq:redterm}. (dashed) exponential fit with $C\simeq2.4\cdot10^{-14}$, $E_{\rm gap}\simeq3.1\cdot10^{-6}$. The gap's order of magnitude is roughly in line with typical small gaps of effective DWP, see Fig.~\ref{fig:exp}(top). (c) (red solid line) TTLS-WKB domain $U_{\rm TTLS}/N$. (solid black line) $T^2$ (dashed) is an exponential fit with $C\simeq1.4\cdot10^{-14}$, $E_{\rm gap}\simeq1.2\cdot10^{-7}$. The gap's order of magnitude is the lowest of all regions, as expected from typical double-well gap values, see Fig.~\ref{fig:exp}(bottom). Both $U_{\rm Q}$ and $U_{\rm TTLS}$ go to a constant value for high temperature, which is easily seen for a two-level system with finite gap -- of course, the large-temperature limit within the present low-temperature scheme has no physical meaning. 
 The relative values of each energy term is also important, the $T\to0$ value is \textit{(i)} for HO, $-0.090$ \textit{(ii)} for  quartic, $-1.8\cdot 10^{-10}$  \textit{(iii)} for TTLS-WKB, $-2.3\cdot 10^{-6}$.}
 \label{fig:fixed}
\end{figure}
 
In Fig.~\ref{fig:T=0}(c) the replicon becomes negative at $\hbar<\hbar_c\simeq0.15$, pointing towards a quantum critical point announcing a RSB phase. At lower $\hbar$, although $\partial_H^2 f_m$ appearing in the integrand~\eqref{eq:frsbrep} becomes noisier in the TTLS and quartic regimes, the replicon clearly goes very quickly towards $-\io$, as expected from the classical RS case without pseudogap in the $H$ distribution~\cite{FU22}. We elaborate on this point in~\secref{sec:pseudogap}.  The quantum critical point  $\hbar_c$ approximately coincides with the disappearance of TTLS in the system (measured at $\hbar\approx0.05$), as $m_m\gtrsim-\hbar^{2/3}$, see Fig.~\ref{fig:T=0}(a). Only HO and quartic potentials remain. Interestingly, this hints at a fullRSB phase physically dominated by TTLS physics, while the RS phase for larger $\hbar$ is dominated by Debye or quartic oscillator physics (see also Fig.~\ref{fig:recap}).

\begin{figure}[!htbp]
\centering
 \includegraphics[width=7cm]{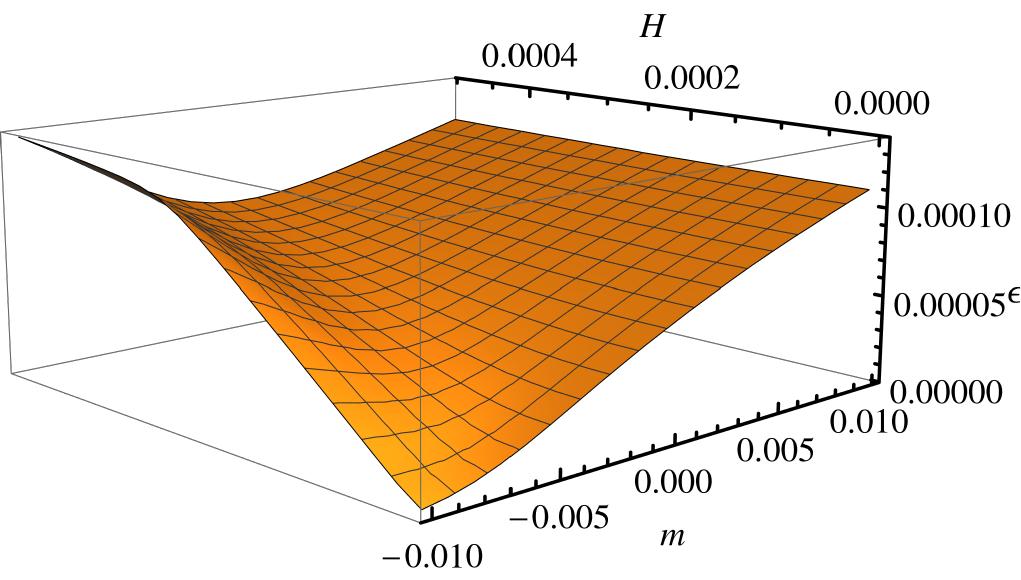}\\
\includegraphics[width=7cm]{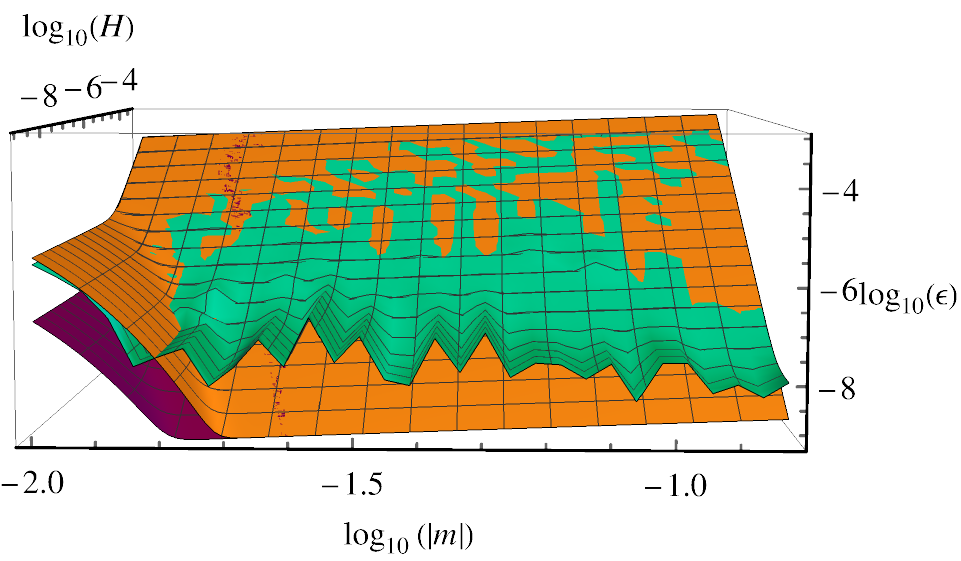}
 \caption{$\hbar=10^{-3}$. (top) Values of the gap between the first two levels $\varepsilon$ in the quartic region, obtained from direct solution of the static Schr\"odinger equation.  (bottom) Values of the gap $\varepsilon$ in the TTLS region obtained from the WKB approximation. The scales are logarithmic in order to focus on the $H\to0$ part, which varies exponentially. The $m$ axis goes from $m=-\hbar^{2/3}=-0.01$ (left) to $m=-0.17<m(\c<3)$, so as to focus on the effectively sampled region for $\hbar=10^{-3}$ at low temperatures (see Fig.~\ref{fig:T=0}(a)). (cyan) direct numerical solution of the static Schr\"odinger equation (orange) WKB approximation~\eqref{eq:TLS}-\eqref{eq:HeffTLS} with $E=E_s$ (purple) with $E=v_s$ . As expected, the direct numerical solution for $H\to0$ is more unstable. For $\abs{H}\gg10^{-8}$ all methods match quite well.}
 \label{fig:exp}
\end{figure}

As in the classical case we expect the ensuing RSB phase to be marginally stable, which entails power-law behavior of the specific heat due to a vanishing many-body gap. Nonetheless a crucial comment is in order: we went beyond the assumption of replica symmetry and made an additional variational approximation which turns the problem into a single-particle one, albeit very close in spirit to the full mean-field one. The latter approximation, as discussed in~\secref{sub:gapCv}, introduces a necessarily finite gap of the system, a flaw of the simplification. For temperatures below this gap, the energy is exponentially damped, which may forbid the observation of power laws: 
\textit{(i)} The absence of TTLS at the critical point implies a large value of the (HO or quartic) gaps, as these scale as a power law in $\hbar$ and not exponentially small in $\hbar$. These large gaps impede the observation of power laws right at the critical point. 
Indeed fitting the specific heat contribution of the HO and quartic regions at the transition $\hbar_c$ with an exponential form 
$\propto e^{-E_{\rm gap}/T}$ we obtain respectively $E_{\rm gap}^{\rm HO}=0.11$, $E_{\rm gap}^{\rm Q}=0.044$.  Notice that the quartic gap scale~\eqref{eq:numtruegap} $\hbar_c^{4/3}\simeq0.08$ or the HO scale $\hbar_c\simeq0.15$ are indeed a correct order of magnitude.  
\textit{(ii)}~For $\hbar\ll\hbar_c$ nevertheless the gaps become small enough to observe the (pre-exponential) power laws in \eg the energy. Note that below this critical value of $\hbar$ the RS ansatz is strictly speaking an approximation, although we argue in~\secref{sec:pseudogap} that this should not modify the present conclusions. We show the contribution of each domain in Fig.~\ref{fig:fixed}, fixing $(\c,q)$ to their $T\to0$ values. This is because their temperature dependence at very low $T$ becomes dominated by numerical errors due to the convergence procedure and is thus not reliable. We observe the scaling of the energy in each domain as $T^2$ for TTLS, $T^6$ for HO and a crossover value in between $\approx T^3$ for the quartic region. 
These laws are predicted by the analysis of the full energy (including $(\c,q)$'s temperature dependence) in~\secref{sub:gapCv} above a domain-dependent threshold temperature given by the small gap of the (HO, quartic or TTLS) domain $E_{\rm gap}$. This gap is determined by a fit and agrees with the expected order of magnitude for each domain, as detailed in Fig.~\ref{fig:exp} for comparison.

The specific heat scaling, as calculated numerically by temperature derivation of the energy, is displayed in Fig.~\ref{fig:CV}(top) for $\hbar=10^{-3}$. The noisy convergence data gives rise to a spurious negative specific heat below $T\sim10^{-5}$, and does not allow to resolve the very low $T$, likely dominated by the TTLS-WKB region, expected to occur for $T\lesssim10^{-5}$ from the cyan curve in which $(\c,q)$ are instead fixed to their $T\to0$ values. The latter curve exhibits the expected behavior (see also Fig.~\ref{fig:scalings}): at $T\ll E_{\rm gap}$, the specific heat is gapped, at higher temperature we have the TTLS linear $T$ dependence and then a crossover to the HO $T^5$ scaling. 
When $\hbar$ is increased in Fig.~\ref{fig:CV}(bottom), the TTLS linear behavior is wiped out close to $\hbar_c$ (actually at the previously mentioned threshold $\hbar\approx0.05$). This is because $\hbar$ becomes so large that the $m_m$ boundary hits the quartic region: for $\hbar=\abs{m_m}^{3/2}$ the $m<0$ axis is entirely contained in the quartic region and the TTLS-WKB region does not exist anymore.  It is replaced by an exponential behavior (gapped) at low $T$, the quartic gap being roughly $\propto \hbar^{4/3}$. For $\c=\c(\hbar=0.05,T\to0)\simeq2.58$ (see Fig.~\ref{fig:T=0}(a)) this indeed happens for $\hbar=(J^2\c-\k_m)^{3/2}\simeq 0.048$, showing that this TTLS threshold value $\hbar=0.05$ and the self-consistent value of $\c$ provided by the algorithm are coherent. 
The quartic behavior $\propto T^2$ for $\hbar<0.04$ is not seen, it is either dominated by the TTLS-WKB behavior at low $T$ or by the HO behavior at higher $T$. For $\hbar > 0.04$ it is hidden by the $\hbar^{4/3}$ gap or dominated by HO. Only very close to $\hbar=0.04$ is it seen at intermediate $T$ values.

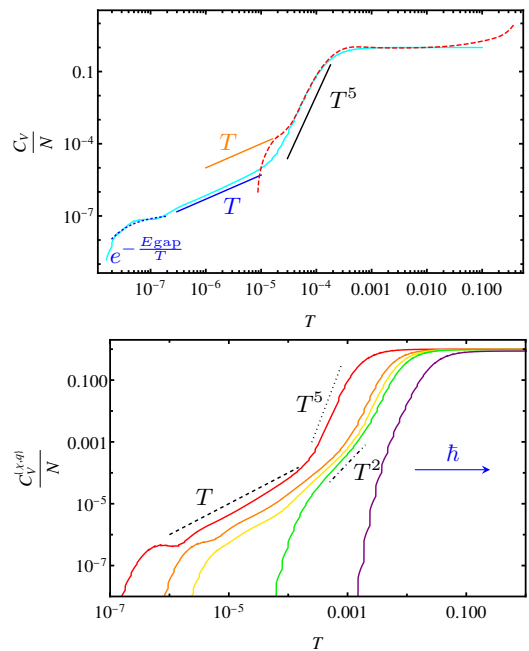
\begin{figure}[!htbp]
\centering
\begin{lpic}[]{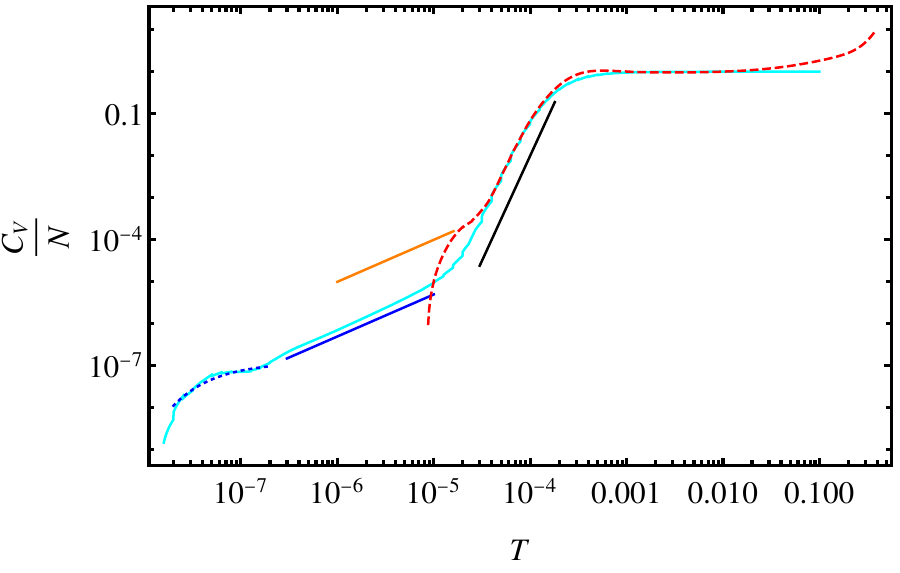(0.45)}
\lbl[]{65,57;$\textcolor{orange}{T}$ }
\lbl[]{65,38;$\blue{T}$}
\lbl[]{40,25;$\blue{e^{-\frac{E{\rm gap}}{T}}}$}
\lbl[]{98,70;$T^5$}
     \end{lpic}
     \begin{lpic}[]{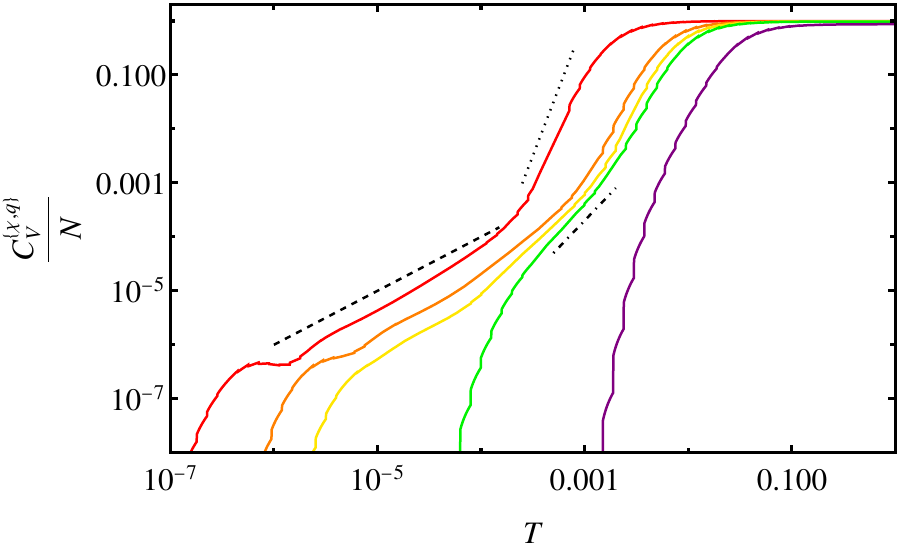(0.45)}
\lbl[]{58,47;$T$ }
\lbl[]{105,55;$T^2$}
\lbl[]{88,75;$T^5$}
\lbl[]{130,60;$\blue{\hbar}$}
\lbl[]{130,55;\begin{tikzpicture}\draw[-stealth,color=blue] (0,0) -- (1,0);\end{tikzpicture}}
     \end{lpic}
 \caption{Total $C_V(T)/N$ for (top) $\hbar=10^{-3}$ (bottom) various $\hbar$. (top) (cyan) $(\c,q)$ are fixed to their $T\to0$ values. Solid blue, orange and black lines are respectively $T$, $T$ and $T^5$. The dashed blue line is an exponential fit $C e^{-E_{\rm gap}/T}$ with $C\simeq1.2\cdot10^{-7}$, $E_{\rm gap}\simeq4.8\cdot10^{-8}$ (dashed red) data for the full $C_V(T)/N$ including the temperature dependence of $(\c(T),q(T))$. Due to numerical imprecision of the self-consistent algorithm used, the latter dependence plateaus at low $T$ instead of tracking the approach to the $T=0$ values. Consequently at temperature below the $T^5$ regime, this curve spuriously drops sharply to zero.   
 (bottom) Increasing $\hbar$ from left to right. $(\c,q)$ are fixed to their $T\to0$ values. Solid curves have respectively $\hbar=0.009, 0.03, 0.04, 0.05, 0.15$. Dashed curve is $T$, dot-dashed $T^2$ and dotted $T^5$. The linear behavior is not seen for $\hbar\gtrsim 0.05$ due to the absence of the TTLS-WKB region (see Fig.~\ref{fig:T=0}(a)). }
 \label{fig:CV}
\end{figure}

\subsection{Analytical study of the specific heat scaling at low temperature}\label{sub:gapCv}

Here we complement the previous numerical study of the RS-DWP phase with an analytical examination of the specific heat from the variational approximation of the thermodynamic energy~\eqref{eq:Uvar}. 

\begin{figure}[!htbp]
\centering
 \includegraphics[width=7cm]{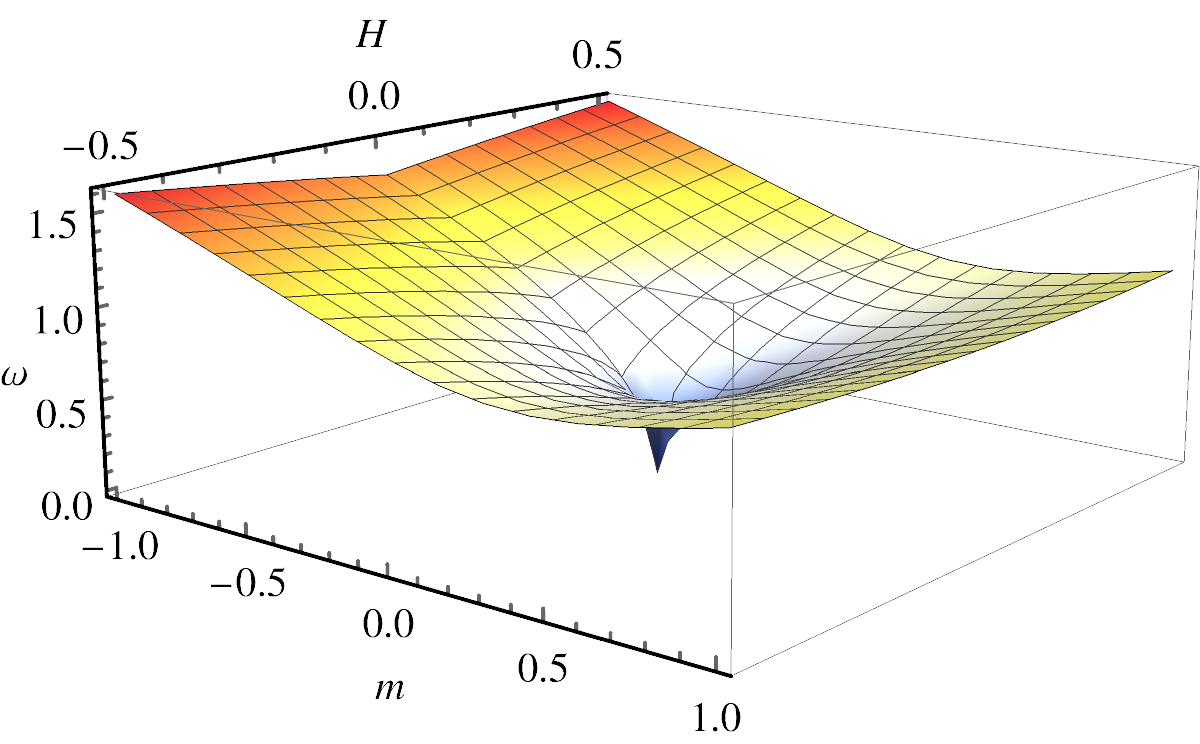}\\
 \includegraphics[width=7cm]{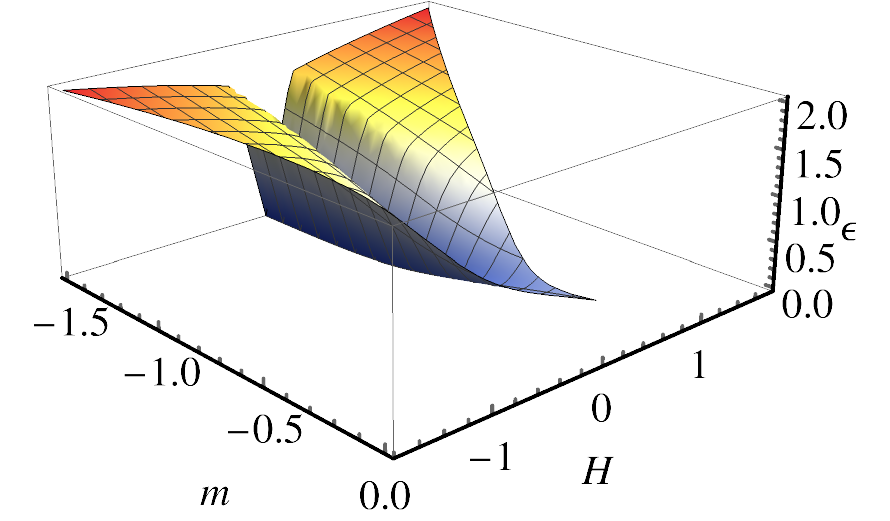}
 \caption{$\hbar=1$. (top) Debye harmonic gap $\hbar\om_a$~\eqref{eq:fHO}. (bottom) ``Exact'' numeric gap. 
}
 \label{fig:Debye_gap}
\end{figure}

\subsubsection{Finiteness of the gap}

To understand the thermodynamic properties at ${T\to0}$, one must first examine  what happens to the gap of the system. In the HO region the gap $\hbar\om_a$ goes to zero as $(m,H)$ goes to the origin, see Fig.~\ref{fig:Debye_gap}(a). Eventually this gap is finite if $\hbar$ is not negligible: it is of order $\hbar^{4/3}$ around $(m,H)=(0,0)$~\eqref{eq:numtruegap}, when $\om_a\sim\hbar^{1/3}$ (quartic region). 
In the TTLS domain the gap $\varepsilon=\sqrt{\D^2+\D_0^2}$ is minimum when both $\D$ and $\D_0$ are minimized. This occurs for $H=0$ and $\abs m$ as large as possible (note that $m$ is bounded from below by $m_m$). Thus strictly speaking the gap cannot vanish in this setting, yet it can be made exponentially small in $\abs m^{3/2}/\hbar$~\eqref{eq:D0H0}. Therefore we expect the specific heat to display a gapped scaling $C_V\sim e^{-\b E_{\rm gap}}$, but the exponential can be very close to 1 if the gap $E_{\rm gap}\ll T$. This is already what was argued in the seminal TLS papers~\cite{AHV72,Ph72}. In the soft-potential model and early TLS papers~\cite{BGGS91,AHV72,Ph72,Ph87,Pa94}, the lowest value of $m$ (or any equivalent mechanism) was provided by a physical argument: the experimental time is limited and tunneling must be faster, imposing an upper bound on the DWP barrier. 

The fact that there is a finite gap is not surprising here as the second term of~\eqref{eq:Uvar} is the energy of an effective one-dimensional quantum mechanical particle: in one dimension there are no degenerate bound states\footnote{Here is the Wronskian argument: let $\psi_1$ and $\psi_2$ be wavefunctions at energy $E$ both fulfilling the static Schr\"odinger equation 
\begin{equation}
 -\frac{\hbar^2}{2M}\psi_i''+V\psi_i=E\psi_i\ ,\qquad i=1,2
\end{equation}
then $\psi_1\psi_2''-\psi_2\psi_1''=0$, so that $\psi_1\psi_2'-\psi_2\psi_1'=$ constant $=0$ (the last equality being true if we consider bound states for a discrete spectrum, as wavefunctions must vanish at infinity), implying $(\psi_1/\psi_2)'=0$, \ie in the end $\psi_1\propto\psi_2$ both represent the same physical eigenstate of the system.} for normalizable wavefunctions~\cite[p.98-106]{messiah}. Consequently, even though there is a numerically predicted RSB transition from the present variational approximation (Fig.~\ref{fig:T=0}(c)), by construction the energy gap cannot vanish. This is an inconsistency of this approximation, which we discuss further in~\secref{sec:spinboson}. Note in addition that the energy~\eqref{eq:Uvar} contains another term $J^2q\c/2$, not clearly in the form of an effective energy,  whose scaling is studied later on. 

\subsubsection{Chain-rule expansion}

Let us now analyze the $T\to0$ scaling of the energy~\eqref{eq:Uvar}. With the help of the variational equation for $\c$~\eqref{eq:frsbc}, integrated twice by parts, we get
\begin{equation}\label{eq:Uab}
 \begin{split}
   \frac{U_{\rm RS}^a}{N}=&\int\dd p(m)\dd H\argc{\z_q(H)\frac{f_m(H)}{\b}-\G_q(H)\frac{\partial f_m(H)}{\partial\b}}\\
 \end{split}
\end{equation}
where the integrands have been symmetrized $H\to-H$ by introducing
\begin{equation}
 \begin{split}
  \G_{q}(H)\equiv&\g_{J^2q}(H+h)+\g_{J^2q}(H-h)\\
  \z_{q}(H)\equiv&\frac{J^2q}{2}\partial_H^2\g_{J^2q}(H-h)+(H\to-H)
 \end{split}
\end{equation}
$f_m(H)$ is given on each domain  respectively by~\eqref{eq:fHO} and~\eqref{eq:fTLS}. 
In the limit $\b\to\io$ we can analyze each integrand, providing
\begin{equation}\label{eq:Uaa}
 \begin{split}
  \frac{U_{\rm RS}^a(T=0)}{N}=&\int_{\rm TLS}\argp{\G_{q_\io}(H)-\z_{q_\io}(H)}\argp{\bar\varepsilon-\frac{\varepsilon}{2}}\\&+\int_{\rm HO}\argp{\G_{q_\io}(H)-\z_{q_\io}(H)}\argp{v_a+\frac{\hbar\om_a}{2}}
 \end{split}
\end{equation}
which is the ground-state energy. We noted ${q_\io\equiv q(\b\to\io)}$. The integrals without measure mean $\int_{H\geqslant0}\dd p(m)\dd H$ over the HO or TLS domain (or both if unspecified). The latter includes TTLS but many steps can be applied verbatim on the quartic region in which only the first two levels are considered ($\varepsilon$ is the gap and ${\bar\varepsilon-\varepsilon/2}$ the ground state); nonetheless we comment in~\secref{sub:anscal} specifically on this domain. Subtracting~\eqref{eq:Uaa} to~\eqref{eq:Uab},
\begin{widetext}
\begin{equation}\label{eq:subtract}
 \begin{split}
   \frac{U_{\rm RS}^a}{N}\underset{T\to0}{=}& \frac{U_{\rm RS}^a(T=0)}{N}
   +\int_{\rm HO}\left[\blue{\G_{q_\io}(H)\frac{\hbar\om_a}{2}\argp{\coth{\frac{\b\hbar\om_a}{2}}-1}}-{\color{orange}\z_{q_\io}(H)\frac{\ln\argp{1-e^{-\b\hbar\om_a}}}{\b}}\right]\\
   &+ \int_{\rm TLS}\left[\red{\G_{q_\io}(H)\frac{\varepsilon}{2}\argp{1-\tanh{\frac{\b\varepsilon}{2}}}}+\green{\z_{q_\io}(H)\frac{\ln\argp{1+e^{-\b\varepsilon}}}{\b}}\right]
   +\argp{q-q_\io}{\frac{\partial}{\partial {q}}\frac{U_{\rm RS}^a}{N}}+\argp{\c-\c_\io}{\frac{\partial}{\partial {\c}}\frac{U_{\rm RS}^a}{N} }\\
   &+\frac{\argp{q-q_\io}^2}{2}\frac{\partial^2}{\partial q^2}\frac{U_{\rm RS}^a}{N}+\argp{q-q_\io}\argp{\c-\c_\io}{\frac{\partial^2}{\partial q\partial {\c}}\frac{U_{\rm RS}^a}{N} }
   +\frac{\argp{\c-\c_\io}^2}{2}\frac{\partial^2}{\partial \c^2}\frac{U_{\rm RS}^a}{N}+\dots
 \end{split}
\end{equation}
\end{widetext}
This is a chain rule expansion for $T\to0$, distinguishing explicit $\b$ dependence and the one coming from $(\c(\b),q(\b))$. In the last two lines the partial derivatives are evaluated at $T=0$ where $(\c,q)=(\c_\io,q_\io)$.

\subsubsection{Three low-energy excitations}

\begin{figure}[!hbtp]
\centering
 (a)\\
 \includegraphics[width=7cm]{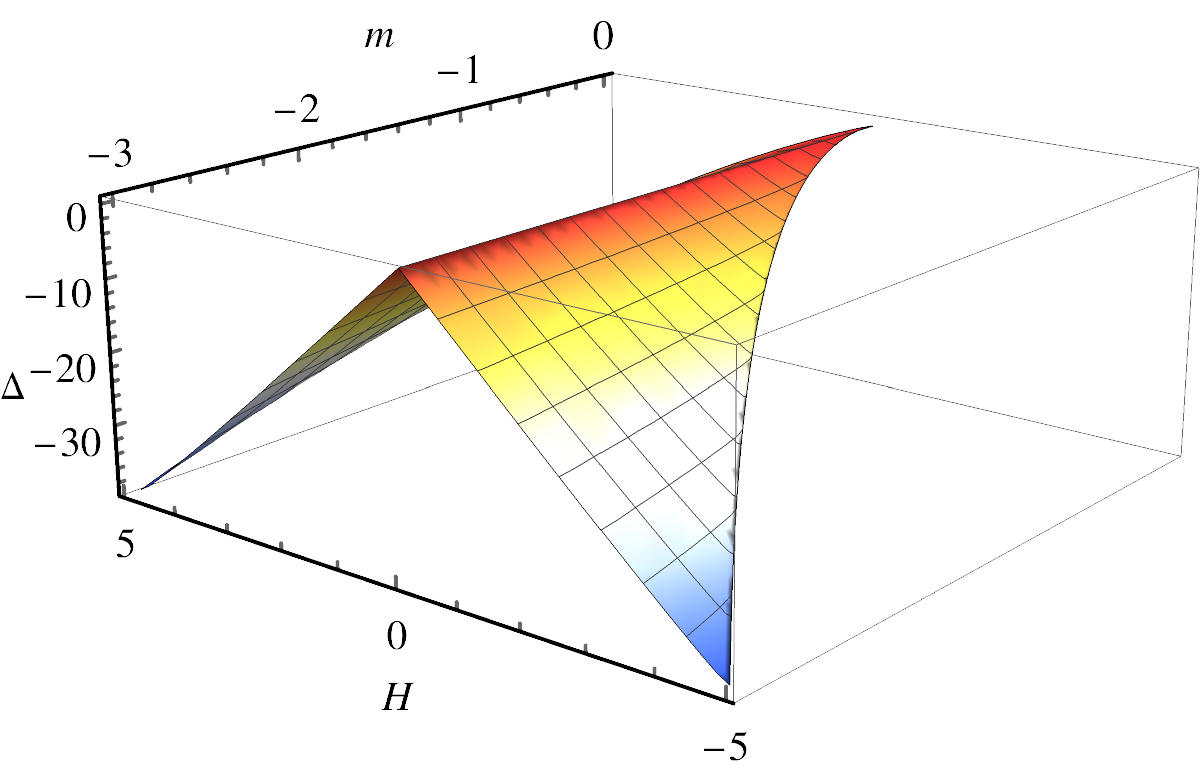}
 \includegraphics[width=7cm]{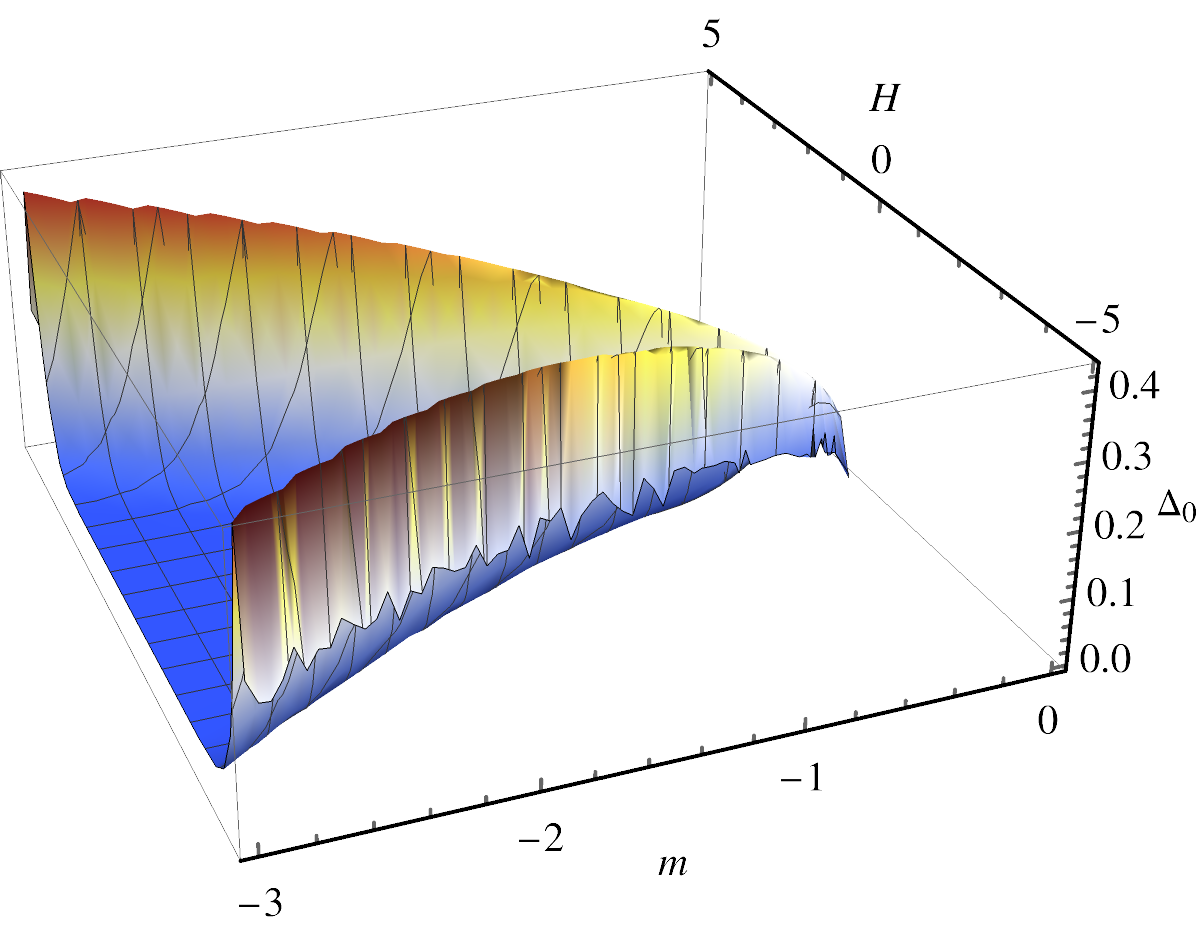}\\
 (b)\\
 \includegraphics[width=7cm]{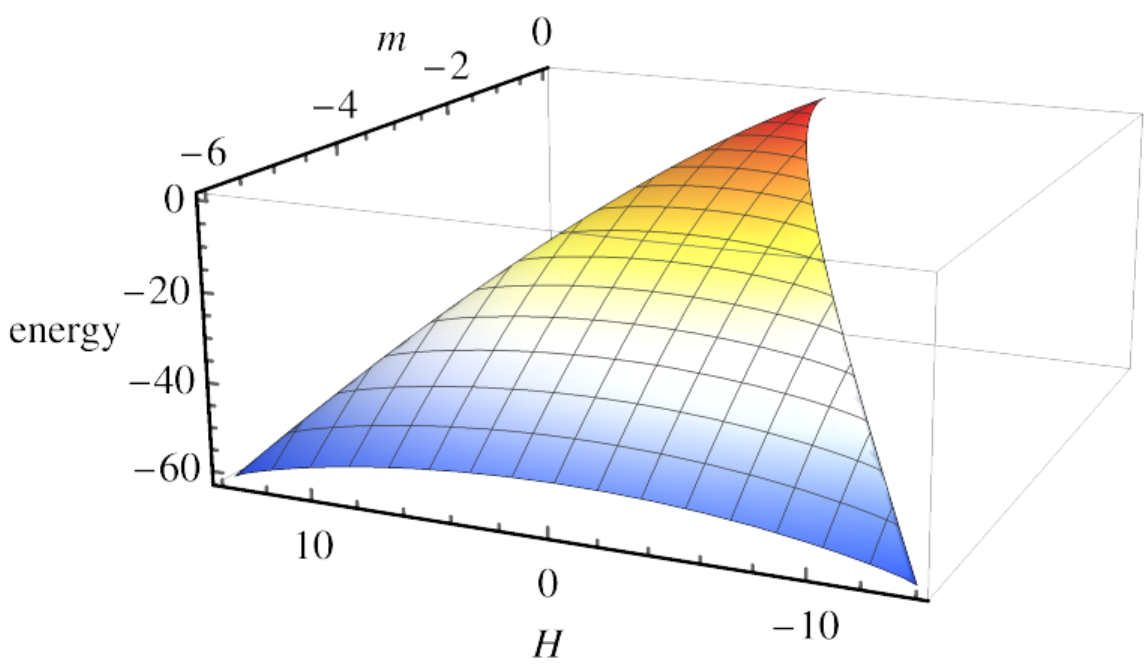}\\
 (c)
 \caption{TTLS parameters for $\hbar=1$. (a) Ground-state energy difference between the wells $\D$~\eqref{eq:HeffTLS}. (b) Tunneling amplitude $\D_0$~\eqref{eq:TLS} from WKB. (c) TTLS parameter $\bar\varepsilon$~\eqref{eq:HeffTLS}.}
 \label{fig:D}
\end{figure}

The low-temperature scaling of each term can now be assessed. 
The $q$ dependence is contained in the functions $(\G_q,\z_q)$ while the $\c$ dependence is in the integral boundaries, as $m=\k-J^2\c$. 
In the following we call $\d q=q-q_\io$ and $\d \c=\c-\c_\io$. 
Dividing the $(m,H)$ plane in three regions (see Fig.~\ref{fig:DWP}), we obtain some basic scalings:\\
 1. In the HO domain, the parameters are $(v_a,\om_a)$~\eqref{eq:fHO} whose $(m,H)$ dependence is set by~\eqref{eq:cubicdepressed}. In the following we note $\th(y)$ any function of ${y\propto H/\abs{m}^{3/2}}$ whose precise details are irrelevant to the scaling argument. We have that $x_a=\sqrt{\abs{m}}\th_1(y)$, therefore $v_a=v_m(x_a)=m^2\th_2(y)$ as each of its three terms scales that way. Besides $\om_a=\sqrt{\abs{m}}\th_3(y)$ where $\th_3(y)$ cannot vanish. Similarly, $H$ derivatives scale like $\partial_H=\abs{m}^{-3/2}\partial_y$. For $\b\to\io$ the critical region of HO integrals is where $\hbar\om_a\sim T$, which is bound to happen close to $(m,H)=(0,0)$, see Fig.~\ref{fig:Debye_gap}(a). This last criterion means that the important scaling variable is $m\sim (\b\hbar)^{-2}$, whereas $y$ turns out to be irrelevant.   $\abs m$ is bounded by $\hbar^{2/3}$~\eqref{eq:lowerm}, implying there is a gap.  We call this lowest HO gap (on the verge of quarticity)  $\varepsilon_0\sim\hbar^{4/3}$ as given by the scaling argument leading to~\eqref{eq:numtruegap}, or noticing that the exponential temperature dependence in the following calculations are given by (exponential minus) $\b\hbar\sqrt{\abs m}\sim\b\hbar^{4/3}$. Note that this is nevertheless vanishing in the semiclassical limit $\hbar\to0$ with $\b\hbar$ fixed, recovering the Debye power laws previously discussed in~\secref{sub:Debye}.\\
 2. In the TLS domains, the parameters are $(\bar\varepsilon,\varepsilon)$~\eqref{eq:TLS}-\eqref{eq:HeffTLS}. The important scaling variable is  $\varepsilon$ and the critical region is $\varepsilon\sim T$. Recall that $(m=0,H=0)$ does not yield a small $\varepsilon$ due to the finite gap~\eqref{eq:numtruegap} at finite $\hbar$. We need to distinguish two cases: \textit{(i)} Quartic region $\abs m \lesssim \hbar^{2/3}$ (around the origin $m=H=0$ dominated by the purely quartic potential). In this region the harmonic approximation fails and, for negative $m$, the WKB approximation fails as well, and $(\bar\varepsilon,\varepsilon)$ was computed numerically in~\secref{sec:num}. The gap $\varepsilon_0\sim\hbar^{4/3}$ is rather large in this region, compared to the TTLS gaps. In the following we shall thus rather focus on the TTLS-WKB excitations; we comment in~\secref{sub:anscal} on the quartic case. \textit{(ii)} TTLS-WKB region $\abs m \gg \hbar^{2/3}$. The $H$ dependence of $\varepsilon$ on a very narrow interval close to $H=0$ is quadratic in $H$ (see~\secref{sub:hbarscal}), and beyond it is linear in $H$  up to the crossover to HO behavior. 
This can be seen from Figs~\ref{fig:various_hbar}(a)(b) although the quadratic behavior is only visible for the highest value of $\hbar=5$. The quadratic zone is so tiny (exponentially small in $\abs{m}^{3/2}/\hbar$) that it does not impact scalings numerically, and in most of the $\argc{0,H_c}$ interval the gap is  linear in $H$. We have (see Fig.~\ref{fig:D}(c)   for $\bar\varepsilon$)
\begin{equation}\label{eq:chgvar}
 \begin{split}
   \bar\varepsilon=&\bar\n_0(m)+\bar\n_2(m)H^2+O(H^4)\,,\\
   \varepsilon\simeq&\n_0(m)+\n_1(m)\abs H+O(H^2)\,,\\
  \n_0(m)=&\D_0(H=0)\,,\quad  \bar\n_0(m)=\hbar\sqrt{\frac{\abs m}{2}}-\frac32m^2\,,\\
  \bar\n_2(m)=&\frac{1}{2m}-\frac78 \frac{\hbar}{\argp{2\abs m}^{5/2}}\,,\quad \n_1(m)>0
%
 \end{split}
\end{equation}
which are the leading-order expressions in terms of $(m,H)$. An explicit expression for $\n_1(m)$ is cumbersome from the WKB expression~\eqref{eq:TLS}, although it is clear from the numerics that it is positive. 
Therefore for $\b\to\io$ the critical region of TTLS integrals is where $\b\varepsilon\sim 1$, meaning the important scaling variable is $H\sim T$. 

\subsubsection{Scalings}

Let us begin with the terms from $\moy{\hat H_{\rm eff}}$ in~\eqref{eq:subtract}. First we note that at finite $\omega_a$ (HO region) or finite $\varepsilon$ (TTLS region), all integrals go exponentially to zero when $\b\to\io$. The scaling regions $\b\hbar\omega_a\sim1$ or $\b\varepsilon\sim1$ provide the dominant contributions. Therefore
\begin{equation}\label{eq:blueterm}
 \begin{split}
  \blue{{\rm blue} \int}=&
  \int_{\rm HO}\dd p(m)\abs{m}^{3/2}\dd y\,\G_{q_\io}(y\abs{m}^{3/2})\\
  \times&\hbar\sqrt{\abs m}\th_3(y)\argc{\coth\argp{{\b\hbar}\sqrt{\abs m}\th_3(y)}-1}\\
  &=O\argp{(T/\hbar)^6e^{-\b\varepsilon_0}}
 \end{split}
\end{equation}
as for $\b\to\io$, the integral is dominated by $\abs m\sim\argp{\b\hbar}^{-2}$, we use this rescaling around $m=0$. We have indicated in the last line the gap $\varepsilon_0\sim\hbar^{4/3}$ to remind of its finiteness. Thus for $T\gtrsim\varepsilon_0$ the behavior is power law but exponentially damped for $T\ll\varepsilon_0$. 

For the corresponding TTLS term, we change variables $H\leftrightarrow\varepsilon$ and then set $u\equiv \b\argp{\varepsilon-\n_0(m)}\in[0,+\io[$ :
\begin{equation}\label{eq:redterm}
 \begin{split}
  \red{{\rm red} \int}\underset{T\to0}{\simeq}&\int\frac{\dd p(m)}{\n_1(m)}\int_{\n_0(m)}\dd\varepsilon\, \G_{q_\io}\argp{\frac{\varepsilon-\n_0(m)}{\n_1(m)}}\\
  &\qquad\qquad\qquad\qquad\times\frac{\varepsilon}{2} \argp{1-\tanh\frac{\b\varepsilon}{2}} \\
&\left\{\begin{split}
  &\underset{T\to0}{\propto}T^2\qquad\textrm{if~}\b\n_0(m)\lesssim 1\\
  &=T\,\G_{q_\io}(0)\int_m\frac{\n_0(m)}{\n_1(m)}e^{-\b\n_0(m)}\  \textrm{if~}\b\n_0(m)\gg 1
  \end{split}
  \right.
 \end{split}
\end{equation}
We sometimes write $\int_m\equiv \int \dd p(m)$ for notational convenience. 
With the same reasonings, the orange and green terms from $J^2q\c/2$ in~\eqref{eq:subtract} respectively scale similarly.
We conclude that these terms yield typical scalings as expected: apart from the quartic gap which brings $e^{-\b\varepsilon_0}$ for $T\ll\varepsilon_0$, the HO domain gives $O\argp{(T/\hbar)^6}$ in agreement with the Debye approximation for a Hessian with eigenvalue density $\r(\l)\sim\l^{3/2}$, whereas the TTLS domain yields $O(T^2)$, apart from the TTLS gap bringing $e^{-\b\n_0(m)}$ if $ T\ll\n_0(m)$.

We now analyze the $(\d q,\d \c)$ contributions using~\eqref{eq:frsbq}-\eqref{eq:frsbc}. The variational equation for $\c$ can be rewritten as done in~\eqref{eq:subtract} via the chain rule: $\c-\c_\io=\d\c_{\rm HO}+\d\c_{\rm TLS}+(\c-\c_\io)\dots+(q-q_\io)\dots$, same for the equation for $q$: $q-q_\io=\d q_{\rm HO}+\d q_{\rm TLS}+(\c-\c_\io)\dots+(q-q_\io)\dots$. Once we have studied the HO and TLS explicit contributions of the integrands, we get a linear system for  $(\d q,\d \c)$. We have
\begin{equation}\label{eq:orangreen}
 \begin{split}
  \d\c_{\rm HO}=&
\frac{2}{J^2q_\io}\times {\color{orange}\rm orange~term}\propto (T/\hbar)^6e^{-\b\varepsilon_0}\\
  \d\c_{\rm TLS}=&
  \frac{2}{J^2q_\io}\times\textrm{\green{green~term}}\propto T^2\int_m\frac{e^{-\b\n_0(m)}}{\n_1(m)}
   \end{split}
\end{equation}
as these come from the $J^2q\c/2$ part of the energy, and
\begin{equation}\label{eq:dq}
 \begin{split}
  \d q_{\rm HO}=&\int_{\rm HO}\G_{q_\io}(H)\hbar \partial_Hv_a\partial_H\om_a\argc{\coth\argp{\frac{\b\hbar\om_a}{2}}-1}\\
  &+\int_{\rm HO}\G_{q_\io}(H)\argc{\frac{\partial_H\ln\tanh\argp{\frac{\b\hbar\om_a}{4}}}{\b}}^2\\
  =&O\argp{(T/\hbar)^4e^{-\b\varepsilon_0}}+O\argp{(T/\hbar) e^{-\b\varepsilon_0}}
   \end{split}
\end{equation}
  \begin{equation}\label{eq:dq2}
 \begin{split}
  \d q_{\rm TLS}=&\int_{\rm TLS}\G_{q_\io}(H)\left\{\argc{
  -\partial_H\bar\varepsilon+\frac{\partial_H\varepsilon}{2}\tanh\frac{\b\varepsilon}{2}}^2\right.\\
  &\qquad\left.-\argc{ -\partial_H\bar\varepsilon+\frac{\partial_H\varepsilon}{2}}^2\right\}
  \\
  \sim&\int_m\frac{\n_1(m)}{4}\int_{\n_0(m)}\dd\varepsilon\argc{\tanh^2\argp{\frac{\b\varepsilon}{2}}-1}\\
  \propto &-T\int_m\n_1(m)e^{-\b\n_0(m)}\\
 \end{split}
\end{equation}
A crucial order of magnitude is that the gaps $\varepsilon_0\sim\hbar^{4/3}\gg\n_0(m)$, the latter being exponentially small in $\hbar$. Hence for $T\ll\varepsilon_0$ the HO contributions are damped and we drop them here\footnote{Note however that the $\d q_{\rm HO}$ contribution brings a linear term~\eqref{eq:dq}, as well as $\d q_{\rm TLS}$ does.}. All in all, the linear system for $T\to0$ is of the form
\begin{equation}\label{eq:MM}
 \begin{pmatrix}
  \d\c\\\d q
 \end{pmatrix}
 =\MM^{-1}\begin{pmatrix}
  O\argp{T^2\int_m e^{-\b\n_0(m)}}\\ 
  O(T\int_me^{-\b\n_0(m)})
 \end{pmatrix}
\end{equation}
$\MM$ is a $2\times2$ matrix independent of the temperature; thus we expect\footnote{
The derivative coefficients constituting the matrix $\MM$ (meaning the $\dots$ coefficients in the $(\d\c,\d q)$ equations above~\eqref{eq:orangreen}) tend to a non-zero constant for $T\to0$. This can be seen for $q$ derivatives using the same diffusion equation argument as in Eq.~\eqref{eq:diffeq}, giving an identical expression to the $T\to0$ equations for $\c$ and $q$, albeit with a second $H$ derivative of $\G_q(H)$. For $\c$ derivatives the $T\to0$ constants are $J^2\int\G_{q_\io}(H)\moy{\hat x}\partial_H\moy{\hat x^2}$ for $\d\c$ and $\int\partial_H^2\G_{q_\io}(H)\moy{\hat x^2}$ for $\d q$. Hence the linear $T$ term in~\eqref{eq:MM} dominates.} both $\d\c,\d q=O(T\int_me^{-\b\n_0(m)})$. 

Finally we look at the partial derivatives over $(\c,q)$ in~\eqref{eq:subtract}. 
The $q$ derivatives involve only the Gaussian function $\g_{J^2q}$. One uses that it satisfies a diffusion equation 
${\partial_q\g_{J^2q}(H)=\frac{J^2}{2}\partial_H^2\g_{J^2q}(H)}$. As a consequence
\begin{equation}\label{eq:diffeq}
 \begin{split}
\frac{\partial}{\partial {q}}\frac{U_{\rm RS}^a}{N}=
& \frac{J^2}2\int\partial_H^2\G_q(H)\argc{\frac{f_m(H)}{\b}-\frac{\partial f_m(H)}{\partial\b}}
 \end{split}
\end{equation}
where we used the variational equation for $\c$~\eqref{eq:frsbc} integrated by parts twice.
The integrand is the same as the explicit $\b$ dependence (colored terms) in~\eqref{eq:subtract}, apart from the global factor $\partial_H^2\G_{q_\io}(H)$ instead of $\G_{q_\io}(H)$ or $\z_{q_\io}(H)$.
This does not impact the scalings as all these functions tend to a constant for $H\to0$ (\ie in the scaling region important for $T\to0$). Therefore the term $\partial_q U_{\rm RS}^a/N$ vanishes with the same scaling as the explicit $\b$ dependence of~\eqref{eq:subtract}. Incidentally this happens for higher $q$ derivatives as well, using repeatedly the diffusion equation for the Gaussian $\g_q$.  As it gets multiplied by $\argp{q-q_\io}\to0$, it is subdominant with respect to the latter explicit $\b$-dependent terms.

Concerning the $\c$ derivative, it is readily done from the shift  $m\to \k-J^2\c$ in the last term of~\eqref{eq:Uvar}, generating an effective-particle average $\propto\moy{x^2}$.
Such  averages of powers of $\hat x$ are also present in the saddle-equation for $q$~\eqref{eq:frsbq} and we express them with $H$ derivatives of $f_m(H)$. Combined, they read
\begin{equation}\label{eq:ApB}
\begin{split}
  \frac{\partial}{\partial\c}\frac{U_{\rm RS}^a}{N}&=A+B\\
  =-\frac{J^2}{2}&\int\G_{q}(H)\argc{\frac{\partial}{\partial\b}\argp{\frac{\partial_H^2f_m}{\b}}+\b\frac{\partial}{\partial\b}\argp{\frac{\partial_Hf_m}{\b}}^2}
\end{split}
\end{equation}
We analyze both contributions separately. All integrands go to 0 in the $\b\to\io$ limit, and  the same scalings dominate as before. The second derivative ($A$) can be integrated twice by parts:
\begin{equation}
 \begin{split}
  A_{\rm HO}=&\frac{J^2}{2}\int_{\rm HO}\partial_H^2\G_q(H)\frac{\partial}{\partial\b}\frac{\ln\argp{1-e^{-\b\hbar\om_a}}}{\b}\\
  =&O\argp{\argp{T/\hbar}^7e^{-\b\varepsilon_0}}\\
  A_{\rm TLS}\propto&
  \left\{\begin{split}
  & T^3\qquad\textrm{if~}\b\n_0(m)\lesssim 1\\
  & T^2\int\dd p(m)e^{-\b\n_0(m)} \quad\textrm{if~}\b\n_0(m)\gg 1
   \end{split}
  \right.
 \end{split}
\end{equation}
The other term ($B$) is akin to $\d q$ in~\eqref{eq:dq}:
\begin{equation}\label{eq:eqB}
 \begin{split}
  B_{\rm HO}
  =&O\argp{(T/\hbar)^4}+O\argp{T/\hbar}\\
   B_{\rm TLS}
\propto&\left\{\begin{split}
 &\  T\quad\textrm{if~}\b\n_0(m)\lesssim 1\\
  &\ \int\dd p(m)\n_1(m)e^{-\b\n_0(m)} \quad\textrm{if~}\b\n_0(m)\gg 1
  \end{split}
  \right.
 \end{split}
\end{equation}
Hence it  turns out that the $B$ term, linear in $T$, dominates $\partial_\c U_{\rm RS}^a/N$. 

Furthermore, using the earlier diffusion equation argument for $q$ derivatives and applying it to~\eqref{eq:ApB} makes the crossed derivative $\partial^2_{q\c}U_{\rm RS}^a/N $ (or higher $q$ derivative)  scale identically to  $\partial_\c U_{\rm RS}^a/N$. 

Concerning $\c$ derivatives of at least second order, again with  the shift  $m\to \k-J^2\c$ we have from~\eqref{eq:Uvar}
\begin{equation}
\begin{split}
  \frac{\partial^n}{\partial\c^n}\frac{U_{\rm RS}^a}{N}&=(-1)^{n+1}J^{2n}\frac{\partial}{\partial\b}\int \frac{\partial^n }{\partial m^n}f_m(H)\\
  \frac{\partial^n }{\partial m^n}f_m(H)&=\argp{-{\b}/{2}}^n \CC_n(x^2)
\end{split}
\end{equation}
with $\CC_n(x^2)$ the cumulant\footnote{For example $\CC_2(y)=\moy{\argp{y-\moy{y}}^2}$, $\CC_3(y)=\moy{\argp{y-\moy{y}}^3}$, ...} of order $n$ in the variable $x^2$, due to the free-energy nature of $f_m$. In the low-temperature limit the biggest contribution from these cumulants is close to the symmetric case $H=0$ in the TLS domain and close to the origin in the HO domain, as further away the energy gaps get larger. In this limit all cumulants vanish (at least linearly in $T$) implying ${\partial_\c^n}{U_{\rm RS}^a}/{N}=O\argp{\b^{n-2}}$ for $n\geqslant2$. Again with the diffusion equation argument any cross derivative $\partial_q^l\partial_\c^k{U_{\rm RS}^a}/{N}=O\argp{\b^{k-2}}$ with $k\geqslant2$. Note that in the expansion such a term gets multiplied by $(q-q_\io)^l(\c-\c_\io)^k=O(\b^{-k-l})$ so that this total contribution is negligible with respect to $T^2$.

\subsubsection{Conclusion}

In the previous calculations, we only considered scalings from the HO and TTLS domains, but not of the quartic region. This stems from the numerical evidence that it is a small crossover region between the both regimes, its extension being of order $\hbar^{5/3}$~\eqref{eq:Qext}. Thus it has a non-negligible value only for large enough $\hbar$, which happens only close to the quantum critical point or beyond it in the RS phase. In the present variational approximation it nonetheless implies a gap of order $\hbar^{4/3}$, which then becomes rather large and damps the energy as $e^{-\b\varepsilon_0}$ for $T\ll\varepsilon_0$.  
In particular, in the above computations~\eqref{eq:blueterm}-\eqref{eq:redterm} the scaling partly relies on the $H$ dependence of the gap. In the quartic region, increasing $m$ it crosses over from the linear dependence of $\varepsilon$ of the TTLS-WKB region~\eqref{eq:chgvar} to the quadratic behavior\footnote{For $m>0$ one has $\om_a\underset{H\to0}{=}\sqrt m+\frac{H^2}{4m^{5/2}}+O(H^4)$.} of $\hbar\om_a$ for $m>0$. As a consequence one expects the energy in this quartic domain to cross over as well and lie in between the TTLS $T^2$ scaling and the $T^6$ Debye scaling -- exponential gap dependences aside. This crossover has been studied in the numerical results of~\secref{sec:num}; see also~\secref{sec:pseudogap} for a further discussion of the gap's $H$ dependence and its influence on the specific heat.

\begin{figure}[!hbtp]
\centering
\begin{lpic}[]{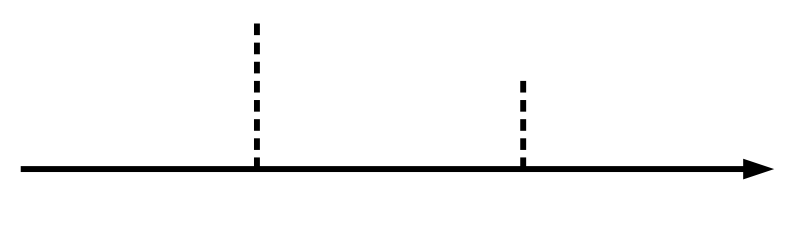(0.6)}
\lbl[]{45,6;$E_{\rm gap}\approx \int \dd p(m) \,e^{-{\abs{m}^{3/2}}/{\hbar}}$}
\lbl[]{40,-4;$(\abs{m}^{3/2}\gg\hbar)$}
\lbl[]{95,6;$E_{\rm gap}\approx \hbar^{4/3}$}
\lbl[]{135,13;$T$}
\lbl[]{20,35;$U_{\rm HO}$, $U_{\rm Q}$, $U_{\rm TTLS}$}
\lbl[]{20,30;exponentially}
\lbl[]{20,25;suppressed}
\lbl[]{20,20;(gapped)}
\lbl[]{66,23;$U_{\rm HO}$, $U_{\rm Q}$ gapped }
\lbl[]{80,33;$U_{\rm TTLS}$ power law}
\lbl[]{115,23;$U_{\rm HO}$, $U_{\rm Q}$ power law }
     \end{lpic}
 \caption{Summary of the energy gap scales and corresponding temperature scaling of the energy in each domain: HO, quartic (Q), TTLS. There is yet another higher energy gap scale, the harmonic one $E_{\rm gap}\propto \hbar$ which we do not display as it is rather close to the quartic gap, semiclassically. In practice in the numerics, the gap separation may not be very sharp, see Fig.~\ref{fig:fixed}. For each domain, as the temperature rises much above the typical gap, higher energy levels could start to be excited, and two-level truncations should break down. For example at high temperature the two-level approximation implies a constant energy.}
 \label{fig:scalings}
\end{figure}
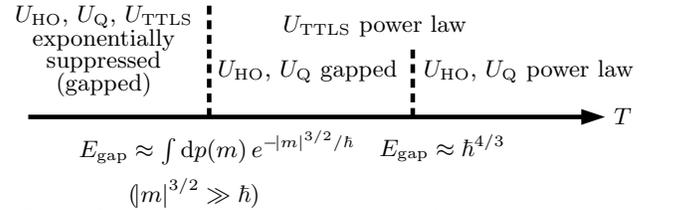

In conclusion, we gather all the previous scalings in~\eqref{eq:subtract}:
\begin{widetext}
\begin{equation}\label{eq:conclU}
 \begin{split}
   \frac{U_{\rm RS}^a}{N}\underset{T\to0}{=}& \frac{U_{\rm RS}^a(T=0)}{N} 
   + \overbrace{\textrm{explicit $\b$ dependence (colored terms)}}^{O(T^2)+\textrm{Debye~}O\argp{(T/\hbar)^6}} +\overbrace{\argp{q-q_\io}}^{O(T)}\overbrace{\frac{\partial}{\partial {q}}\frac{U_{\rm RS}^a}{N}}^{\textrm{same as explicit}}+\overbrace{\argp{\c-\c_\io}}^{O(T)}\overbrace{\frac{\partial}{\partial {\c}}\frac{U_{\rm RS}^a}{N} }^{O(T)}\\
   &+\underbrace{\frac{\argp{q-q_\io}^2}{2}}_{O(T^2)}\underbrace{\frac{\partial^2}{\partial q^2}\frac{U_{\rm RS}^a}{N}}_{\textrm{same as explicit}}+\underbrace{\argp{q-q_\io}}_{O(T)}\underbrace{\argp{\c-\c_\io}}_{O(T)}\underbrace{{\frac{\partial^2}{\partial q\partial {\c}}\frac{U_{\rm RS}^a}{N} }}_{O(T)}
    +\underbrace{\frac{\argp{\c-\c_\io}^2}{2}}_{O(T^2)}\underbrace{\frac{\partial^2}{\partial \c^2}\frac{U_{\rm RS}^a}{N}}_{\rm const.}+\dots
 \end{split}
\end{equation}
\end{widetext}
As shown above higher-order terms in the expansion are subdominant. 
In~\eqref{eq:conclU} we deliberately left out the exponential dependences stemming from the finite gaps $E_{\rm gap}$. These exponentials are negligible when $T\gg  E_{\rm gap}$ and one observes the power laws. Yet if $T$ becomes too high, one starts to excite the rest of the spectrum and the present low-temperature approximation breaks down. We have found out three types of gaps: the TTLS-WKB one that is exponentially small in $\hbar$, the quartic gap $\sim \hbar^{4/3}$ and the harmonic one $\sim\hbar$. In the present semiclassical analysis the former two are the smallest. A summary of the scales is given in  Fig.~\ref{fig:scalings}.
The analytical computation of the specific heat, relying on a low-temperature expansion and critical scalings of each integration domain, corresponds to the complementary numerical assessment of~\secref{sec:num} for the ``explicit'' temperature dependence. However the contribution due to the temperature dependence of $(\c(T),q(T))$ could not be properly probed in the numerics. It is studied further here, revealing that they do not spoil the linear scaling of the specific heat, as they bring a similar linear contribution.

\subsection{Conclusion}\label{sub:anscal}

\begin{figure}[!hbtp]
\centering
\begin{lpic}[]{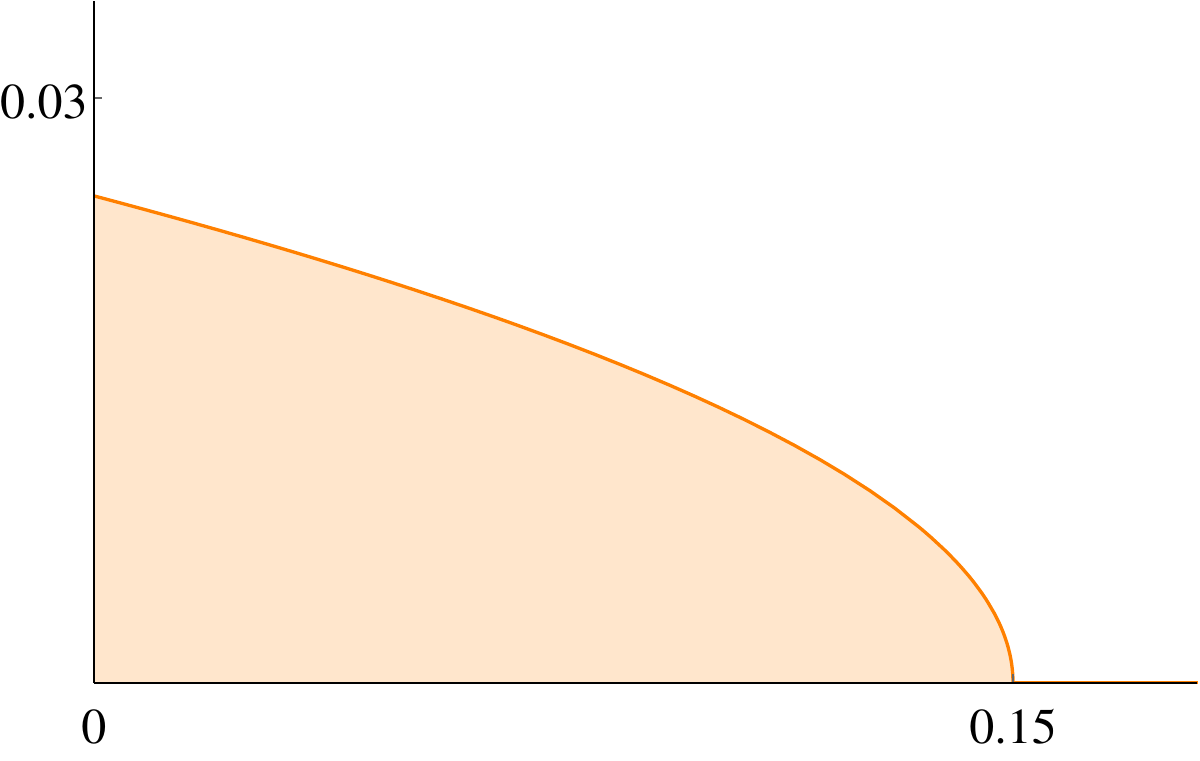(0.27)}
\lbl[]{5,60;$T$}
\lbl[]{85,0;$\hbar$}  
\lbl[]{75,100;{\small Convex phase (RS)}}
\lbl[]{75,50;{\small Glassy (fullRSB)}}
\lbl[]{150,22;\begin{tikzpicture}\draw[dashed,thick,color=blue] (0,0) -- (0,0.6);\end{tikzpicture}}
\lbl[]{150,6;{\tiny \blue{$0.05$}}}
\lbl[]{35,6;{\tiny \blue{$10^{-3}$}}}
\lbl[]{125,20;{\small \blue{TTLS}}}
\lbl[]{91,13;\begin{tikzpicture}\draw[color=blue,thick] (0,0) -- (3.1,0);\end{tikzpicture}}
\lbl[]{91,14;\begin{tikzpicture}\draw[color=blue,thick] (0,0) -- (3.1,0);\end{tikzpicture}}
\lbl[]{175,13;\begin{tikzpicture}\draw[color=yellow!80!red,thick] (0,0) -- (1.3,0);\end{tikzpicture}}
\lbl[]{175,14;\begin{tikzpicture}\draw[color=yellow!80!red,thick] (0,0) -- (1.3,0);\end{tikzpicture}}
\lbl[]{220,13;\begin{tikzpicture}\draw[color=violet,thick] (0,0) -- (1,0);\end{tikzpicture}}
\lbl[]{220,14;\begin{tikzpicture}\draw[color=violet,thick] (0,0) -- (1,0);\end{tikzpicture}}
\lbl[]{200,22;\begin{tikzpicture}\draw[dashed,thick,color=violet] (0,0) -- (0,0.6);\end{tikzpicture}}
\lbl[]{172,13;{\small \orange{$\bullet$}}}
\lbl[]{200,6;{\tiny \textcolor{violet}{$2$}}}
\lbl[]{175,45;{\tiny \textcolor{yellow!80!red}{SWP\&DWP}}}
\lbl[]{225,20;{\small  \textcolor{violet}{SWP}}}
\lbl[]{225,100;{\small \textcolor{violet}{only}}}
\lbl[]{225,90;{\small \textcolor{violet}{harmonic}}}
\lbl[]{225,80;{\small \textcolor{violet}{oscillators}}}
\lbl[]{225,68;{\small \textcolor{violet}{(Debye)}}}
\lbl[]{200,-10;{\scriptsize   \textcolor{yellow!80!red}{   $E_{\rm gap}>0$}}}
\lbl[]{185,5;\begin{tikzpicture}\draw[-stealth,color=yellow!80!red] (0,0) -- (-0.6,0.4);\end{tikzpicture}}
\lbl[]{225,45;\begin{tikzpicture}\draw[-stealth,color=violet] (0,1.5) -- (0,0.6);\end{tikzpicture}}
\lbl[]{175,35;\begin{tikzpicture}\draw[stealth-stealth,color=yellow!80!red] (-0.2,0) -- (1,0);\end{tikzpicture}}
\lbl[]{175,55;{\small \textcolor{yellow!80!red}{ quartic}}}
\lbl[]{175,66;{\small \textcolor{yellow!80!red}{ and}}}
\lbl[]{175,75;{\small \textcolor{yellow!80!red}{ Debye}}}
\end{lpic}
\caption{Sketch of the quantum phase diagram displayed in Fig.~\ref{fig:qphd}(inset), within the variational approximation.  Excitations dominating the specific heat for $T\to0$, with the boundaries of their regime, are shown close to the $\hbar$ axis. From right to left, new excitations appear on top of the previous (on the right) ones. The incorrect finiteness of the gap at the quantum critical point is mentioned. }
\label{fig:recap}
\end{figure}

In this section we have applied the semiclassical analysis of~\secref{sec:semi} but starting from the classical RSB phase bearing effective DWP. The presence of these DWP changes radically the structure of the expansion: in particular it cannot be perturbative anymore as the semiclassical saddle-point solutions are instantons. We then approached the RSB phase from the RS phase by letting $\hbar$ finite, large enough. We simplified the self-consistent problem by building a variational approximation, that retains the qualitatively new features of this phase and remains coherent with the classical limit. It amounts to replace the full dynamical impurity problem by a static one, different from the usual ``static approximation'' of quantum mean-field glasses~\cite{BM80b,CGSS01,BK10}.  The impurity problem becomes standard single-particle quantum mechanics. A further simplification was considered by investigating directly the low-temperature limit. The collection of (effective) single particles induces, depending on the parameters of the effective potentials, three types of elementary excitations: \textit{(i)} Debye excitations,  \ie quantum harmonic oscillators in the bottom of a well \textit{(ii)} tunneling two-level systems (TTLS) \textit{(iii)} quartic excitations, \ie particles in a potential energy well (or double well) very close to a purely quartic potential, that cannot be approximated by a harmonic well and where the wavefunction is widely spread spatially, contrary to TTLS exhibiting localized probability of presence only close to the bottom of each well through tunneling. Numerical checks of the low-temperature approximation for the quartic excitation and the TTLS one (WKB method) have been performed. 
Then we solved numerically the self-consistent variational equations. We found a RSB transition, which for $T=0$ amounts to a quantum critical point (see the inset in Fig.~\ref{fig:qphd}). Much above this critical point, only Debye excitations are present, and only SWP populate the effective potentials. Lowering $\hbar$, close to the quantum critical point there are both Debye and quartic excitations; DWP are found around the quantum critical point but no tunneling two-level systems exist yet. One has to be within the RSB phase to find occurences of TTLS, which dominate the specific heat. The specific heat has been only partially computed numerically, but we provided analytic arguments for its low-temperature scaling. TTLS bring a dominant $C_V\sim T$ contribution, as can be expected from these degrees of freedom, with the caution that the total energy of the system is not simply a sum of individual energies of effective degrees of freedom. We showed that other contributions to the thermodynamic energy do not impair the linear scaling. The Debye behavior is $C_V\sim T^5$ seen at larger temperatures, and the quartic excitations yield a crossover exponent in between (evaluated numerically around 2). Fig.~\ref{fig:recap} points out the principal features of the phase diagram found in this approach. 

However, a first issue concerns the variational approximation itself. Although it predicts a RSB transition to a marginal glassy phase, at the quantum critical point the many-body energy gap does not vanish, and one ultimately finds at $T\to0$ exponentially damped specific heats. This may be expected from a semiclassical approach, where small gaps may be missed. Here the rationale is within the variational approximation that ascribes the many-body gap to a single-particle one, which always remains finite. The transition occurs for a sizeable value of $\hbar$ where gaps are large enough, so that in addition the power-law scaling right at the transition is numerically unclear. 
This is related as well to the critical frequency scaling of the propagator~\eqref{eq:varapprox}, which is necessarily incorrect in this approximation. For example if one hits the transition to the RSB-DWP phase from the RS phase where only single wells are present (as in~\secref{sec:semi}), plugging the present variational approximation and making the analytical analysis of~\secref{sub:gapCv} (getting rid of TTLS contributions) would provide the Debye contribution $C_V\sim T^5$ but likely another spurious\footnote{This comes from the term $\d\c \partial_\c U^a_{\rm RS}$ in~\eqref{eq:subtract}, combining the HO contribution to $\d\c$ from \eqref{eq:dq} and \eqref{eq:MM}, with the HO contribution in \eqref{eq:ApB},\eqref{eq:eqB}. } linear-in-$T$ term from the temperature dependence of $\c(T)$. Note that when TTLS are present, at variance such contributions do not harm the overall linear scaling, as the individual TTLS contribution is robustly linear in $T$, a contribution difficult to avoid, that is presumably decisive.  
A possible way to resolve these issues is presented in~\secref{sec:spinboson}.

A second issue concerns the RSB phase. Strictly speaking the previous analysis assumes the RS ansatz. In~\secref{sec:pseudogap} we discuss the impact of RSB on the above physical picture.

\section{RSB-DWP phase: specific heat, pseudogap and marginal stability condition}\label{sec:pseudogap}

In~\secref{sec:DWP} we applied a variational approximation to simplify the impurity problem in the RS phase, \ie at large enough $\hbar$. In this section we discuss what happens lowering $\hbar$ so that we enter the RSB-DWP phase, still under this simplifying variational scheme for $T\to0$.  
A crucial quantity appears in this fullRSB phase: the field distribution  $P_\k(x,H)$, especially at level $x=1$, see the energy~\eqref{eq:URSBvar}. 
We can reproduce the same analysis as in Sec.~\ref{sub:gapCv} for the $T\to0$ behavior of the energy; the essential difference here is that the field distribution may not be Gaussian anymore, which may change the temperature scalings with respect to the RS case, if this distribution vanishes at $H=0$. 
 Thus we need to understand better the behavior of $P_\k(1,H)$ for $H\to0$. Indeed in the classical $T\to0$ case~\cite{FU22},  $P_\k(1,H)$ is regular for $m>0$ while for $m<0$ it has a linear pseudogap $P_\k(1,H)\sim\g_\k\abs H$ for $\abs H\sim T$. This pseudogap is mandatory to cure the singular behavior of the classical replicon $\l_R\to-\io$, giving it a finite value that vanishes at the RSB-$\om^4$ transition. 

Here we argue that around $H=0$, $P_\k(1,H)$ is regular for finite $\hbar$ but vanishes in the limit $\hbar\to0$ in order to satisfy the marginal stability condition.  As a consequence the RS-DWP $T\to0$ scaling of the specific heat~\eqref{eq:conclU} holds within the marginal phase at finite $\hbar$ as well. 

To see this, we consider the marginal stability condition~\eqref{eq:frsbrep}. It must be verified in the fullRSB phase. At $T\to0$ we have that
\begin{equation}
 \frac{f_\k(1,H)}{\b}\underset{\b\to\io}{=}-E_g(\k,H)
\end{equation}
with $E_g(\k,H)$ the ground-state energy of the impurity problem. 
This means one has to understand the behavior of the ground state energy to get the analytic behavior of the field distribution. 
At finite $\hbar$, the ground-state behaves far from classically and remains smooth with $H$, so that $f_\k''(1,H)$ is smooth too, hence nothing dramatic happens in the replicon's integral. 
 Thus we expect a smooth dependence as well of $P_\k(1,H)$ to satisfy the marginal condition, notably that it goes to a finite constant for $H\to0$. This is akin to what happens in the quantum SK model in a transverse field at finite quantum fluctuations (the transverse field itself)~\cite{AM12,KZL24}. Nonetheless when $\hbar$ is decreased a singularity could emerge in $f''_\k(1,H)$ near symmetric DWPs as the degeneracy gets unlifted: indeed in the classical model for $m<0$  $f''_\k(1,H)$ is very singular in $H=0$ and there is a low-temperature linear pseudogap in $P_\k(1,H)$  to counterbalance it~\cite{FU22}. 
 We now focus on the emergence of this singularity as $\hbar$ is lowered, as it may affect thermodynamic quantities. For instance, the replicon has an odd diverging behavior when $\hbar\to0$, see discussion around Fig.~\ref{fig:T=0}(c).

\subsection{Ground-state analyticity}\label{sub:analytic}

We need first to understand the behavior of $f''_\k(1,H)/\b$ close to $H=0$, appearing in the marginal stability condition~\eqref{eq:frsbrep}. It is easier to investigate it examining
\begin{equation}\label{eq:fEg}
 \frac{f'_\k(1,H)}{\b}=\frac{\Tr \,\hat x\, e^{-\b \hat H_{\rm eff}}}{\Tr \, e^{-\b \hat H_{\rm eff}}}\underset{\b\to\io}{\to}-E_g'(m,H)
\end{equation}
Note that~\eqref{eq:fEg} is singular in the classical limit at $H=0$, $T=0$. Indeed in this case for DWP ($m<0$), $f'_\k(1,H)/{\b}\to x_a(m,H)=\sign(H)\argp{\sqrt{6\abs m}+O(H)}$. $x_a$ is the absolute minimum and satisfies ${\partial_x v_m(x_a(m,H),H)=0}$. By differentiating over $H$ the latter equation, one has~\cite{BLRUZ21}[App.B] 
\begin{equation}\label{eq:classdxadH}
 \frac{f''_\k(1,H)}{\b}=\frac{\partial x_a}{\partial H}=\frac{1}{\tilde a}+\argc{x_a(m,0^+)-x_a(m,0^-)}\d(H)
\end{equation}
\ie the singular \textit{classical} $T=0$ behavior of $f''_\k(1,H)/\b$. The regular part corresponds to the static susceptibility $f''_\k(1,H)/\b=\b\argc{\moy{x^2} -\moy{x}^2}= 1/\tilde a=1/\argp{m+x_a^2/2}$ we encountered in the single-well case ($m\geqslant0$) of~\secref{sec:SGLD}, see~\eqref{eq:c_0} and~\eqref{eq:tildeamass}.

Now in the quantum case, let us note that for $m>0$ (SWP) $E_g\simeq v_a+\hbar\om_a/2$ is analytic: $v_a$ and $\omega_a$ are  analytic functions of $x_a$, which from~\eqref{eq:cubicdepressed} is an analytic odd function of $H$, $x_a(m,H)=2\sqrt{2\abs{m}}\sinh\argc{\frac13\textrm{arcsinh}\argp{{3 H}/{\argp{2\abs{m}}^{3/2}}}}$. No issue is suspected here, and this should hold irrespective of the harmonic approximation. Therefore this $m>0$ region does not need any regularization mechanism and we expect smoothness of the corresponding $P_\k(1,H)$. Similarly, in the DWP-HO region (Fig.~\ref{fig:DWP}) the HO expressions are smooth due to smoothness of the functions in~\eqref{eq:cubicdepressed}. We need to assess what can happen for $m\leqslant0$ close to $H=0$, \ie within the quartic and TTLS regions.

Let us then consider almost symmetric DWP at $\hbar>0$. The singular behavior seen in~\eqref{eq:classdxadH} is cutoff, as can be seen numerically. This stems from, in the limit $T\to0$,\begin{equation}\label{eq:wavepsi}
 \frac{f'_\k(1,H)}{\b}=\frac{\Tr \,\hat x\, e^{-\b \hat H_{\rm eff}}}{\Tr \, e^{-\b \hat H_{\rm eff}}}\underset{\b\to\io}{\to}\bra{-}\hat x\ket{-}=\int \dd x\, x \abs{\psi_-(x)}^2
\end{equation}
where the analyticity and $\hbar$ dependence is now buried into the wavefunction $\psi_-(x)=\braket{x|-}$. $\ket -$ is the ground state and equivalently the lowest level in the TLS approximation for $\hat H_{\rm eff}$. It satisfies $-(\hbar^2/2)\partial_x^2\psi_-+v_m(x)\psi_-=E_g(m,H)\psi_-$, and is smooth due to the Laplacian term. For $H=0$, due to parity of the Hamiltonian, $\psi_-(x)$ is an even function (with no node, a bump within each well, going to zero at infinities, see bottom-left inset in Fig.~\ref{fig:DWP})~\cite[Chap.III]{messiah}. Then from~\eqref{eq:wavepsi} using $x\to -x$ one gets\footnote{Incidentally, these general properties are recovered within the WKB approximation (in the TTLS region) for the ground-state energy $E_g=\bar\varepsilon-\varepsilon/2$ only for an analytic choice of energy $E$ inside the WKB integral~\eqref{eq:TLS}. Indeed, $\bar\varepsilon$ is analytic and $\D^2\propto H^2$ as well (Fig.~\ref{fig:D}). Therefore, as $\D_0(H=0)>0$, $\varepsilon$ will be analytic unless $\D_0$ isn't. In~\eqref{eq:TLS}, the product $\om_a\om_s$ is an even analytic function of  $H$. Finally, we numerically see that the WKB integral is analytic if and only if $E$ is analytic (note that $E$ also enters into the defintion of $a$ and $b$). However the choices $E=v_s$ (made in the numerics) or $E=E_s$ imply a non-analytic dependence of the WKB integral, yielding $\l_R\underset{\hbar\to0}{\to}-\io$ -- which occurs for small $\hbar$ in Fig.~\ref{fig:T=0}(c). This is akin to the classical RS case where the pseudogap mechanism curing the singularity is absent~\cite{BLRUZ21,FU22}.} $f'_\k(1,0)/\b=0$. This is in contrast with the classical $T=0$ result discussed above.

\subsection{Ground-state scalings in $\hbar$ for almost symmetric double wells}\label{sub:hbarscal}

The previous observations mean that there is an $\hbar$ dependent regime, going to zero in the classical limit, in which analyticity of $f_\k(1,H)/\b$ is reinstated. 

\begin{figure}[h!]
\centering
\begin{lpic}[]{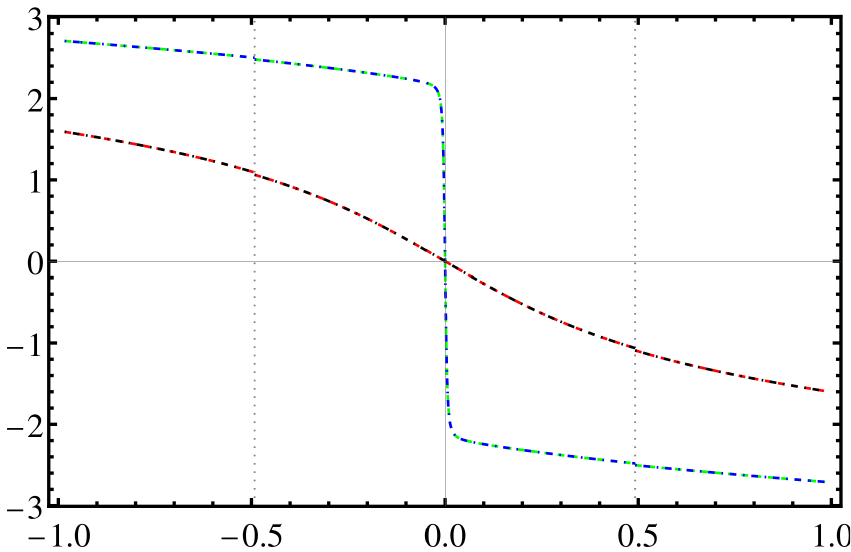(0.42)}
\lbl[]{-10,50,90;$\frac{\partial_HE_g}{\hbar^{1/3}}\argp{\frac{m}{\hbar^{2/3}},\frac H\hbar}$ }
\lbl[]{75,-5;$H/\hbar$}
     \end{lpic}\\~~\\~~\\
     \begin{lpic}[]{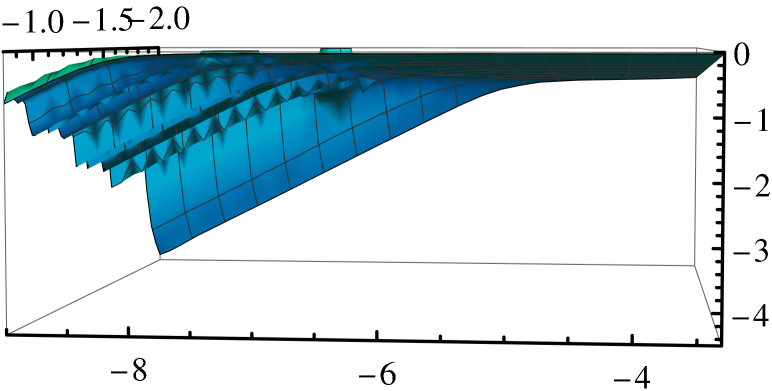(0.5)}
\lbl[]{140,30,-90;$\log_{10}\abs{E_g'}$ }
\lbl[]{15,70;$\log_{10}\abs m$}
\lbl[]{65,-5;$\log_{10} H$}
     \end{lpic}
 \caption{(top) Scaling of $E_g'=\partial_H E_g=-\bra{-}\hat x\ket{-}$ given by \eqref{eq:Egscal}, exactly obeyed $\forall\hbar$. Black/red: $m/\hbar^{2/3}=-10^{-4}$ (close to $m=0$) with $\hbar=0.1$ (black), $\hbar=10$ (red). Blue/green: $m=-\hbar^{2/3}$ (boundary with TTLS-WKB region) with $\hbar=0.1$ (green), $\hbar=10$ (blue). 
 Away from the vertical dotted lines signaling $H=\pm H_c\simeq \pm0.49\hbar$ (\eqref{eq:quartic} and Fig.~\ref{fig:mHlowT}), the lines are continued by their HO approximations, showing the correct connection between the two regimes.
 When $\hbar\to0$, the interpolating regime $H\propto \hbar$ shrinks and $E_g'\sim-x_a$ becomes discontinuous. 
 Note that at the threshold value $m=-\hbar^{2/3}$ (blue/green) we start to see that the scaling $H\propto \hbar$ is not anymore appropriate to account for the variation of $E_g'$. 
 (bottom) Log-plot of the first derivative of the ground-state energy with respect to $H$, obtained by solving the static Schr\"odinger equation, for $\hbar=10^{-3}$ (hence some numerical noise). The range of $m$ is in the TTLS-WKB regime, $-0.15<m<-0.01$. As $m$ decreases ($\ln_{10}\abs m$ increases), the range of $H$ that interpolates from the classical value $E_g'\simeq-x_a$ to 0 decreases exponentially.}
 \label{fig:quarticscaling}
\end{figure}
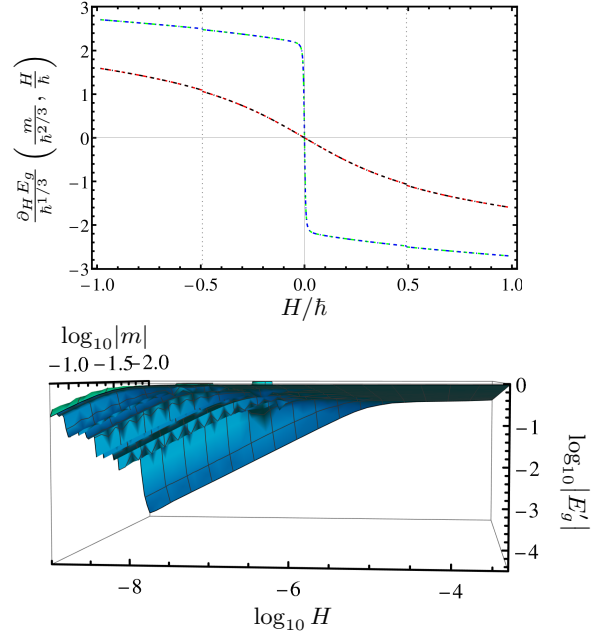

In the quartic region, for the scaling reasons presented in~\secref{sub:boundaries}, we have the following scaling
\begin{equation}\label{eq:Egscal}
 E_g(m,H)=\hbar^{4/3}\wt E_g\argp{\frac{m}{\hbar^{2/3}},\frac H\hbar}
\end{equation}
This is shown numerically in Fig.~\ref{fig:quarticscaling}(top), as well as the analyticity for $H\to0$. The $H$ interval that interpolates the quantum to classical crossover is therefore linear in $\hbar$ in this region. In the TTLS-WKB domain $ m \lesssim -\hbar^{2/3}$, numerically there is an exponentially small regime of $H$ that interpolates from $H=0$ where $E_g'=0$ to the classical values at larger $H$ (Fig.~\ref{fig:quarticscaling}(bottom)). 

The rationale for different scalings according to the order of magnitude of $m$ lies in the fact that close to the quartic potential, the action 
\begin{equation}\label{eq:kinkaction}
S=\int_0^{\b\hbar} \dd t \argc{\frac{\dot x^2}{2}+v_m(x)}\propto\hbar^{4/3}\ll\hbar\ , 
\end{equation}
in other words there are no instantons (finite-action solution of the saddle-point equation)~\cite{coleman_erice,rajaraman} and the quartic scalings prevail. For $ m \lesssim -\hbar^{2/3}$ these scalings break down, instanton solutions become possible (the action becomes order one). Summing over an infinite number of bounces drastically changes the $\hbar$ dependence from perturbative to non-perturbative (compare  the HO gap scaling $E_{\rm gap}\propto\hbar$, or quartic $E_{\rm gap}\propto\hbar^{4/3}$, to the instantonic/WKB $E_{\rm gap}\propto e^{-S_0/\hbar}$, with $S_0$ the kink solution~\eqref{eq:TLS})~\cite{coleman_erice}.

We now assess what is this exponentially small in $\hbar$ regime in the TTLS domain. For the vanishing of the replicon and from~\eqref{eq:fEg}, we are interested in the $(H,\hbar)$ dependence of the second derivative of $E_g$. From the previous analyticity discussion and the symmetry $H\to-H$ it follows that we can expand, fixing $m$ and $\hbar$,
\begin{equation}\label{eq:Egexpand}
\begin{split}
  E_g(H)=&\bar\varepsilon-\frac\varepsilon2=E_g(0)+\frac{H^2}{2}E_g''(0)+O(H^4)\\
  \Rightarrow\ E_g'(H)&\underset{H\to0}{\sim}HE_g''(0)=H\frac{\partial^2}{\partial H^2}\restriction{\argp{\bar\varepsilon-\frac\varepsilon2}}{H=0}
 \end{split}
\end{equation}
The $\hbar$ dependence of $\bar\varepsilon$ and $\D$ is linear. However $\varepsilon(H=0)=\wh{\D_0}\sim e^{-S_0/\hbar}$~\eqref{eq:TLS} (we put wide hats on quantities taken at $H=0$) is exponentially small in $\hbar$, see~\eqref{eq:D0H0} for an approximation. Expanding $\varepsilon=\sqrt{\D^2+\D_0^2}$~\eqref{eq:HeffTLS} one has 
\begin{equation}\label{eq:Dhats}
 \varepsilon=\wh{\D_0}+\frac{H^2}{4}\argp{\frac{\wh{({\D_0^2})''}}{\wh{\D_0}}+\frac{\wh{({\D^2})''}}{\wh{\D_0}}}+O(H^4)
\end{equation}
Now notice that the tunneling amplitude $\D_0$~\eqref{eq:TLS}, after any $H$ derivation taken in $H=0$, retains an overall exponentially small factor $e^{-S_0/\hbar}$ ; other factors are mild. Thus exponentially speaking
$ \wh{({\D_0^2})''}\sim(\wh{\D_0})^2$. So, the quadratic term in $H$ in~\eqref{eq:Dhats} is dominated by the exponentially large ${\wh{({\D^2})''}}/{\wh{\D_0}}\sim1/{\wh{\D_0}}$. We conclude from~\eqref{eq:fEg},\eqref{eq:Egexpand},\eqref{eq:Dhats} that $E_g''(0)\sim-\varepsilon''(0)/2\sim-1/\wh{\D_0}$ is exponentially large and
\begin{equation}
f_\k'(1,H)/\b\sim -E_g'\underset{H\to0}{\sim}{H}/{\wh{\D_0} }
\end{equation}
Therefore $f_\k'(1,H)/\b$ goes from 0 at $H=0$ to a scale at large enough $H$ roughly given  by matching the HO approximation, $f_\k'(1,H)/\b\simeq-\argp{v_a+\hbar\om_a/2}'\underset{H\to0}{=}\sqrt{6\abs m}+\hbar\sqrt3/(8m)$ which is of order 1. This means that the $H$ scale over which $f_\k'(1,H)/\b$ vanishes is exponentially small and given by the tunnel splitting:
 \begin{equation}
  H\sim\wh{\D_0}\sim e^{-S_0/\hbar}=\varepsilon(H=0)
 \end{equation}
 This is confirmed numerically in Fig.~\ref{fig:quarticscaling}.
Note that the same analysis  holds for $\varepsilon$, and explains that the quadratic regime $\varepsilon-\varepsilon(H=0)\propto H^2$ happens on this exponentially small scale, hence it is seen only at fairly large $\hbar$ in Fig.~\ref{fig:various_hbar}(top). Then it crosses over to a linear regime $\varepsilon\propto H$ on a much larger linear scale in $\hbar$, up to  $H\simeq H_c(m)\propto\hbar$, to eventually reach  the HO value $\varepsilon\sim\hbar\om_a$. This explains why the quadratic regime is irrelevant for the tunneling two-level system contribution to the specific heat, which is dominated by the much larger portion of the $(m,H)$ plane where the linear regime $\varepsilon\propto H$ is valid.

\subsection{Tunneling two-level systems imply a pseudogap at zero temperature for $\hbar\to 0$}

We now have everything at hand to show that singularities in the marginal stability condition~\eqref{eq:frsbrep} for $T=0$ and $\hbar\to0$ are avoided due to a pseudogap in the field distribution for $m\lesssim-\hbar^{2/3}$, leading to the possibility of verifying this condition and thus the persistence of a fullRSB phase  when lowering $\hbar$. 
Indeed,  in this limit for $m<0$, $f''_\k(1,H)$ takes the classical value given by the regular part of~\eqref{eq:classdxadH} everywhere, except in a small region scaling with $\hbar$ close to $H=0$. This regular part can be safely extended for $\hbar\to0$ to $H=0$, thus the integral becomes $\int\dd p(\tilde a)/\tilde a^2 $ in the notations of~\secref{sec:SGLD} ($\tilde a=v_m''(x_a)$). This is convergent even in $(m,H)\to(0,0)$ where $x_a=0$, owing to $p(\tilde a)\sim \tilde a^{3/2}$, see~\secref{sub:lowMat}~\cite{BLRUZ21,FU22}. Yet for the $m<0$ region close to $H=0$, we need to separate the quartic regime $ m\gtrsim -\hbar^{2/3}$ where $H\propto\hbar$ from the TTLS-WKB regime $ m\lesssim -\hbar^{2/3}$ where $H\propto\wh\D_0(m)$. The replicon then reads
\begin{widetext}
 \begin{equation}\label{eq:Hzero}
 \begin{split}
  \l_R\propto&\, \overbrace{1-J^2\int_{0}^{m_M}\dd p(m)\int_{-\io}^\io\dd H\,P_\k(1,H)\argc{v_m''(x_a)}^{-2}}^{\l_R^{\rm regular}>0}
    -\hbar^{1/3}J^2\int_{-1}^0\dd p(\tilde m)\int_{-\io}^\io\dd \tilde H\, P_\k(1,\hbar\tilde H)\argc{\partial_{\tilde H}^2\wt E_g(\tilde H,\tilde m)}^2\\
 &-J^2\int_{m_m}^{-\hbar^{2/3}}\dd p(m)\int_{-\io}^\io\dd \hat H\,\frac{ P_\k(1,\wh\D_0(m)\hat H)}{\wh\D_0(m)}\argc{\partial_{\hat H} \bra{-}\hat x\ket{-}}^2
 \end{split}
\end{equation}
\end{widetext}
The first contribution to the replicon is the regular part of the classical replicon for positive $m$. In the quartic region we factor out the $\hbar$ dependence using the quartic scalings of~\eqref{eq:Egscal}, \ie $\tilde m=m/\hbar^{2/3}$, $\tilde H=H/\hbar$. Note that $\wt E_g$ is a smooth function. This quartic contribution vanishes as $\hbar^{1/3}$ in the classical limit, and there is no need for a regularization mechanism in the field distribution at these scales. 
Instead in the TTLS-WKB domain, we use the scaling variable $\hat H =H/\wh\D_0(m)$. The divergence that gives birth to the singular part of the classical equation~\eqref{eq:classdxadH} manifests itself in the factor $1/\wh\D_0(m)$ inside the integrand, quickly diverging when $\hbar\to0$. To balance it, we need $P_\k(1,0)$ to vanish and postulate a pseudogap for $m\lesssim -\hbar^{2/3}$:
\begin{equation}
 P_\k(1,H)\underset{H\to0}{\propto}\abs{H}^\n
\end{equation}
Then the last integral of~\eqref{eq:Hzero} is proportional to $\wh\D_0(m)^{\n-1}$. If $\n>1$, this TTLS-WKB  integral vanishes for $\hbar\to0$ and the replicon is positive. This contradicts our hypothesis of a marginal stability transition; we thus need this $H\to0$ region to contribute so that $\l_R=0$. But if $\n<1$, this term diverges for $\hbar\to0$ and $\l_R\to-\io$ as in the classical RS limit~\cite{BLRUZ21,FU22}, recovering the brutal effect of the singularity of~\eqref{eq:classdxadH}. This cannot be and we conclude that the pseudogap must be linear, $\n=1$, which makes contact and agrees with the classical $T\to0$ analysis~\cite{FU22}. This pseudogap induces a crossover of the specific heat scaling in temperature, from the linear quantum one to the classical one. 

\subsection{Conclusion}

We have examined how taking into account fullRSB within the variational approximation of~\secref{sec:DWP}, \ie in the phase where TTLS appear, changes the conclusions of~\secref{sec:DWP}. We argued that at finite $\hbar$ this modification would enforce marginal stability while not impairing the linear scaling of the specific heat. This is due to the regularity of the field distribution, as opposed to the classical case where it is pseudogapped.  However in the limit $\hbar\to0$, if TTLS are present,  the same linear pseudogap as in the classical $T\to0$ regime is mandatory for  $m\lesssim -\hbar^{2/3}$ to preserve marginal stability. Indeed these excitations create a singularity in the marginal stability condition, regularized by such a pseudogap in the field distribution. This singularity extends over an exponentially small scale of the field $H$, given by the tunnel splitting.  This is the same small scale over which the TTLS gap $\varepsilon$ is not anymore linear in $H$ but quadratic, as found numerically in~\secref{sec:DWP}. The narrowness of this quadratic regime implies its irrelevance on the specific heat scaling. The specific heat then crosses over to its classical regime. 

\newpage

\section{Towards a consistent solution in the RSB-DWP phase}\label{sec:spinboson}

\subsection{Inconsistent imaginary-time dependence and many-body gap under the variational approximation}

In the previous sections~\ref{sec:DWP}-\ref{sec:pseudogap}, we have studied the RS-DWP phase through a variational approximation for the impurity problem, which captures the essential semiclassical physics of tunneling through potential barriers.
In the RS-DWP phase, the many-body gap should be finite in agreement with the variational approximation. For $T\to0$, we expect instead the many-body gap to vanish at the continuous RSB transition and within the RSB-DWP phase. The variational scheme is thus inconsistent as it predicts a zero-temperature continuous RSB transition without a vanishing energy gap. In this section we point out that this discrepancy must be resolved by a proper treatment of the non-local dynamical equations of the impurity problem, and mention a possible way via an analogy with the spin-boson model, as well as its consequences. 
We now work in the conventional time unit. 
%

An internal inconsistency of the RS variational approximation can be guessed in the time dependence of $G(t)$.  Here we  investigate it further, computing it self-consistently at low temperature within the variational approximation $\b G(t)=\c\d(t)$. 
Recall that $G(t)$ is given self-consistently by~\eqref{eq:G},\eqref{eq:varq}:
\begin{equation}\label{eq:selfG}
 G(t)=\int\dd p(\k)\dd H\, P_\k(1,H) \moy{x(t)x(0)}_c
\end{equation}
For $(m,H)$ outside the HO region, for $\b\to\io$ we truncate to the first two levels, one can then extract the correlation function\footnote{In the pseudospin representation, we approximate ${\hat x \approx\begin{pmatrix}
                                                         x_a &0\\0&x_s
                                                        \end{pmatrix}  }$. The prefactor of $e^{-\varepsilon\abs{t-s}/\hbar} $ in~\eqref{eq:zinngap} is then
\begin{equation}
 \abs{\bra{-}\hat x\ket+}^2=\frac{(x_a-x_s)^2\D_0^4}{4\argc{\D_0^2-\D(\varepsilon-\D)}^2}
\end{equation}
with $\ket{\pm}$ the ground and excited states of $\hat H_{\rm eff}$~\eqref{eq:HeffTLS}.
} defining $G(t)$: 
\begin{equation}\label{eq:zinngap}
\begin{split}
 \moy{x(t)x(s)}=&\frac{\Tr \, e^{-\b\hat H_{\rm eff}}  {\cal T}\hat x(t)\hat x(s)}{\Tr\,e^{-\b\hat H_{\rm eff}}}
\underset{\abs{t-s}\to\io}{\propto}\,e^{-\varepsilon\abs{t-s}/\hbar} 
\end{split}
\end{equation}
$\cal T$ is the time-ordering operator. 
In the low-temperature limit, as expected from general arguments in presence of a gap~\cite[Sec.2.4]{Zinn-Justin}, one recovers an exponential relaxation for the TTLS part of $G(t)$ due to the non-zero gap $\varepsilon$. 
Now for $(m,H)$ in the HO region dominated by a single well, there is a finite $O(\hbar)$ gap, in the quartic region where it is $O(\hbar^{4/3})$\footnote{In~\secref{sec:SGLD}, within the semiclassical expansion this quartic gap goes to zero, in agreement with the gapless Debye frequency scalings of~\secref{sub:lowMat}.}. 
 Putting both two-level and HO regions together, $G(t)$ is obtained by an integral over exponential decays with a featureless weight $P_\k(1,H)$. 
 
 This result differs from the original Dirac delta~\eqref{eq:varapprox}. Worst, at the $T\to0$ RSB transition the replicon should vanish with a concomitant power-law-tailed  $G(t)$ (critical scaling)~\cite{SZ81,sachdev,CM22}, as happens in~\secref{sub:lowMat} when approaching the RSB transitions.
 This calls for a proper solution to the imaginary-time problem. 
 If this impurity problem is properly handled, then the RS or RSB  off-diagonal structure of the overlap order parameter is inferred from the replica equations requiring the dynamical content as an input (see~\secref{sub:fRSB}). These equations are  readily solved in the RS case or efficiently computed numerically from the partial differential equations in the RSB case. 

\subsection{Approximate mapping to the spin-boson model}

For a fully consistent solution of the model, the crucial improvement over the variational approximation is thus to tackle the non-locality of the dynamical impurity problem. In particular, one needs to compute the effective partition function $f_\k(1,H)$ and connected correlation functions in imaginary time (to get $G(t)$~\eqref{eq:selfG} or the self-energy). 
In the spirit of the previous sections, one could again resort to compute the saddle-point trajectories (instantons) from~\eqref{eq:spsingle} (written in the RS assumption, but extendable almost verbatim to the RSB phase), which are time dependent, resum all instanton contributions and perform an asymptotic expansion around these saddle points as in~\secref{sec:SGLD}. A kink trajectory retaining the time non-locality of~\eqref{eq:spsingle} does not have an analytic expression, which makes this strategy a difficult task. 
In the following, for generality we consider the fullRSB equations although the focus is on the dynamical impurity problem. 

Let us consider the problem from a different angle. Fixing $G(t)$, $\AA[x]$ in~\eqref{eq:actionAfull} describes the action of a single degree of freedom with non-Markovian self-interaction. This is very reminiscent of the action of a particle linearly coupled to a bath of oscillators~\cite{BP02}. We can unfold the action $\AA[x]$ \`a la Caldeira-Leggett~\cite{CL81,CL83a}, its Hamiltonian reading
\begin{equation}\label{eq:spinbosonH}
 \wt H =\frac{\hat p^2}{2M}+v_\k(\hat x)+\sum_\a\frac{\hat p_\a^2}{2m_\a}+\frac{m_\a\om_\a^2}{2}\hat x_\a^2-c_\a \hat x_\a \hat x
\end{equation}
Writing the Feynman path integral for the partition function $\wt Z=\Tr\,e^{-\b\wt H}$, we get the dynamical input of the replica equations:
\begin{equation}\label{eq:sbdefs1}
 \begin{split}
 f_\k(1,H)=&\ln\oint \mathrm{D}x\, e^{\AA[x]}=\ln\argp{\wt Z/Z_{\rm bath}}\ ,\\
  Z_{\rm bath}=&\prod_\a\frac{1}{2\sinh\argp{\b\hbar\om_\a/2}}
 \end{split}
\end{equation}
$Z_{\rm bath}$ is the partition function of the collection of harmonic oscillators alone. The coupling to the bath brings a renormalization of the quadratic term of the potential~\cite{CL83a,LCDFGZ87,weiss}, which is apparent when writing the action in terms of the self-energy~\eqref{eq:selfdef}\footnote{We could have equivalently considered the Hamiltonian in the counter-term form:
\begin{equation}\label{eq:Htilde}
 \wt H =\frac{\hat p^2}{2M}+v_m(\hat x)+\sum_\a\frac{\hat p_\a^2}{2m_\a}+\frac{m_\a\om_\a^2}{2}\argp{\hat x_\a-\frac{c_\a}{m_\a\om_\a^2}\hat x}^2
\end{equation}}
\begin{equation}\label{eq:actionI}
  \begin{split}
  \AA[x]=&-\frac1\hbar\int_0^{\b\hbar}\dd t \argc{\frac M2 {\dot x}^2+v_m(x)}\\
  &- \frac{1}{2\b\hbar^2}\int_0^{\b\hbar}\dd t\dd s\, x(t) I(t-s)x(s)
  \end{split}
 \end{equation}
 with   $m=\k-J^2\c$ the renormalized quadratic coefficient. Defining the bath spectral function~\cite{weiss}
 \begin{equation}\label{eq:defJ}
 \wt J(\om)=\frac\p2 \sum_\a\frac{c_\a^2}{m_\a\om_\a}\d(\om-\om_\a)\ , 
 \end{equation}
the KHGPS quantities are then identified by the mapping:
\begin{equation}\label{eq:selfchi}
 \begin{split}
   J^2\c=&\frac2\p \int_0^\io\dd\om\frac{\wt J(\om)}{\om}\\
   \wt I(\om_n)=&\frac2\p \om_n^2\int_0^\io\frac{\dd\om}{\om}\frac{\wt J(\om)}{\om_n^2+\om^2}
 \end{split}
\end{equation}
 In the spin-boson model, the bath spectral function is assumed gapless from the outset: $\wt J(\om)\underset{\om\to0}{\sim}\om^s$. In the KHGPS model,  as a consequence of~\eqref{eq:selfchi}, this scaling corresponds to a critical RSB transition where the self-energy $\wt I(\om)\underset{\om\to0}{\sim}\om^s$ may become non-analytic at zero frequency. For this reason this `mapping' may be well suited to study the RSB transition. 

A possible program for a solution through this framework is to start from a guess (given \eg by the  variational approximation of~\secref{sec:DWP}) for the susceptibility $\c$ and replica parameters $q(x)$. An ansatz for the self-energy $I$ could be provided by setting 
\begin{equation}\label{eq:Jansatz}
 \wt J(\om)=\CC(T)\om^se^{-\om/\om_{\rm cut}}
\end{equation}
 Then one has to enforce the self-consistent equations (such as the replica partial differential equations, ~\eqref{eq:frsbq},\eqref{eq:frsbc} and~\eqref{eq:selfG}) and obtain their fixed point for all these parameters. For the self-energy this amounts to match~\eqref{eq:selfG} and~\eqref{eq:selfchi}. As the replica equations are coupled, an iterative way to get the fixed point is to update the  values at each iteration, compute the dynamical functions (correlations, $f_\k(1,H)$ and its  $H$ derivatives, see~\eqref{eq:frsbq},\eqref{eq:frsbc})  and repeat until convergence, as in~\secref{sec:num}. In doing so, one has to average over $(\k,H)$ dynamical averages (for $q(x)$, $\c$ or $\wt I(\om)$) or to get the effective free energy (for $f_\k(1,H)$). In the case where the single-particle potential $v_m(x)$ is a SWP, one can put to work methods from the  damped harmonic oscillator problem~\cite[Chap.6]{weiss} to get these dynamical observables. Similarly, if the potential is a DWP, one truncates first the model~\eqref{eq:spinbosonH},\eqref{eq:actionI} to the spin-boson one~\cite{LCDFGZ87,weiss} (up to a constant energy)
\begin{equation}\label{eq:trunc}
 \wt H\sim \frac{\D_0}{2}\hat \s_x+\frac{\D}{2}\hat \s_z-\frac12\hat \s_z\sum_\a\hbar\l_\a(\hat b_\a+\hat b_\a^\dagger)+\sum_\a\hbar\om_\a \hat b_\a^\dagger \hat b_\a
\end{equation}
using $\hat x_\a=\sqrt{\frac{\hbar}{2m_\a\om_\a}}(\hat b_\a+\hat b_\a^\dagger)$, $\hat p_\a=i\sqrt{\frac{m_\a\om_\a\hbar}{2}}(\hat b_\a^\dagger-\hat b_\a)$ and $\l_\a=x_a c_\a\sqrt{\frac{2}{\hbar m_\a\om_\a}}$. $\hat b_\a^\dagger$, $\hat b_\a$ are usual bosonic creation/annihilation operators $\argc{\hat b_\a,\hat b_\g}=0$, $\argc{\hat b_\a,\hat b_\g^\dagger}=\d_{\a\g}$. Then one extracts dynamical observables using methods from the dissipative two-level system problem~\cite[Part.IV]{weiss}. Self-consistency of the dynamical equation~\eqref{eq:selfG} is a priori preserved in the crucial low-frequency limit, as in either harmonic or two-state dissipative quantum models,  the imaginary-time position correlation function decays as $\moy{x(t)x(0)}_c\sim t^{-(s+1)}$, matching the above $\om^s$ low-frequency behavior of $\wt I(\om)$ and $\wt J(\om)$. Regarding the effective free energy $f_\k(1,H)$, note that in the above partition function~\eqref{eq:sbdefs1}, one would have to subtract the contribution of the fictive bath $Z_{\rm bath}$.

This mapping is approximate, as unlike the KHGPS model the spin-boson model has no self-consistent structure and takes the bath spectral functions as given (corresponding to the critical regime in the KHGPS model). Here one has to enforce the self-consistent equations to determine all quantities, including the bath spectral function. Note in addition that the latter function may contain a temperature dependence (beyond the one incorporated by the Matsubara frequencies $\om_n=2\p n/(\b\hbar)$). This is absent from the spin-boson model and may affect some scalings. To take this into account we proposed the ansatz~\eqref{eq:Jansatz} with an extra temperature dependence in the prefactor. 
Nonetheless note that in the results of~\secref{sec:SGLD}, where only SWP are present, there is no such explicit $T$ dependence in the self-energy at first semiclassical order. 

The marginal condition~\eqref{eq:replicon0} provides the glass transition line, \ie for $T=0$ the quantum critical point $\hbar_c$ discussed in~\secref{sec:num}.  Finally, the specific heat should be revealed by, from~\eqref{eq:UfullRSB}:\\
\textit{(i)} computing the extra interaction terms (first line) and comparing it to the other term \textit{(ii)}, as was done in the variational approximation in~\secref{sub:gapCv} \\
\textit{(ii)} an average over $(\k,H)$ of the impurity specific heat provided by the spin-boson model (second line). Only $(\k,H)$ regions with the dominant specific heat contributes to the critical scaling, \ie only the dominant quantum excitations should be taken into account.  
For instance, expecting TTLS to yield the dominant contribution, in the dissipative model~\eqref{eq:spinbosonH} their specific heat scales as $C_V\sim T^s$~\cite{weiss,GW88} in the symmetric DWP region, 
meaning that a linear specific heat corresponds to the Ohmic case $s=1$. 
Therefore, taking into account the non-locality in time of the problem induces the vanishing of the gap as expected. 

Interestingly, at zero temperature,  if $s<1$ (sub-Ohmic case), or at strong enough system-bath coupling (here embodied by the prefactor of the bath spectral function or the self-energy) in the Ohmic case $s=1$, the two-level system gap $\D_0=0$ vanishes. This corresponds to a dissipation-induced suppression of tunneling at $T=0$, \ie the localization phenomenon first found by Chakravarty and Bray \& Moore~\cite{Ch82,BM82}. This is a polaronic effect coming from dressing of the tunneling amplitude by the coupling to high-frequency bath modes  (known as adiabatic renormalization~\cite{Ch82,BM82,LCDFGZ87,weiss})\footnote{Wegner flow equations~\cite[Sec.18.1]{weiss} supply a self-consistent equation for the renormalized tunneling amplitude $\D_0^r$ :
 \begin{equation}\label{eq:Wegner}
  \D_0^r=\D_0\exp\argp{-\frac{x_a^2}{2\p\hbar}\int_0^\io\dd\om\,\frac{\wt J(\om)}{\om^2-(\D_0^r)^2}}
 \end{equation}
 with $\D_0\propto e^{-S_0/\hbar}$ the bare tunneling amplitude (see~\eqref{eq:TLS},\eqref{eq:kinkaction} and footnote~\footref{footinst}), which can be calculated through instanton or WKB methods. For the sub-Ohmic case, $\D_0^r=0$ comes consistently from the divergence of the frequency integral, whereas it remains finite in the super-Ohmic case.}. In the variational approach of~\secref{sec:DWP}, gaplessness is missing due to the non-vanishing of this single-particle tunneling amplitude, which is resolved by this dissipation-induced localization. Yet it is not the only route to a vanishing gap of the many-body system: if $s>1$ (super-Ohmic case), the effective two-level system is tunneling and has a finite gap $\D_0>0$. Here the criticality of the bath is enough to induce the one of the whole (system \textit{and} bath). Such a program is left for future work.

\section{Concluding discussion}\label{sec:conclu}

The standard tunneling model (STM)~\cite{AHV72,Ph72} (or its extension as the Soft Potential Model, SPM), relying primarily on TTLS excitations, successfully accounts for low-temperature properties of real glasses around $1$ K, except for essential questions that remain to clarify, such as the physical nature of tunneling entities and the role of interactions. Mean-field models are more amenable  to an analytic solution and critical scalings are usually sharper than in finite dimension. The main motivation of this work is the curiosity that the relevance of TTLS in mean-field quantum glasses is far from established, whereas it is a pivotal concept in finite-dimensional ones. Here we studied the low-temperature quantum thermodynamics in the KHGPS model, a mean-field model whose low-energy excitations are similar to the ones of actual glasses.  

Because the problem remains difficult to solve analytically, following earlier works~\cite{SGLD04,SGLD05,S05,FMPS19} we chose as angle of attack to perform a semiclassical limit of $\hbar\to0$ with $\b\hbar$ fixed. 
We understood that this scheme technically corresponds to a standard loop expansion around the semiclassical saddle-point trajectories, which helps to grasp the prevailing low-energy excitations.  If this trajectory is trivial (static classical zero-temperature solution), the underlying physics is ruled to first order by a Debye approximation on a disordered energy landscape. Instead the trajectories may be instantons, which connects with different non-perturbative physics (here of tunneling in double-well potentials). 

In the phase where the classical energy landscape is convex (replica-symmetric phase at $T=0$), the specific heat exhibits a gapped scaling. Approaching the glass phases where replica symmetry gets broken, the many-body energy gap vanishes and a critical power-law scaling of the specific heat is obtained. This scaling depends on the underlying dominant physical excitations of the system. When the mean-field impurity problem contains only a collection of single-well potentials, as in the non-interacting model, the scaling is cubic while when the interaction between the bare degrees of freedom (harmonic oscillators in a single well) is on the verge of creating effective double-well potentials, the scaling is quintic. This mirrors the predominant harmonic excitations and the nature of their instability towards replica-symmetry breaking. 
Beyond the transition line where the scaling is cubic, one finds a marginal spin-glass phase where replica symmetry is continuously broken, hosting only single-well potentials in the impurity problem (RSB-SWP phase). The semiclassical expansion there hints at the same disordered Debye behavior as on the transition line at the level of the self-energy, yet the cancellation of non-Debye terms in the specific heat could not be entirely resolved. This is a similar situation to the fullRSB phase of the quantum spherical perceptron~\cite{FMPS19}. 

We identified a generic argument for such exponents, which corresponds to a disordered version of the Debye approximation well known for crystals. It rationalizes previous perturbative results from such a semiclassical scheme~\cite{FMPS19}. Namely, the mechanism is criticality (implying a gapless phase, here provoked by the GPS instability) combined with random matrix theory for the density of classical vibrational modes of the system. It generalizes (somewhat more physically) the SGLD argument previously put forward in the literature~\cite{SGLD04,SGLD05,S05} in which only the vanishing of the replicon eigenvalue would determine the scaling of the specific heat. This left out other cases, such as the quintic scaling above where the spin-glass susceptibility is finite at the transition (finite replicon), or the case of jamming where the replicon cancels but the first perturbative order yields a different (linear) Debye scaling~\cite{FMPS19}, or even cases where this framework becomes non perturbative, here in presence of two-level tunneling impurities. 

Indeed, we next studied the physically most interesting phase where effective double-well potentials start populating the impurity problem (RSB-DWP phase), through the same semiclassical strategy. There, the semiclassical expansion looses its perturbative nature as  non-Debye excitations appear due to the double wells. Instantons need to be considered in the dynamical impurity problem, contrary to the Debye case. Through a variational approximation designed to retain this instantonic character, we investigated both numerically and analytically the phase diagram and the specific heat for $T\to0$. Tuning $\hbar$ one goes from a RSB phase (small enough $\hbar$) to a RS phase (high enough $\hbar$) separated by a quantum critical point. At low temperature, TTLS dominate the inner RSB phase, bringing a linear scaling of the specific heat, while they are absent in the RS phase or close to the quantum critical point where harmonic and quartic modes prevail. In the inner RSB phase, Debye quintic scaling  is then seen only as a crossover at higher temperature. The global physical picture is similar to the one of the phenomenological soft-potential model, in which a mixture of soft harmonic excitations and of TTLS determine the behavior of the specific heat. It points towards the emergence of TTLS inside the marginal quantum glass phase, a different scenario from the ones displayed by other mean-field quantum glass models. We showed robustness of the linear specific heat scaling in the RSB phase within the variational approximation, by studying the marginal stability condition and the emergence of a pseudogap in the field distribution for $\hbar\to0$. Overall the quantum regime of the KHGPS model within a `local' variational approximation shows faithful mean-field transposition of the SPM designed for the phenomenology of actual low-temperature glasses, connecting these  phenomenological concepts to mean-field glass theory ones. Although the low-temperature expansion of the specific heat displays some robust features, our variational approximation is not a fully consistent solution, which may ultimately jeopardize the obtained linear scaling.  We finally proposed a way to fix its main inconsistence through a fuller solution of the dynamical impurity problem, by considering the similarity of the Matsubara impurity action with the one of the spin-boson model. This would give a scenario for the vanishing of the many-body gap at the quantum critical point, missed by the variational approximation, either through super-Ohmic criticality of the effective bath provided by the system itself, or by bath-induced quantum localization of the TTLS degrees of freedom. 

The main progresses from the present model in the approach of solving mean-field models bearing more proximity with low-temperature excitations in actual glasses, pioneered by K\"uhn and collaborators, concern \textit{(i)} the classical spectrum of the vibrational modes exhibiting the universal low-frequency scaling of the density of modes $D(\om)\sim\om^4$ seen in finite dimensions, concomitant with a finite spin-glass susceptibility at $T=0$ \textit{(ii)} a physical understanding of the approach to the quantum marginally stable phases and the dominant low-energy excitations \textit{(iii)} a direct study of the quantum thermodynamics. Regarding this last point, previous works by K\"uhn and collaborators considered either \textit{(i)} the total thermodynamic energy as only the part consisting in the effective single-particle problem resulting from mean-field decoupling, quantized a posteriori~\cite{KH97,Ku00,Ku03}, \textit{(ii)} the first-principles quantum thermodynamic energy within the usual static approximation~\cite{BM80b,CGSS01}, \textit{(iii)} further field-theoretical loop approximations (in the order parameter $Q_{ab}(t)$)~\cite{BK10}. 
\textit{(i)} is unsatisfying from a first-principle viewpoint. For \textit{(ii)}, we note that in the KHGPS model we avoided this static approximation as it would wipe out the emergence of double-well potentials in the system\footnote{Indeed in the static approximation, $G(t)\to q_d-q$ which makes the non-local term of the impurity action only renormalize the single-particle linear (field) term, after Hubbard-Stratonovitch uncoupling. The result is only a slight change of the $H$ field distribution, while the instability of the single wells is driven by the distribution of the quadratic term.}, and relied instead on a variational method. Both the usual static and the present variational approximations cannot truly address the imaginary-time impurity problem, essential for critical scalings. \textit{(iii)}, at variance, tackles the field theory, though with mitigated results and without considering crucial effects of RSB marginality.  

In phenomenological models, the ground-state energy cannot be predicted. The focus there is on the energy gaps that are enough to compute \eg the specific heat. A related basic assumption is the behavior of the distribution of TTLS parameters $P(\D,\D_0)$.  In the STM one has that $P(\D,\l)=$ constant, where $\l=e^{-\D_0}$, meaning that $P(\D,\D_0)\propto1/\D_0$. 
In the KHGPS model, we correspondingly get\footnote{
 Simplifying in the limit $H\to0$ relevant to low temperatures:
\begin{equation}
  \D\underset{H\to0}{=}-\argc{2\sqrt{6\abs m}+\frac{\sqrt3}{4}\frac\hbar m }  H+O(H^3)
\end{equation}
and $\D_0$ given at first order by~\eqref{eq:TLS}. This is enough to compute the Jacobian at small $H$ for the change of variables $(m,H)\to(\D,\D_0)$.}:
\begin{equation}
\begin{split}
  P(\D,\l)=&\D_0P(\D,\D_0)\\
  \underset{H\to0}{\sim}&\frac{P_\k(1,H)}{\k_M-\k_m}\frac{\D_0(m,H)}{\abs{\frac{\partial\D_0}{\partial m}(m,H)\argc{2\sqrt{6\abs m}+\frac{\sqrt3}{4}\frac\hbar m }}}
\end{split}
\end{equation}
which is valid only close to $H=0$ within the variational approximation. $\D_0$ is given by~\eqref{eq:TLS} and approximately given by~\eqref{eq:D0H0} for $H=0$. This is not constant in terms of $(\D,\l)$, as in the SPM~\cite[Chap.9]{Esquinazi} and in K\"uhn's work~\cite{Ku03}. TTLS get slightly depleted as $m$ decreases and more strongly so in the classical limit $\hbar\to0$ in the vicinity of the marginal glassy phase, due to the pseudogap in the field distribution $P_\k(1,H)$ at small $H$. Such a depletion was indeed anticipated by K\"uhn as an effect of replica symmetry breaking in~\cite{Ku03}. 

The quantum KHGPS model may behave differently from the quantum Sherrington-Kirkpatrick model  in a transverse field (TFSK). Indeed, naively, the former may be thought as a soft-spin version of the latter,  composed of interacting two-level systems (Pauli spins $\s^\m$) from the outset.  The replicated free energy is formally identical to the KHGPS one~\eqref{eq:FAio} and the impurity problem is ruled by a similar action to the one studied here, except that the double wells are `hard' (quantum half spins):
\begin{equation}
 \begin{split}
  \SS_{\rm eff}(h)&=\frac{J^2}{ 2}\int_0^\b\dd t\dd t'\,\hat\s^z(t)G(t-t')\hat\s^z(t')\\
  &+\int_0^\b\dd t\,\argc{h\hat\s^z(t)+\G\hat\s^{x}(t)}\\
  G(t-t')&=\int\dd h\, P(1,h)\la\TT\hat\s^z(t)\hat\s^z(t')\ra_{ \SS_{\rm eff}(h)}-q(1)
 \end{split}
\end{equation}
where $\G$ is the transverse field, $J$ the disordered coupling strength and $\hbar=1$. Notice that the time-local part of the action directly represents a TTLS with the longitudinal field as the only fluctuating variable due to interactions, instead of a disordered collection anharmonic oscillators in the KHGPS model where both the field $H$ and the elastic constant $m$ are fluctuating. 
Nevertheless, in the TFSK, both close to the quantum critical point or in the marginal glass phase, previous studies found no trace~\cite{RSY95,AM12,TSS23,KZL24} of the TTLS behavior universal in finite-dimensional glasses (same situation in the related SU$(2)$ Heisenberg spin glass~\cite{KMGP23}). Following~\cite{AM12}'s arguments, its specific heat should read  $C_V\sim T^3$, governed only by disordered Debye excitations with a density of states $D(\om)\sim\om^2$, as in the marginal glass phase of most quantum mean-field models~\cite{S05,FMPS19}. This is  analogous to the physical picture we advocate in the replica-symmetric or RSB-SWP phases of the KHGPS model. The TFSK was not part of the models analyzed by Schehr, Giamarchi and Le Doussal. It would then certainly be beneficial to perform a similar semiclassical expansion in order to make contact with the Debye interpretation of the collective low-energy modes in the TFSK~\cite{AM12,CM22}. 
 The transverse field hybridizes low-energy classical states, giving rise to collective (delocalized) spin waves~\cite{AM12}. In contrast, within the variational approximation, it seems that the single- and double-well degrees of freedom in the KHGPS model, created by virtue of the interactions between the microscopic single-site potentials, do not hybridize once dressed with the quantized momentum. The system behaves as a collection of independent such degrees of freedom, echoing the original STM and the SPM.  Among them, the double wells corresponding to TTLS degrees of freedom thus maintain their TTLS nature. Whether this is a side effect of the variational approximation, of `single-particle' character, remains to be understood. However note that such locality is also found classically in \eg the localized eigenvectors corresponding to the soft modes of the Hessian of the classical Hamiltonian in minima~\cite{BLRUZ21}, related to the softest single-well potentials. They  appear through the GPS instability that produces the double wells at the origin of the TTLS. Together with the avoided classical pseudogap in the field distribution $P_\k(1,H)$ for finite quantum fluctuations, these TTLS would be responsible for the linear specific heat. A proper solution, validating or not this scenario, remains to be found; we described a potential fuller solution in~\secref{sec:spinboson}.

We proved validity of the disordered version of Debye approximation at first order in the semiclassical loop expansion in the replica-symmetric phase. An important perspective would be to adapt the arguments of Schehr~\cite{S05} for higher orders, which in particular leave untouched the power-law scaling but only  renormalize perturbatively the prefactors, to the general case of mean-field models described by a self-consistent impurity problem (such as the present one and~\cite{FMPS19}). In the RSB-DWP phase, an analytical solution through the approximate spin-boson mapping of~\secref{sec:spinboson} could be very interesting to better understand the quantum critical point and the (possibly suppressed) tunneling behavior in the marginal phase. Quantum Monte Carlo algorithms may be complementarily much welcome for the comprehension of the quantum phase diagram and the low-temperature behavior of thermodynamic observables such as the specific heat, although we expect  accessing low temperatures to be challenging.

\begin{acknowledgments}
I thank Laura Foini, Giampaolo Folena, Nikita Kavokine, Markus M\"uller and Marco Schir\'o for useful discussions, as well as Sebastian P\"ogel for his help on the Mathematica cluster of IPhT. I warmly thank Pierfrancesco Urbani for sparking my interest in the KHGPS model, important discussions and hints throughout this work.
\end{acknowledgments}

\bibliographystyle{mioaps}
\bibliography{HSMBL}

\end{document}